\documentclass[superscriptaddress,twocolumn,showpacs,
amssymb,amsmath,nobibnotes,aps,prd,
nofootinbib]{revtex4-1}
\pdfoutput=1
\usepackage{graphicx,subfigure,bm,color,psfrag,hyperref}
\usepackage{amsfonts}
\usepackage{lipsum}
\usepackage{mathtools}
\usepackage{verbatim}
\newtheorem{remark}{Remark}[section]
\usepackage[normalem]{ulem}
\usepackage[dvipsnames]{xcolor}

\definecolor{blue}{rgb}{0.36, 0.54, 0.66}
\definecolor{amaranth}{rgb}{0.9, 0.17, 0.31}
\definecolor{pink}{rgb}{0.87, 0.56, 0.81}
\definecolor{ao}{rgb}{0.0, 0.5, 0.0}
\definecolor{maroon}{rgb}{0.76, 0.13, 0.28}
\definecolor{cardinal}{rgb}{0.77, 0.12, 0.23}
\definecolor{lightcardinal}{rgb}{0.97, 0.42, 0.53}
\definecolor{frenchlila}{rgb}{0.53, 0.38, 0.56}
\definecolor{yellow}{rgb}{1.0, 1.0, 0.87}
\definecolor{lightseagreen}{rgb}{0.7, 0.92, 0.68}
\definecolor{gray}{rgb}{0.9, 0.9, 0.9}
\definecolor{lightblue}{rgb}{0.66, 0.84, 0.96}

\hypersetup{colorlinks,linkcolor={maroon},citecolor={magenta},urlcolor={ao}}

\begin{document}

\title{Gravitational reheating formulas and bounds in oscillating backgrounds}

\author{Jaume de Haro}
\email{jaime.haro@upc.edu}
\affiliation{Departament de Matem\`atiques, Universitat Polit\`ecnica de Catalunya, Diagonal 647, 08028 Barcelona, Spain}

\author{Llibert Arest\'e Sal\'o}
\email{l.arestesalo@qmul.uc.edu}
\affiliation{School of Mathematical Sciences, Queen Mary University of London,
Mile End Road, London, E1 4NS, United Kingdom}

\author{Supriya Pan}
\email{supriya.maths@presiuniv.ac.in}
\affiliation{Department of Mathematics, Presidency University, 86/1 College Street,  Kolkata 700073, India}
\affiliation{Institute of Systems Science, Durban University of Technology, PO Box 1334, Durban 4000, Republic of South Africa}

\begin{abstract}

In this article we calculate the reheating temperature in the cosmological scenarios where heavy scalar particles are gravitationally produced, due to a conformally coupled interaction between a massive scalar quantum field and the Ricci scalar,  during the oscillations of the inflaton field. We explore two distinct cases, namely the one in which these particles decay during the domination of the inflaton's energy density and the other one where the decay occurs after this phase. For each scenario, we have derived formulas to calculate the reheating temperatures based on the energy density of the produced particles and their decay rate. We establish bounds for the maximum reheating temperature, defined as the temperature reached by the universe when the decay of gravitationally produced particles concludes at the onset of the radiation-dominated epoch.  Finally, we use the Born approximation to find analytic formulas for the reheating temperature.

\end{abstract}

\vspace{0.5cm}

\pacs{04.20.-q, 98.80.Jk, 98.80.Bp}

\maketitle

\section{Introduction}

According to the theory of cosmological inflation, our Universe underwent a phase of accelerated expansion for a very short period of time driven by some scalar field known as ``inflaton''~\cite{Guth:1980zm,Sato:1980yn,Linde:1981mu}\footnote{It should be noted that in 1980, Starobinsky (sadly he passed away recently) proposed a specific inflationary model widely known as the $R^2$ inflation (where $R$ denotes the
Ricci scalar) \cite{Starobinsky:1980te}, however, unlike Guth's paper \cite{Guth:1980zm}, the notable qualities of inflation were not directly indicated in his seminal paper \cite{Starobinsky:1980te}. } (also see \cite{Barrow:1981pa,Lucchin:1984yf,Hawking:1984kj,Lyth:1984gv,Belinsky:1985zd,Khalfin:1986ui,Silk:1986vc,Mijic:1986iv,Burd:1988ss,Olive:1989nu,Ford:1989me,Adams:1990pn,Freese:1990rb,Wang:1991ww,Linde:1993cn,Barrow:1993hn,Barrow:1994nt,Vilenkin:1994pv,Peter:1994dx,Sasaki:1995aw,Barrow:1995xb,Lidsey:1995np,Parsons:1995ew,Liddle:1998wc,Guth:2000ka,Riotto:2002yw,Feinstein:2002aj,Boubekeur:2005zm,Conlon:2005jm,Ferraro:2006jd,Cheung:2007st,Baumann:2008bn,Pal:2009sd,Martin:2013nzq,Martin:2013tda,Sebastiani:2013eqa,Hamada:2014iga,Freese:2014nla,Rubio:2018ogq,Odintsov:2023weg,Giare:2023wzl,Jinno:2023bpc}). As the inflationary phase comes to an end, the Universe must transit from this state of accelerated expansion to a hot, radiation-dominated phase consistent with the Big Bang Nucleosynthesis (BBN) and the subsequent evolution would lead to the current observable Universe. This crucial transition is achieved through a process known as reheating~\cite{Kofman:1994rk,Kofman:1997yn,Felder:1998vq,Bassett:2005xm,Allahverdi:2010xz, Martin:2014nya}. In this context, gravitational reheating \cite{Ford:1986sy, Spokoiny:1993kt,Lankinen:2019ifa} is a compelling mechanism by which the interaction of a quantum field coupled with gravity  generates massive particles which decay into standard model (SM) particles and other forms of matter and radiation, thereby reheating the Universe. Unlike conventional reheating scenarios that involve specific couplings between the inflaton and other fields, gravitational reheating relies on the dynamics of the spacetime itself, making it a universal process that does not depend on the details of particle interactions.

The essential idea behind gravitational reheating is that, as the inflaton oscillates around the minimum of its potential after inflation, it generates an oscillating  time-varying gravitational field. This varying field can produce particles via the mechanism of gravitational particle production, 
i.e., at the Lagrangian level, there is an interaction term between a massive quantum field—in our case, a scalar field—and the Ricci scalar, which leads to the production of spin-$0$ particles. The energy density of these produced particles eventually needs to dominate over the energy density of the inflaton to ensure a successful reheating process \cite{Dorsch:2024nan}. This transition is critical for establishing the thermal bath required for the standard Big Bang model to proceed.

The efficacy of gravitational reheating depends significantly on the behavior of the inflaton field and its potential. For instance, the rate at which the energy density of the inflaton decreases plays a crucial role. If the inflaton's energy density decreases slower than that of radiation, it could dominate the energy budget of the Universe once again, leading to scenarios incompatible with the concordance model of cosmology.  To avoid this issue, it is often required that the energy density of the inflaton decreases faster than that of radiation, which imposes constraints on the effective Equation of State (EoS) parameter 
$w_{\rm eff}$ of the inflaton field. Specifically, for successful gravitational reheating, the effective EoS parameter must satisfy 
$w_{\rm eff}>1/3$,  implying that certain conditions on the inflaton potential need to be met, such as 
$n>2$ for a potential of the form $\varphi^{2n}$.

Gravitational reheating has been found particular relevance in the models of quintessential inflation~\cite{Peebles:1998qn, Chun:2009yu, Lankinen:2016ile,DeHaro:2017abf,AresteSalo:2017lkv,Hashiba:2018iff, Haro:2018zdb, deHaro:2019oki,Salo:2021vdv, deHaro:2022ukj,deHaro:2023gho},\footnote{We refer to an incomplete list of works in the category of quintessential inflation~\cite{Giovannini:1999bh,Peloso:1999dm,Kaganovich:2000fc,Yahiro:2001uh,Dimopoulos:2001ix,Giovannini:2003jw,Sami:2004xk,Rosenfeld:2005mt,Rosenfeld:2006hs,Neupane:2007mu,Bento:2008yx,Bento:2009zz,Hossain:2014coa,deHaro:2016hpl,deHaro:2016hsh,deHaro:2016ftq,deHaro:2016cdm,Geng:2017mic,Dimopoulos:2017zvq,Agarwal:2017wxo,Haro:2015ljc,Haro:2019peq,Ahmad:2019jbm,Akrami:2020zxw,AresteSalo:2021lmp,AresteSalo:2021wgb,deHaro:2021swo,Dimopoulos:2022rdp,Alho:2023pkl,Das:2023nmm,Inagaki:2023mxv,Giare:2024sdl} which have gathered notable attention since its introduction by Peebles and Vilenkin~\cite{Peebles:1998qn}. } where the Universe transits from inflation to a phase dominated by kinetic energy (kination), characterized by an effective EoS, 
$w_{\rm  eff}=1$. 
In this scenario, the energy density of the inflaton scales with the scale factor of the Friedmann-Lema\^{i}tre-Robertson-Walker universe as 
$a(t)^{-6}$, ensuring it diminishes rapidly and does not interfere with the radiation-dominated phase necessary for the subsequent thermal history of the Universe

In the present work, our main goal,  in the context of gravitational particle production, using the so-called 
{\it Bogoliubov approach} \cite{Kaneta2022},
is to derive analytic expressions for the  reheating temperature and establish  upper bounds for them. The key component in obtaining these formulas is the energy density of the produced particles at the end of inflation, which must be calculated numerically. For potentials that are quadratic near the minimum, analytic formulas have been derived from numerical calculations, depending on the mass of the produced particles and approximating this energy density \cite{Ema:2018ucl,Kolb:2023ydq}. Then, assuming that these formulas also apply to viable potentials $-$ those which behave like $\varphi^{2n}$ near the minimum, with $n>2$    $-$ we derive the analytic expression for the maximum reheating temperature.  Focusing on the conformally coupled case  and dealing with scalar particles, 
we carry out our numerical calculations for the specific case of 
$n=3$, obtaining essentially the same maximum reheating temperature as that provided by the formulas in \cite{Kolb:2023ydq}.
Finally, by applying the Born approximation, which leads to the {\it Boltzmann approach} in gravitational particle production
\cite{Kaneta2022,Clery2022,Kaneta2021, Kaneta2020}, 
we are able to obtain analytic formulas for the reheating temperature.

In summary, gravitational reheating provides a robust and model-independent mechanism to produce a hot thermal bath, setting the stage for the Universe's evolution post-inflation. Understanding and correctly implementing this process is crucial for any inflationary model to be consistent with the observed properties of the Universe.

The article is structured as follows. In section~\ref{sec-2} we offer a concise description of reheating through gravitational production of heavy particles. In section~\ref{sec-3} we present the bounds on the maximum reheating temperature. In section~\ref{sec-4},  we present our numerical calculations and apply them to determine the reheating temperature.
In section~\ref{sec-5}, 
we use the Born approximation to find 
the analytic formulas and
bounds on the reheating temperature.
Finally, in the last section~\ref{sec-summary}, we present our conclusions. 

Throughout  the manuscript we  use natural units, i.e., 
 $\hbar=c=k_B=1$,  and the reduced Planck's mass is denoted by $M_{\rm pl}\equiv \frac{1}{\sqrt{8\pi G}}\cong 2.44\times 10^{18}$ GeV.

\section{Reheating via gravitational production of heavy particles}
\label{sec-2}

Considering the flat Friedman-Lema\^{i}tre-Robertson-Walker (FLRW) space-time with $a(t)$ as the expansion scale factor of the universe,  
we begin by denoting $\rho_B(t)$ and $\langle\rho(t)\rangle$ as the energy densities of the background and  the scalar heavy confomally coupled  particles  produced gravitationally, respectively.
{Assuming that close to the minimum of the potential, which for simplicity we take $\varphi_{min}=0$, the potential behaves like $\varphi^{2n}$, for example the one studied in 
\cite{Drewes:2017fmn}
\begin{eqnarray}\label{potential}
    V_n(\varphi)=\lambda M_{\rm pl}^4\left(1- e^{-\sqrt{\frac{2}{3}}\frac{\varphi}{M_{\rm pl}}}\right)^{2n}, 
\end{eqnarray}
 where $\lambda$ is a dimensionless parameter which has to be computed from the power spectrum of scalar perturbations. 
Using the virial theorem, one can conclude that, when the inflaton oscillates, the effective EoS  parameter is given by $w_{\rm eff}=(n-1)/(n+1)$ \cite{Turner:1983he}.
Since after inflation, the energy density of the produced particles (or that of their decay products, e.g. light relativistic particles)  eventually has to dominate that of the inflaton for the Universe to become reheated, we must demand that the energy density of the inflaton decreases faster than that of radiation. This requires
$w_{\rm eff} > 1/3\Longrightarrow n>2$.
Otherwise, if the energy density of the inflaton decreases more slowly than that of radiation, it could dominate again during this period, which is incompatible with the concordance model.
Therefore, potentials appearing in  Starobinsky ($n=1$) or Higgs inflation \cite{Rubio:2018ogq}, cannot contain gravitational reheating as a mechanism to reheat the universe.

\begin{remark}
    It is important to realize that when the potential behaves like a quartic one, namely,     
    $\varphi^4$, i.e., $n=2$, the viability of the model is possible if the decay of the massive produced particles occurs well after the domination of the inflaton's energy density. In that case, both energy densities, as a function of the scale factor, $a (t)$,  decrease at the same rate, as     
 $a^{-4}(t)$, but the inflaton's energy density will never dominate in the future.

\end{remark}

During the oscillations of the inflaton, which  take place shortly after the end of inflation, prior to the decay of the produced particles, and denoting by ``END'' the end of inflation, on average,  the evolution of its  energy density is,
\begin{align}
    \rho_B(t)=\rho_{\rm B, END}\left( \frac{a_{\rm END}}{a(t)}\right)^{3(1+w_{\rm eff})}=\rho_{\rm B, END}\left( \frac{a_{\rm END}}{a(t)}\right)^{\frac{6n}{n+1}}.
    \end{align}

Dealing with the gravitational production of particles, we
consider
the Lagrangian density for a real massive scalar field interacting with the Ricci scalar
\cite{birrell_davies_1982}:
\begin{eqnarray}\label{Lagrangian-density}
    {\mathcal L}=\frac{1}{2}\sqrt{|g|}(g^{\mu\nu}\partial_{\mu}\phi
    \partial_{\nu}\phi-m_{\chi}^2\phi^2-\xi R\phi^2),\end{eqnarray}
where  $m_{\chi}$ is the mass of the field,  $R$  is the Ricci scalar and $\xi$ is the dimensionless coupling constant.  The Euler-Lagrange equation leads to the following dynamical equation 
\begin{eqnarray}\label{dynamical-eq-fields}
(\nabla^{\mu}\nabla_{\mu}+m_{\chi}^2+\xi R)\phi=0,
\end{eqnarray}
where $\nabla_{\mu}$ denotes the covariant derivative.
To facilitate the procedure, we consider the  conformally coupled case
$\xi=1/6$, and thus, the Lagrangian becomes  
\begin{eqnarray}\label{lagrangian}
    {\mathcal L}=\frac{a^2}{2}\left\{(\phi')^2-(\nabla\phi)^2-a^2\left[m^2_{\chi}+\frac{R}{6}\right]\phi^2
    \right\},
\end{eqnarray}
where  the prime denotes the derivative with respect to the conformal time $\eta$ and 
 $\nabla$ is the gradient operator. 
To simplify this expression we perform the change of variable $\phi=\chi/a$, obtaining
\begin{align}\label{lagrangian1}
    {\mathcal L}=\frac{1}{2}\bigg\{(\chi')^2-(\nabla \chi)^2-a^2m^2_{\chi}\chi^2\bigg\}-\frac{1}{2}\frac{d}{d\eta}\left(\chi^2\frac{a'}{a}\right),
\end{align}
and since the Lagrangian (\ref{lagrangian1}) contains a total derivative, it is dynamically equivalent to 
\begin{eqnarray}\label{lagrangian2}
    \bar{\mathcal L}=\frac{1}{2}\bigg\{(\chi')^2-(\nabla \chi)^2-a^2m^2_{\chi}\chi^2\bigg\},
\end{eqnarray}
thus,
from the Euler-Lagrange equations we get its dynamical equation
\begin{eqnarray}
    \chi''-\Delta \chi+ a^2m_{\chi}^2\chi=0,
\end{eqnarray}
where in Fourier space takes the form 
\begin{eqnarray}
    \chi_k''+\omega_k^2(\eta)\chi_k=0,
\end{eqnarray}
being $\omega_k(\eta)=\sqrt{k^2+m_{\chi}^2a^2(\eta)}$ the frequency of the $k$-mode. 
Next, we will expand the modes in the following way,
\begin{align}\label{zs}
\chi_{k}(\eta)= \alpha_k(\eta)\phi_{k,+}(\eta)+
\beta_k(\eta)\phi_{k,-}(\eta),
\end{align}
where $\alpha_k(\eta)$ and $\beta_k(\eta)$ are the time-dependent Bogoliubov coefficients and 
we have introduced the   positive ($+$) and negative ($-$)  frequency {\it adiabatic modes} 
\begin{eqnarray}
    \phi_{k,\pm}(\eta)=\frac{e^{\mp i\int^{\eta}_{\eta_i} \omega_k(\tau)d\tau}}{\sqrt{2\omega_k(\eta)}}.
    \end{eqnarray}

Imposing that the modes satisfy   the condition
\begin{align}\label{zs_1}
\chi_{k}'(\eta)= -i\omega_k(\eta)\left(\alpha_k(\eta)\phi_{k,+}(\eta)-
\beta_k(\eta)\phi_{k,-}(\eta)\right),
\end{align}
one can show that   the Bogoliubov coefficients 
are subjected to the differential equation
\cite{Zeldovich:1971mw}
\begin{eqnarray}\label{Bogoliubov}
\left\{ \begin{array}{ccc}
\dot{\alpha}_k(t) &=& \frac{\dot{\omega}_k(t)}{2\omega_k(t)}e^{-2i\int^{t} \frac{\omega_k(t)}{a(t) }dt}
\beta_k(t)\\
\dot{\beta}_k(t) &=& 
\frac{\dot{\omega}_k(t)}{2\omega_k(t)}e^{2i\int^{t}
\frac{\omega_k(t)}{a(t)}dt }
\alpha_k(t), \end{array}\right.
\end{eqnarray}
where   
an overdot denotes the derivative with respect to the cosmic time, and the following relation $|\alpha_k(t)|^2-|\beta_k(t)|^2=1$, holds.  
It is crucial to recognize that the $\beta$-Bogoliubov coefficients encapsulate both vacuum polarization effects and particle production. Shortly after the initiation of the oscillations, the polarization effects become negligible. This implies that when they stabilize at a value denoted by $\beta_k$, the coefficients only reflect the contribution of the produced particles \cite{Salo:2021vdv}, and its energy density, after the stabilization of the $\beta$-Bogoliubov coefficients, is given by
\cite{birrell_davies_1982}:
\begin{eqnarray}\label{vacuum-energy2}
\langle\rho(t)\rangle&&= \frac{1}{2\pi^2a^4(t)}\int_0^{\infty} k^2\omega_k(t)|\beta_k|^2 dk\nonumber \\&& \cong \frac{m_{\chi}}{2\pi^2a^3(t)}\int_0^{\infty} k^2|\beta_k|^2 dk.
\end{eqnarray}

After understanding the evolution of these energy densities before  the decay of the heavy particles into lighter ones, two distinct scenarios unfold: 
decay during and after the end of the inflaton's domination. Consequently, we will delve into both situations with  attention to detail.

\subsection{Decay before the end of the background domination}\label{decay_before}

Let $\Gamma$ be the decay rate of the heavy massive particles, 
and it is worth noting that the decay process ends when $\Gamma$ is of the same order as the Hubble rate. 
Denoting $\rho_{\rm B, \rm dec}$ and $\langle \rho_{\rm dec}\rangle$ as the energy density of the background and that of the produced particles at the end of the decay, respectively, after the decay, which we will assume nearly instantaneous and without a substantial drop of energy, the corresponding energy densities 
evolve as:
\begin{align}\label{afterdecay}    \rho_{\rm B}(t)=\rho_{\rm B, \rm dec}\left( \frac{a_{\rm dec}}{a(t)}\right)^{\frac{6n}{n+1}} ~\mbox{and}~
\langle \rho(t)\rangle= \langle \rho_{\rm dec}\rangle   \left( \frac{a_{\rm dec}}{a(t)}\right)^4,
\end{align}
because, after the decay,  the particles are currently  relativistic.

\begin{remark}
Note that a more detailed analysis  of the decay of massive particles into lighter ones $-$ see the Appendix-A $-$ requires solving the Boltzmann equations:
\begin{eqnarray}\label{Boltzmann}
    \frac{d \langle \rho(t)\rangle}{dt}+3H\langle \rho(t)\rangle=-\Gamma \langle \rho(t)\rangle\nonumber\\
    \frac{d \langle \rho_r(t)\rangle}{dt}+4H\langle \rho_r(t)\rangle=\Gamma \langle \rho(t) \rangle,   \end{eqnarray}
where $\langle \rho_r(t)\rangle$ denotes the energy density of the decay products, that is, of radiation.  Since $\langle \rho(t)\rangle=
\langle \rho_{\rm dec}\rangle
\left(a_{\rm dec}/a(t)\right)^3e^{-\Gamma t}$
is a solution of the first equation, inserting it into the second and,  using the method of variation of parameters,   one gets:
\begin{eqnarray}
\langle \rho_r(t)\rangle=\langle \rho_{\rm dec}\rangle
\left( \frac{a_{\rm dec}}{a(t)}\right)^4
\int_0^t\frac{a(s)}{a_{\rm dec}}\Gamma e^{-\Gamma s}ds,\end{eqnarray}
which we will integrate in the static case, i.e., by approximating $a(s) \cong a_{\rm dec}$ within the integral, and thus obtaining:
\begin{eqnarray}
\langle \rho_r(t)\rangle\cong\langle \rho_{\rm dec}\rangle
\left( \frac{a_{\rm dec}}{a(t)}\right)^4\left(1-e^{-\Gamma t}\right),\end{eqnarray}
which shows that for $t\gg 1/\Gamma$
 the energy density of radiation evolves as
in equation (\ref{afterdecay}).
    \end{remark}

Hence, given the virtually instantaneous nature of the thermalization process, the universe undergoes reheating at the conclusion of the inflaton's domination, denoted by the sub-index ``${\rm end}$'', that is, when both energy densities are of the same order. This leads to the relation:
\begin{eqnarray}
    \left( \frac{a_{\rm dec}}{a_{\rm end}}\right)^{\frac{2(n-2)}{n+1}}
    =\frac{\langle \rho_{\rm dec}\rangle}{\rho_{\rm B, \rm dec}} \Longrightarrow
\langle \rho_{\rm end}\rangle= 
{\langle \rho_{\rm dec}\rangle^{\frac{3n}{n-2}}}{\rho_{\rm B, \rm dec}^{-\frac{2(n+1)}{n-2}}}.  
\end{eqnarray}
Therefore, from the Stefan-Boltzmann law, the reheating temperature has the following expression:  
\begin{eqnarray}\label{reheating0}
 T_{{\rm reh}}(n)&&\equiv \left(\frac{30}{\pi^2g_{\rm reh}} \right)^{1/4}
 \langle\rho_{\rm end}\rangle^{\frac{1}{4}}\nonumber \\&&= 
 \left(\frac{30}{\pi^2g_{\rm reh}} \right)^{1/4}
 \langle\rho_{\rm dec}\rangle^{\frac{1}{4}}
 \left(\frac{\langle\rho_{\rm dec}\rangle}{\rho_{\rm B,\rm dec}}\right)^{\frac{n+1}{2(n-2)}}, 
 \end{eqnarray}
 where $g_{\rm reh}=106.75$, is the effective number of degrees of freedom for the Standard Model. 
At this juncture, we can enhance this formula by considering the evolution of the corresponding energy densities before the decay. They follow the expressions:
\begin{eqnarray}
\rho_{\rm B}(t)
\cong
3H_{\rm st}^2M_{\rm pl}^2 \left( \frac{a_{\rm st}}{a(t)} \right)^{\frac{6n}{n+1}}=
3H_{\rm END}^2M_{\rm pl}^2 \left( \frac{a_{\rm END}}{a(t)} \right)^{\frac{6n}{n+1}}
\end{eqnarray}
\begin{eqnarray}
\langle \rho(t)\rangle=\langle \rho_{\rm st}\rangle \left( \frac{a_{\rm st}}{a(t)} \right)^3=\langle \rho_{\rm END}\rangle \left( \frac{a_{\rm END}}{a(t)} \right)^3,
\end{eqnarray}
where the sub-index `${\rm st}$' means the moment of the stabilization of the Bogoliubov coefficients,
which occurs approximately in $10$ $e$-folds after the end of inflation, as has been showed in \cite{Dorsch:2024nan}, and
we have taken into account that $
\langle \rho_{\rm st}\rangle \ll \rho_{\rm B,st}$, implying
$\rho_{\rm B,st}\cong 
3H_{\rm st}^2M_{\rm pl}^2$. In addition, we have defined $\langle \rho_{\rm END} \rangle \equiv\langle \rho_{\rm st}\rangle \left(a_{\rm st}/a_{\rm END}\right)^3 $.
Note that,  
one can calculate analytically 
$\rho_{\rm B, END}$,
because since the end of inflation occurs when the slow roll parameter $\epsilon=\frac{M_{\rm pl}^2}{2}\left(\frac{dV}{d\varphi}/V\right)^2$ 
is equal to $1$, one can calculate analytically $\rho_{\rm B, END}=\frac{3}{2}V(\varphi_{\rm END})\equiv\frac{3}{2}V_{\rm END}$.
Then,  when the heavy particles have completely  decayed, which occurs when $H\sim \Gamma$,  the semi-classical Friedmann equation
$H^2=\dfrac{1}{3M_{\rm pl}^2}\left(\rho_{\rm B}+\langle \rho\rangle\right)$,
becomes, 
\begin{eqnarray}\label{semiclassical}
3\Gamma^2M_{\rm pl}^2=\rho_{\rm B,END}x^{\frac{2n}{n+1}}+\langle\rho_{\rm END}\rangle
    x,\end{eqnarray}
where we have introduced the notation $x=\left(a_{\rm END}/a_{\rm dec}\right)^3$. The solution has the form $x=F_n(\Theta_1,\Theta_2)$, 
where $F_n$ is a function which depends on $n$ and the parameters, 
$\Theta_1=\langle \rho_{\rm END}\rangle/\rho_{\rm B,END}$ and $\Theta_2=3\Gamma^2M_{\rm pl}^2/\rho_{\rm B, END}$. For example, when $n=\infty$, i.e. when $w_{\rm eff}=1$,  ones has
\begin{eqnarray}
    F_{\infty}(\Theta_1,\Theta_2)=
    \frac{1}{2}\left(\sqrt{\Theta_1^2+4\Theta_2}-\Theta_1  \right),
\end{eqnarray}
and for $n=3$ we obtain $x^{3/2}+\Theta_1x-\Theta_2=0$, which is a cubic equation when one introduces the variable $x=z^2$. This equation can be solved using the Cardano's formulas, obtaining:
\begin{align}
F_3(\Theta_1,\Theta_2)=\left[\sqrt[3]{
\frac{1}{2}\left(\Theta_2-\frac{2\Theta_1^3}{27}+\sqrt{\Theta_2
\left(\Theta_2-\frac{4\Theta_1^3}{27}\right)
}
\right)}+
\right. \nonumber\\ \left.
\sqrt[3]{
\frac{1}{2}\left(\Theta_2-\frac{2\Theta_1^3}{27}-\sqrt{\Theta_2
\left(\Theta_2-\frac{4\Theta_1^3}{27}\right)
}
\right)
    }-\frac{\Theta_1}{3}\right]^2,    \end{align}
provided that, 
$\Theta_2-\frac{4\Theta_1^3}{27}>0$, which always happens. Effectively, from the equation $x^{3/2}+\Theta_1x-\Theta_2=0$ we have the bound $\Theta_2>\Theta_1 x$. On the other hand, since the decay occurs during the domination of the inflaton's energy density, one has $\langle \rho_{\rm dec}\rangle\leq \rho_{\rm B,dec}$ which is equivalent to $\Theta_1^2\leq x$, then we have:
\begin{eqnarray}
    \Theta_2-\frac{4\Theta_1^3}{27}>
    \Theta_1 x\left(1-\frac{4}{27}\right)>0.
    \end{eqnarray}
Coming back to the reheating formula,
we can find the final  expression of the reheating temperature as a function of the parameters $\Theta_1$ and $\Theta_2$:
\begin{align}\label{temperature}
    T_{\rm reh}(n)=
    \left( \frac{90}{\pi^2 g_{\rm reh}}\right)^{1/4}
    \left( \frac{\Theta_1^3}{F_n(\Theta_1,\Theta_2)} \right)^{\frac{n}{4(n-2)}}\sqrt{H_{\rm END}M_{\rm pl}}
    \nonumber \\
    \sim 5\times 10^{-4}
    \left( \frac{\Theta_1^3}{F_n(\Theta_1,\Theta_2)} \right)^{\frac{n}{4(n-2)}} M_{\rm pl}, 
\end{align}
where we have taken, as usual for many inflationary models,  $H_{\rm END}\sim 10^{-6}M_{\rm pl}$. 
On the other hand, given that the decay occurs during the domination of the inflaton's energy density, we have the constraints $\Gamma \leq H_{\rm st}$ and $\langle \rho_{\rm dec}\rangle\leq \rho_{\rm B,dec}$, which after some algebra,
is equivalent to $\Theta_1^{\frac{n+1}{n-1}}\leq  F_n(\Theta_1,\Theta_2)$.
Finally, we calculate the maximum reheating temperature, which is obtained when the decay occurs at the end of inflaton's domination, that is, when
$\langle\rho_{\rm dec}\rangle=\rho_{\rm B,dec}$, which after some algebra, is equivalent to:
\begin{eqnarray}
    F_n(\Theta_1,\Theta_2)=
    \Theta_1^{\frac{n+1}{n-1}},
\end{eqnarray}
and thus, the maximum reheating temperature has the following expression as a function of the parameter $\Theta_1$:
\begin{eqnarray}\label{Tem_max} 
T_{\rm reh}^{\rm max}(n)&&=
\left( \frac{90}{\pi^2 g_{\rm reh}}\right)^{1/4}
    \Theta_1^{\frac{n}{2(n-1)}} \sqrt{H_{\rm END}M_{\rm pl}} \nonumber\\&& \cong 5\times 10^{-1}\Theta_1^{\frac{n}{2(n-1)}} \sqrt{H_{\rm END}M_{\rm pl}}
    \nonumber\\&&\sim 5\times 10^{-4}
    \Theta_1^{\frac{n}{2(n-1)}}M_{\rm pl}    .
\end{eqnarray}
Note that the maximum reheating temperature increases as 
$n$
 increases because, with higher 
$n$, the energy density of the produced particles overtakes that of the inflaton more quickly. Therefore, the maximum value is obtained for potentials that, when the inflaton oscillates, the universe enters in a stiff matter era, i.e., $w_{\rm eff}=1$. In this situation, the maximum reheating temperature is given by,
\begin{eqnarray}\label{Tem_max1}
    T_{\rm reh}^{\rm max}(\infty)&&=
    \left( \frac{90}{\pi^2 g_{\rm reh}}\right)^{1/4}
\sqrt{\Theta_1H_{\rm END}M_{\rm pl}}\nonumber\\&& \sim 5\times 10^{-4}\sqrt{\Theta_1}M_{\rm pl},
\end{eqnarray}
where, once again,  we have taken $H_{\rm END}\sim 10^{-6}M_{\rm pl}$.

\subsection{Decay after the end of the background domination}

In this subsection, we will examine the scenario where the decay occurs after the end of domination of the inflaton's energy density. In this case, given the instantaneous nature of the thermalization process, reheating is concluded upon the completion of decay. Therefore, the reheating temperature is determined by,
\begin{eqnarray} \label{reheating3}
T_{\rm reh}=\left( \frac{30}{\pi^2 g_{\rm reh}} \right)^{1/4}\langle\rho_{\rm dec}\rangle^{1/4}, 
\end{eqnarray}
where one must ensure that ${\Gamma}\leq H_{\rm end}$.  The value of the Hubble rate at the end of the background domination can be calculated  by considering that, in this scenario, the energy density of the produced particles decays as $a(t)^{-3}$ throughout the entire domination of the inflaton's energy density. Thus, at the end of the background domination:
\begin{align}
\Theta_1=\frac{\langle\rho_{\rm END}\rangle}{\rho_{\rm B,END}}=\left( \frac{a_{\rm END}}{a_{\rm end}} \right)^{\frac{3(n-1)}{n+1}} \Longrightarrow \nonumber\\ 
H_{\rm end}^2=\frac{2\rho_{\rm B,end}}{3M_{\rm pl}^2}= 
\frac{2\rho_{\rm B,END}}{3M_{\rm pl}^2}\left( \frac{a_{\rm END}}{a_{\rm end}} \right)^{\frac{6n}{n+1}}
=
2H_{\rm END}^2
\Theta_1^{\frac{2n}{n-1}},\end{align}
and thus, the constraint $\Gamma\leq H_{\rm end}$,  becomes
$\Gamma\leq  \sqrt{2}
\Theta_1^{\frac{n}{n-1}} H_{\rm END}$.  
To refine the formula for the reheating temperature (\ref{reheating3}), we perform the following calculation:
\begin{align}\label{calculation}
&&\langle\rho_{\rm end}\rangle=\rho_{\rm B,end}=\rho_{\rm B, END}
\Theta_1^{\frac{2n}{n-1}} \Longrightarrow\nonumber\\&&
\langle\rho_{\rm dec}\rangle^{1/4}=
\langle\rho_{\rm end}\rangle^{1/4}\left(\frac{a_{\rm end}}{a_{\rm dec}}\right)^{3/4} =\nonumber\\&&
=\rho_{\rm B,END}^{1/4}
\Theta_1^{\frac{n}{2(n-1)}} \left(\frac{a_{\rm end}}{a_{\rm dec}}\right)^{3/4}. 
\end{align}
Therefore, when the decay is immediately  finished, 
introducing the notation $y=\left(a_{\rm end}/a_{\rm dec}\right)^3$,
the semi-classical Friedmann  is given by,
\begin{eqnarray}
3\Gamma^2M_{\rm pl}^2=\rho_{\rm B,end}(y+y^{\frac{2n}{n+1}})\Longrightarrow\nonumber\\
    y^{\frac{2n}{n+1}}+y-\Theta_2\Theta_1^{-\frac{2n}{n-1}}=0
    ,
\end{eqnarray}
where we have used that
$(3\Gamma^2M_{\rm pl}^2)/\rho_{\rm B,end}=
\Theta_2\Theta_1^{-\frac{2n}{n-1}}$.
Writing the solution as:
\begin{eqnarray}\label{Theta12}
\left(\frac{a_{\rm end}}{a_{\rm dec}}\right)^3=
G_n \left(\Theta_2\Theta_1^{-\frac{2n}{n-1}} \right),
\end{eqnarray}
{where $G_{n}$ is also a function which depends on $n$ and $\Theta_1$, $\Theta_2$,} and thus, using the formulas (\ref{Theta12}) and (\ref{calculation}), 
the reheating temperature takes the following final form:
\begin{align}\label{temperature_after_kination}
T_{\rm reh}(n)=    
\left(\frac{90G_n(\Theta_2\Theta_1^{-\frac{2n}{n-1}})}{\pi^2g_{\rm reh}} \right)^{1/4}
\Theta_1^{\frac{n}{2(n-1)}}
\sqrt{H_{\rm END}M_{\rm pl}}.
\end{align}}

\begin{remark}
    Note that the reheating formula can be simplified when the decay is well after the end of the inflaton's energy density domination. In that case $\left(a_{\rm end}/a_{\rm dec}\right)^3\ll 1$,
    and will have 
    $G_n(\Theta_2\Theta_1^{-\frac{2n}{n-1}})\cong \Theta_2\Theta_1^{-\frac{2n}{n-1}}$. Therefore, the reheating temperature will be approximated by,
\begin{align}
    T_{\rm reh}(n)\cong \left(\frac{90\Theta_2}{\pi^2g_{\rm reh}} \right)^{1/4}
\sqrt{H_{\rm END}M_{\rm pl}}\nonumber\\=\left(\frac{90}{\pi^2g_{\rm reh}} \right)^{1/4}
\sqrt{\Gamma M_{\rm pl}}\nonumber\\
\ll \left(\frac{180}{\pi^2g_{\rm reh}} \right)^{1/4} 
\Theta_1^{\frac{n}{2(n-1)}}
\sqrt{H_{\rm END} M_{\rm pl}},\end{align}
    where we have used that when the decay is well after the end of the inflaton's domination we have 
$\Gamma\ll H_{\rm end}=\sqrt{2}
\Theta_1^{\frac{n}{n-1}}H_{\rm END}$.
\end{remark}
Finally, we can see the simplest expression of this reheating temperature is obtained when $n=\infty$, obtaining
\begin{eqnarray}
    G_{\infty}(\Theta_2\Theta_1^{-2})=
    \frac{1}{2\Theta_1^2}\left(\sqrt{\Theta_1^4+4\Theta_2^2}-\Theta_1^2\right), 
\end{eqnarray}
and thus,
\begin{align}
T_{\rm reh}(\infty)=    
\left(\frac{45(\sqrt{\Theta_1^4+4\Theta_2^2}-\Theta_1^2)}{\pi^2g_{\rm reh}} \right)^{1/4}\sqrt{H_{\rm END}M_{\rm pl}}~.
\end{align}
In addition, when $n=3$, one can calculate analytically 
$G_3(\Theta_1\Theta_2^{-3})$ as we have calculated $F_3(\Theta_1,\Theta_2)$
in section \ref{decay_before}.

\section{Bounds on the maximum reheating temperature}
\label{sec-3}

We start the section recalling that $\Theta_1= \langle\rho_{\rm END}\rangle/\rho_{\rm B,\rm END}$. 
Given that the maximum reheating temperature is contingent on the value of $\Theta_1$, its calculation becomes imperative. The problem is that 
$\langle\rho_{\rm END}\rangle$ has to be calculated numerically, and in the literature only appears analytic formulas in the quadratic case $V(\varphi)=\frac{1}{2}m_{\varphi}^2\varphi^2$. However, for a quadratic model, during the oscillations, $w_{\rm eff}=0$, that is, the energy density of the inflaton scales as matter, which, as we have already explained, makes impossible  a successful reheating. 

\subsection{Analytic formulas for the particle production and the corresponding bounds}

In order to have an analytic formula for the maximum reheating temperature, we will assume that for values of $n$ of the order $1$, the energy density of the produced particles is of the same order than the one obtained in the quadratic case. Therefore, we will start using the  formula obtained in  \cite{Ema:2015dka,Ema:2018ucl,Chung:2018ayg}:
\begin{eqnarray}\label{energy_density_pp}
\langle\rho(t)\rangle\cong CH_*^3m_{\chi}\left(\frac{m_{\chi}}{m_{\varphi}}\right)^4\left( \frac{a_{\rm END}}{a(t)}\right)^3,
\end{eqnarray}
where $m_{\chi}\ll m_{\varphi}$, $H_*$ is the scale of inflation, i.e., the value of the Hubble rate at the horizon crossing and $C\cong 2\times 10^{-3}$ is a dimensionless constant.
Next we will use that, at the horizon crossing, we have 
\begin{eqnarray}
   H_*^2\cong \frac{1}{6M_{\rm pl}^2}m_{\varphi}^2\varphi_*^2,
\end{eqnarray}
and from the formula of the power spectrum of scalar perturbations
$\frac{H_*^2}{8\pi^2\epsilon_* M_{\rm pl}^2}\sim 2\times 10^{-9}$ we get:
\begin{eqnarray}
    16\pi^2\times 10^{-9}\epsilon_*M_{\rm pl}^2\sim \frac{1}{6M_{\rm pl}^2}m_{\varphi}^2\varphi_*^2,
\end{eqnarray}
and taking into account that for a quadratic potential one has
\begin{eqnarray}
    \epsilon_*=\frac{1}{4}(1-n_s),\qquad
    \varphi_*^2=\frac{8M_{\rm pl}^2}{1-n_s},
\end{eqnarray}
one obtains:
\begin{eqnarray}
    m_{\varphi}^2\sim 3\pi^2\times 10^{-9}(1-n_s)^2M_{\rm pl}^2\cong 3\pi^2\times 10^{-12}M_{\rm pl}^2,
\end{eqnarray}
where we have taken $n_s=0.96$.  Inserting the value of $m_{\varphi}$ in the formula (\ref{energy_density_pp}) one finds:
\begin{eqnarray}\label{energy_density_pp1}
\langle\rho_{\rm END}\rangle\sim 2\times 10^{18} H_*^3m_{\chi}\left(\frac{m_{\chi}}{M_{\rm pl}} \right)^4,
\end{eqnarray}
for $m_{\chi}\ll m_{\varphi}\sim 10^{-6} M_{\rm pl}$.
Therefore, we will obtain:
\begin{eqnarray}
    \Theta_1\sim 6\times 10^{17}\frac{H_*^3m_{\chi}}{H_{\rm END}^2M_{\rm pl}^2}\left(\frac{m_{\chi}}{M_{\rm pl}} \right)^4.
        \end{eqnarray}

To find a bound for the maximum reheating temperature, note that, 
as in the majority of inflationary models, $H_{\rm END}\sim 10^{-1}H_*\sim 10^{-6}M_{\rm pl}$,
one gets $\Theta_1\cong 6\times 10^{14}\left(\frac{m_{\chi}}{M_{\rm pl}} \right)^5$. 
Therefore, we bound the maximum reheating temperature (\ref{Tem_max}), taking into account the bound $\Theta_1^{\frac{n}{2(n-1)}}<\sqrt{\Theta_1}$ 
for $n\geq 2$, one gets:
\begin{align}
&& T_{\rm reh}^{\rm max}(n)<\left(\frac{90}{\pi^2g_{\rm reh}}\right)^{1/4}
    \sqrt{\Theta_1
    H_{\rm END}M_{\rm pl}}\nonumber\\
    && \cong  10^4
    \left(\frac{m_{\chi}}{M_{\rm pl}} \right)^{5/2}M_{\rm pl} <3\times 10^{-14}M_{\rm pl}\sim 8\times 10^4 \mbox{GeV},
    \end{align}
    where, once again, we have taken $H_{\rm END}\sim 10^{-6}M_{\rm pl}$.
Next, we will consider another analytic formula for the energy density of the produced particles, the one
presented in \cite{Kolb:2023ydq}:
\begin{eqnarray}\label{particle_kolb}
    \langle \rho(t)\rangle=\frac{H_{\rm END}^2m_{\chi}^2}{16 \pi^2}\left( \frac{a_{\rm END}}{a(t)}\right)^3, 
\end{eqnarray}
 for conformally coupled particles satisfying $m_{\chi}<H_{\rm END}$.  In this case,
\begin{align}
&& \Theta_1 = \frac{m_{\chi}^2}{48 \pi^2M_{\rm pl}^2}
    \sim 2\times 10^{-3}\frac{m_{\chi}^2}{M_{\rm pl}^2}
    \Longrightarrow     \nonumber\\
&& T_{\rm reh}^{\rm max}(n)\cong 5\times 10^{-1}
    \left(\frac{m_{\chi}}{4\sqrt{3}\pi M_{\rm pl}}  \right)^{\frac{n}{n-1}}
\sqrt{H_{\rm END}M_{\rm pl}}.       \end{align}
Finally, we have:
\begin{align}
&& T_{\rm reh}^{\rm max}(n)< 
    \left(\frac{90}{\pi^2g_{\rm reh}}\right)^{1/4}    
    \sqrt{\Theta_1
    H_{\rm END}M_{\rm pl}}\nonumber\\
    && \cong 2\times 10^{-5}{m_{\chi}}<2\times 10^{-12}M_{\rm pl}\sim 5\times 10^6 \mbox{GeV},\end{align}
where we have taken $H_{\rm END}\sim 10^{-6}M_{\rm pl}$ and 
$m_{\chi}<10^{-7}M_{\rm pl}$.
We see that in both cases, i.e., for both expressions of the energy density of the produced particles
(\ref{energy_density_pp}) and (\ref{particle_kolb}), the maximum reheating temperature is bellow $10^9$ GeV, and thus,  over-passing the gravitino problem, that is, it is lower than $10^9$ GeV (see ~\cite{Ellis:1982yb} for details).

\subsubsection{Case: $n=3$}

{ To obtain more accurate bounds we
consider  the potential
\begin{eqnarray}
    V_3(\varphi)=\lambda M_{\rm pl}^4\left(1- e^{-\sqrt{\frac{2}{3}}\frac{\varphi}{M_{\rm pl}}}\right)^{6}. 
\end{eqnarray}
For this potential, as we will see in next section, we will have an effective EoS parameter, $w_{\rm eff}=1/2$, and also
$\lambda\sim 5\times 10^{-11}$,  
 {$H_{\rm END}\sim 9\times 10^{-7}M_{\rm pl}$} and $H_*\cong \sqrt{\frac{\lambda}{3}}M_{\rm pl}\cong 4\times 10^{-6} M_{\rm pl}$.  In this case, the maximum reheating temperature becomes:
\begin{eqnarray}
T_{\rm reh}^{\rm max}(3)\cong 5\times 10^{-4}\Theta_1^{3/4}M_{\rm pl}\cong
 2\times 10^{5}\frac{\langle \rho_{\rm END}\rangle^{3/4}}{M_{\rm pl}^2},
\end{eqnarray}
where we have used that $\Theta_1\cong 4\times 10^{11}\frac{\langle \rho_{\rm END}\rangle}{M_{\rm pl}^4}$.
If one considers, the gravitational particle production given by  the formulas (\ref{energy_density_pp})
and (\ref{energy_density_pp1}), one gets:
\begin{eqnarray}
 T_{\rm reh}^{\rm max}(3)&&\sim  8\times 10^{6}
    \left(\frac{m_{\chi}}{M_{\rm pl}}\right)^{3}(m_{\chi}^{3}M_{\rm pl} )^{1/4} \nonumber\\
&& <5\times 10^{-20}M_{\rm pl}\sim 10^2 \mbox{MeV},
\end{eqnarray}
where we have  the bound $m_{\chi}<10^{-7}M_{\rm pl}$.}
On the other hand, to have a successful reheating one has to demand that $T_{\rm reh}^{\rm max}>1\mbox{ MeV}\sim 5\times 10^{-22} M_{\rm pl}$, what happens for $m_{\chi}>3\times 10^{-8}M_{\rm pl}.$  Therefore, the gravitational particle production given by (\ref{energy_density_pp}), only provides a viable maximum reheating temperature for masses satisfying: 
\begin{eqnarray}
&& 3\times 10^{-8}M_{\rm pl}<m_{\chi}<10^{-7}M_{\rm pl}
\Longleftrightarrow\nonumber\\ 
&& 
7\times 10^{10}\mbox{ GeV}< m_{\chi}<
2\times 10^{11}\mbox{ GeV}.
\end{eqnarray}
Finally,  considering (\ref{particle_kolb}), for $n=3$ and $m_{\chi}<10^{-7}M_{\rm pl}$,  we will have:
\begin{eqnarray}
 T_{\rm reh}^{\rm max}(3)&&\sim 3\times 10^{-6}
    \left(\frac{m_{\chi}}{M_{\rm pl}} \right)^{3/2}M_{\rm pl}\nonumber\\ 
    && < 9\times 10^{-17}M_{\rm pl}\sim  2\times 10^2\mbox{GeV},
\end{eqnarray}
and demanding $T_{\rm reh}> 1\mbox{ MeV}$, we get $m_{\chi}> 3\times 10^{-11} M_{\rm pl}$. Therefore, for the energy density of produced particles given by (\ref{particle_kolb}), a viable maximum reheating temperature is only possible for masses in the range
\begin{eqnarray}
&& 3\times 10^{-11}M_{\rm pl}<m_{\chi}<10^{-7}M_{\rm pl}
\Longleftrightarrow\nonumber\\
&& 
7\times 10^{7}\mbox{ GeV}< m_{\chi}<
2\times 10^{11}\mbox{ GeV}.
\end{eqnarray}
Note also that, making the same kind of calculation,  for not viable potentials, i.e., as we have already discussed, for $n\leq 2$, the maximum reheating temperature  is less than $10$ MeV.

\subsubsection{Case: $n\rightarrow \infty$}

We consider the extreme case, where after inflation the universe enters in an stiff phase.  As we have already shown, in this case the maximum reheating temperature is given by
(\ref{Tem_max1}), and the expression of the parameter $\Theta_1$ is:
\begin{eqnarray}
\Theta_1=\frac{\langle\rho_{\rm END}\rangle}{3H_{\rm END}^2M_{\rm pl}^2}
    \sim 3\times 10^{11}\frac{\langle\rho_{\rm END}\rangle}{M_{\rm pl}^4},\end{eqnarray}
where we have taken $H_{\rm END}\sim 10^{-6}M_{\rm pl}$.
Therefore, recalling that from the analytic formula
(\ref{energy_density_pp}), we have obtained $\langle\rho_{\rm END}\rangle\sim 2\times 10^{18}H_*^3m_{\chi}\left(\frac{m_{\chi}}{M_{\rm pl}}\right)^4$, we will have:
\begin{align}
\Theta_1\sim 6\times 10^{29}\frac{H_*^3m_{\chi}}{M_{\rm pl}^4}\left(\frac{m_{\chi}}{M_{\rm pl}}\right)^4
    \sim 4\times 10^{13}\left(\frac{m_{\chi}}{M_{\rm pl}}\right)^5,   
    \end{align}
where we have chosen $H_*\sim 4\times 10^{-6} M_{\rm pl}$. And the maximum reheating temperature (\ref{Tem_max1}) becomes:
\begin{eqnarray}
 T_{\rm reh}^{\rm max}(\infty)&&\sim 3\times 10^{3}\left(\frac{m_{\chi}}{M_{\rm pl}}\right)^{5/2}M_{\rm pl}\nonumber\\
&& <  10^{-14}M_{\rm pl}\sim 2\times 10^4 \mbox{GeV},
    \end{eqnarray}
where we have used that $m_{\chi}<10^{-7}M_{\rm pl}$. In the same way as in the case $n=3$, now a viable maximum reheating temperature is possible for
\begin{eqnarray}
&& 6\times 10^{-11}M_{\rm pl}<m_{\chi}<10^{-7}M_{\rm pl}
\Longleftrightarrow\nonumber\\
&& 
 10^{7}\mbox{ GeV}< m_{\chi}<
2\times 10^{11}\mbox{ GeV}.
\end{eqnarray}

And for the analytic formula 
(\ref{particle_kolb}), 
we have already seen that $\Theta_1\sim 2\times 10^{-3}\frac{m_{\chi}^2}{M_{\rm pl}^2}$, obtaining:
\begin{eqnarray}
 T_{\rm reh}^{\rm max}(\infty)&&\sim 2\times 10^{-5}\frac{m_{\chi}}{M_{\rm pl}}M_{\rm pl}\nonumber\\
&& <2\times 10^{-12}M_{\rm pl}\sim 5\times 10^6\mbox{GeV},
\end{eqnarray}
and a viable maximum reheating temperature requires:
\begin{eqnarray} 
&& 10^{-17}M_{\rm pl}<m_{\chi}<10^{-7}M_{\rm pl}
\Longleftrightarrow\nonumber \\
&&
 25\mbox{ GeV}< m_{\chi}<
2\times 10^{11}\mbox{ GeV}.\end{eqnarray}

\subsection{Application to $\alpha$-attractors}

We want  to highlight that in \cite{Flores:2024lzv} the authors deal with $\alpha$-attractors whose  potential
is:
\begin{eqnarray}
    V_n(\varphi)=\lambda M_{\rm pl}\left(\sqrt{6}
    \tanh\left(\frac{\varphi}{\sqrt{6}M_{\rm pl}}\right) \right)^{2n},
\end{eqnarray}
being
\begin{eqnarray}
    \lambda\cong \frac{36\pi^2}{6^n N_*^2}10^{-9}\cong 6^{-n}10^{-10},
\end{eqnarray}
where we have chosen as the number of last $e$-folds $N_*=55$.
In the minimally coupled case,  the authors numerically found that 
for masses of the order  $m_{\chi}\sim 10^{-4}H_{\rm END}$, the energy density of the produced particles, when $n=4$,  is (see Fig. $1$ of Ref. \cite{Flores:2024lzv}):
\begin{eqnarray}
\langle\rho_{\rm END}\rangle\cong 10^2H_{\rm END}^3m_{\chi}=10^{-2}H_{\rm END}^4,
\end{eqnarray}
which leads to:
\begin{eqnarray}
    \Theta_1\cong 3\times 10^{-3}\left(\frac{H_{\rm END}}{M_{\rm pl}}\right)^2.
\end{eqnarray}

Consequently, the maximum reheating temperature is:
\begin{eqnarray}
 T_{\rm reh}^{\rm max}(4)&&=\left(\frac{90}{\pi^2g_{\rm reh}}\right)^{1/4}
\Theta_1^{2/3}\sqrt{H_{\rm END}M_{\rm pl}}\nonumber\\
&& \cong 10^{-2}\left( \frac{H_{\rm END}}{M_{\rm pl}}\right)^{11/6} M_{\rm pl}.
\end{eqnarray}

To find $H_{\rm END}$, we solve the equation $\frac{M_{\rm pl}^2}{2}\left(\frac{dV_n}{d \varphi}/V_n\right)^2=1$, obtaining:
\begin{eqnarray}
\sinh^2\left(\frac{2\varphi_{\rm END}}{\sqrt{6}M_{\rm pl}}\right)=\frac{2n^2}{9},
\end{eqnarray}
which after some algebra, one gets:
\begin{eqnarray}
    V_n(\varphi_{\rm END})\cong 10^{-10}M_{\rm pl}^4\left(
    \frac{18+2n^2-6\sqrt{9+2n^2}}{2n^2}
    \right)^n,
\end{eqnarray}
and thus:
\begin{align}
    H_{\rm END}\cong 7\times 10^{-6}
    \left(
    \frac{18+2n^2-6\sqrt{9+2n^2}}{2n^2}  \right)^{n/2}M_{\rm pl}.
    \end{align}
Then, for $n=4$ one has 
 $H_{\rm END}\sim  10^{-6}M_{\rm pl}$, which leads to the following maximum reheating temperature:
\begin{eqnarray}
    T_{\rm reh}^{\rm max}(4)\cong 10^{-13}M_{\rm pl}\cong 2\times 10^5 \mbox{ GeV}.
\end{eqnarray}
In addition, when $n=8$ the authors find~\cite{Flores:2024lzv}:
\begin{eqnarray}
\langle\rho_{\rm END}\rangle\cong 2\times 10H_{\rm END}^3m_{\chi}=2\times 10^{-3}H_{\rm END}^4,
\end{eqnarray}
which leads to:
\begin{eqnarray}
    \Theta_1\cong 6\times 10^{-4}\left(\frac{H_{\rm END}}{M_{\rm pl}}\right)^2.
\end{eqnarray}

Consequently, the maximum reheating temperature is:
\begin{eqnarray}
 T_{\rm reh}^{\rm max}(8)&&=\left(\frac{90}{\pi^2g_{\rm reh}}\right)^{1/4}
\Theta_1^{4/7}\sqrt{H_{\rm END}M_{\rm pl}}\nonumber\\
&& \cong 8\times 10^{-3}\left( \frac{H_{\rm END}}{M_{\rm pl}}\right)^{23/14} M_{\rm pl}.
\end{eqnarray}
Since for $n=8$ one has 
 $H_{\rm END}\sim 8\times 10^{-7}M_{\rm pl}$, then the maximum reheating temperature
 is:
\begin{eqnarray}
    T_{\rm reh}^{\rm max}(8)\cong 8\times10^{-13}M_{\rm pl}\cong 2\times 10^6 \mbox{ GeV}.
\end{eqnarray}

\section{Qualitative study and Numerical calculations}
\label{sec-4}

We will begin this section by following the approach of \cite{Ling:2021zlj}, where the authors study $\alpha$-attractors with a potential that behaves like $\varphi^2$ near the minimum. However, the article~\cite{Ling:2021zlj} considers the minimally coupled case, where the produced particles acquire an effective mass $m_{\rm eff}^2 = m_{\chi}^2 - \frac{R}{6}$, leading to a tachyonic instability during inflation, which may enhance particle production.

We conduct a qualitative study of the $k$-modes in the conformally coupled case, where the effective mass is equal to the bare mass $m_{\chi}$, thus avoiding the tachyonic instability. We will identify the modes influencing the reheating temperature and determine the initial conditions that ensure that the modes are in the adiabatic vacuum.

First of all, it is important to determine when the $\beta$-Bogoliubov coefficients stabilize. For a potential that behaves quadratically near the minimum, as shown in \cite{Kaneta2022}, the $\beta$-Bogoliubov coefficients can be calculated using either the Bogoliubov or Boltzmann approach for modes satisfying $k > a_{\rm END} H_{\rm END}$. For modes in the range $k < a_{\rm END} H_{\rm END}$, an approximate value for the $\beta$-Bogoliubov coefficient can be obtained by assuming an abrupt phase transition from the de Sitter regime to matter domination, which is the scenario when the potential is quadratic. As we have already seen, for a power-law potential $\varphi^{2n}$, the effective equation of state is $w_{\rm eff} = \frac{n-1}{n+1}$.

On the other hand, when $n > 2$, since an analytic calculation is not possible, we must recognize that the stabilization of the $\beta$-Bogoliubov coefficients occurs when the adiabatic condition ${\omega'_k}/{\omega_k^2} \ll 1$ is satisfied, where the prime denotes differentiation with respect to conformal time, indicating that particle production ceases. For example, we take the conservative bound ${\omega}'_k/\omega_k^2 < 10^{-7}$. Then, in the conformally coupled case, this condition is satisfied when:
\begin{align}
\frac{{\omega}'_k}{\omega_k^2}= \frac{m_{\chi}^2a^3H}{\omega_k^3}<\frac{H(t)}{m_{\chi}}=
    \frac{H_{\rm END}}{m_{\chi}}\left(\frac{a_{\rm END}}{a(t)} \right)^{\frac{3n}{n+1}}\cong 10^{-7}. 
    \end{align}
Therefore, the number of $e$-folds after the end of inflation needed to stabilize the $\beta$-Bogoliubov coefficients satisfies:
    \begin{eqnarray}
    \label{stabilization}
       && N \equiv \ln\left(\frac{a_{\rm st}}{a_{\rm END}}\right)\cong\frac{n+1}{3n}\ln\left(10^7\frac{H_{\rm END}}{m_{\chi}}  \right)
\Longrightarrow\nonumber\\&&
        N
        >\frac{1}{3}
    \ln\left(10^7\frac{H_{\rm END}}{m_{\chi}}  \right).
    \end{eqnarray}
Let us note that for $n=3$ and $m_{\chi}\sim 10^{-2}H_{\rm END}$, we have $N\cong 9$.
On the other hand, it is also important to realize that we have to take initial conditions at a given  time $t_i$, namely $\alpha_k(t_i)=1$ and 
$\beta_k(t_i)=0$, when the adiabatic condition is fulfilled.
Since we deal with the case 
$m_{\chi}\ll H_{\rm END}$, we will have
 $m_{\chi}\ll H_i$, 
 where $H_i$ denotes the Hubble rate at the initial time $t_i$.
 Therefore, 
 considering modes satisfying $k\geq a_iH_i$, the adiabatic condition becomes:
 \begin{eqnarray}
     \frac{\omega_{k,i}'}{\omega_{k,i}^2}=
    \frac{m_{\chi}^2a_i^3H_i}{\omega_{k,i}^3}\ll\frac{a_i^3H_{i}^3}{k^3}\leq 1.
 \end{eqnarray}
On the contrary, modes satisfying $k\leq a_iH_i$ are not in the adiabatic vacuum, but for large enough early conditions, they do not contribute to the reheating temperature and thus can be disregarded. Effectively,
the contribution of those modes, to the energy density of produced particles at the end of inflation is: 
\begin{eqnarray}
I_{\rm END}&& \equiv \frac{1}{2\pi^2a^4_{\rm END}}\int_0^{a_iH_i} \omega_{\rm k,END} k^2|\beta_k|^2 dk 
 \nonumber\\&&\leq
\frac{1}{2\pi^2a_{\rm END}^4}\int_0^{a_iH_i} \omega_{\rm k, END} k^2 dk,
\end{eqnarray}
where we have used that $|\beta_k|\leq 1$. Since 
$m_{\chi}\ll H_{\rm END}$, and using that $\omega_{\rm k,END}\ll a_{\rm END}H_i$,  we have, 
 \begin{align}
        I_{\rm END}\ll\frac{H_i}{2\pi^2a^3_{\rm END}}\int_0^{a_iH_i}  k^2 dk 
        = \frac{H_i^4}{6\pi^2}\left( \frac{a_i}{a_{\rm END}}\right)^3,
\end{align}
and its contribution to $\Theta_1$, namely $\bar{\Theta}_1$, is,
\begin{eqnarray}
    \bar{\Theta}_1\ll
    \frac{H_i^4}{18\pi^2H_{\rm END}^2M_{\rm pl}^2}\left( \frac{a_i}{a_{\rm END}}\right)^3.     
\end{eqnarray}
Meaning that, after some algebra,  its contribution to the maximum reheating temperature is bellow:
\begin{eqnarray}
   && 5\times 10^{-1}\sqrt{\bar{\Theta}_1H_{\rm END}M_{\rm pl}}\nonumber\\&&\ll
    4\times 10^{-2}\frac{H_i^2}{\sqrt{H_{\rm END}M_{\rm pl}}}
    \left( \frac{a_i}{a_{\rm END}}\right)^{3/2}
    \nonumber\\&&
    \sim 4\times 10^{-9}
    \left( \frac{a_i}{a_{\rm END}}\right)^{3/2} M_{\rm pl}   ,
    \end{eqnarray}
where we have taken $10^{-1}H_i\sim H_{\rm END}\sim 10^{-6}M_{\rm pl}$. 
Finally, taking initial conditions at more than $22$ $e$-folds before the end of inflation, we will see that this contribution is less than $2\times 10^{-23}M_{\rm pl}\sim 5\times 10^{-2}$ MeV, 
which is not significant to the total reheating temperature. To preserve the success of BBN, the reheating temperature has to be greater than $1$ MeV.

Next, we will find the relevant modes in the particle production. Since particle production ends when the $\beta$-Bogoliubov coefficients stabilize, we see that, during  particle production, for modes satisfying 
$k> a_{\rm st} m_{\chi}$, this dynamical equation  can be approximated by
\begin{eqnarray}
    \chi_k''+k^2\chi_k=0,
\end{eqnarray}
which corresponds to mass-less Minkowskian modes, and thus there is no particle production. As a consequence, the modes that contribute to the particle production are in the range
$a_iH_i\leq k\leq a_{\rm st}m_{\chi}$ where
\begin{align}
    a_i\leq e^{-22}a_{\rm END}\quad \mbox{and}\quad
    a_{\rm st}\geq \left(10^7\frac{H_{\rm END}}{m_{\chi}}\right)^{\frac{n+1}{3n}}a_{\rm END}.
\end{align}
Choosing $a_i=a_{*}$ where the star ``$*$'' denotes  the horizon crossing,
which in all inflationary models occurs for more than
 $22$~$e$-folds to the end of inflation, 
we can conclude that the relevant modes are those that, at the horizon crossing, are inside the Hubble horizon and, at the end of particle production, are outside the mass horizon 
$1/m_{\chi}$, i.e., the ones satisfying $a_*H_*<k<a_{\rm st}m_{\chi}$,
with $a_{\rm st}\sim
\left(10^7H_{\rm END}/m_{\chi} \right)^{\frac{n+1}{3n}}a_{\rm END}$, which for $m_{\chi}\sim 10^{-2}H_{\rm END}$ is of the order $a_{\rm st}\sim e^{\frac{7(n+1)}{n}}a_{\rm END}$.
At this point,  it is important to realize that 
for any individual mode, the initial conditions
$\alpha_k(t_i)=1$ and 
$\beta_k(t_i)=0$
can be chosen well after of the 
$22$~$e$-folds before the end of inflation.
For relevant modes
satisfying $a_{\rm END}H_{\rm END}\leq k\leq a_{\rm st}m_{\chi}$ one can take initial conditions at the end of inflation. Effectively, for these  modes we have: 
\begin{eqnarray}
    \frac{\omega_{k,i}'}{\omega_{k,i}^2}=\frac{m_{\chi}^2a_i^3H_i}{\omega_{k,i}^3}\leq \frac{m_{\chi}^2H_i}{H^3_{\rm END}}
    \left(\frac{a_i}{a_{\rm END}}\right)^3,   
    \end{eqnarray}
and taking $a_i\sim a_{\rm END}$ we conclude that, for modes in the range $a_{\rm END}H_{\rm END}\leq k\leq a_{\rm st}m_{\chi}$, the adiabatic condition $\frac{\omega_{\rm k,END}'}{\omega_{\rm k, END}^2}\leq \frac{m_{\chi}^2}{H_{\rm END}^2}\ll 1$
is satisfied. 
And for modes satisfying $k\leq a_{\rm END}H_{\rm END}$, we write $k=e^{-M}a_{\rm END}H_{\rm END}$ with $M>0$, obtaining:
\begin{eqnarray}
    \frac{\omega_{k,i}'}{\omega_{k,i}^2}&&=\frac{m_{\chi}^2a_i^3H_i}{\omega_{k,i}^3}\leq \frac{m_{\chi}^2H_i}{H^3_{\rm END}}
    \left(e^M\frac{a_i}{a_{\rm END}}\right)^3\nonumber\\&&\sim 10\frac{m_{\chi}^2}{H_{\rm END}^2} \left(e^M\frac{a_i}{a_{\rm END}}\right)^3,    
\end{eqnarray}
where during inflation, since the Hubble rate practically does not change, we have taken $H_i\sim 10 H_{\rm END}$. Then,  taking the number of $e$-folds from the initial time $t_i$ to the end of inflation equal to $(M+1)$, and since we are assuming $m_{\chi}\ll H_{\rm END}$,  we find
\begin{eqnarray}
    \frac{\omega_{k,i}'}{\omega_{k,i}^2}\leq 5\times 10^{-1}\frac{m_{\chi}^2}{H_{\rm END}^2}\ll 1,
\end{eqnarray}    
and thus, the adiabatic condition is fulfilled.

To end this discussion we will find a generic upper bound of the maximum reheating temperature.
From the equations (\ref{Bogoliubov}), and making the approximation $\alpha(t)\equiv 1$, we find the bound: 
\begin{eqnarray}\label{betabound}
|\beta_k|&&\leq \int_{t_i}^{t_{\rm st}}\frac{\dot{\omega}_k(t)}{2\omega_k(t)}dt=
\int_{t_i}^{t_{\rm st}}
\frac{m_{\chi}^2a^2(t)H(t)}{2\omega_k^2(t)}dt\nonumber\\
&&\leq\frac{m_{\chi}}{4k}
\int_{t_i}^{t_{\rm st}}a(t) H(t)dt
\cong \frac{m_{\chi}a_{\rm st}}{4k},
\end{eqnarray}
where we have used the inequality
$\omega_k^2(t)\geq 2km_{\chi}a(t)$.
Then,  taking into account (\ref{stabilization}), we find:
\begin{eqnarray}
\langle \rho_{\rm END}\rangle&&\leq
    \frac{m_{\chi}^3a_{st}^2}{32\pi^2 a_{\rm END}^3}\int_{a_iH_i}^{a_{st}m_{\chi}}1 dk
    \cong \frac{m_{\chi}^4}{32\pi^2}
    \left(\frac{a_{st}}{a_{\rm END}}\right)^3
   \nonumber\\ &&\cong
\frac{m_{\chi}^4}{32\pi^2}\left(10^7\frac{H_{\rm END}}{m_{\chi}}\right)^{\frac{n+1}{n}}.    \end{eqnarray}

Therefore we have:
\begin{eqnarray}
   \Theta_1\leq 
   \frac{1}{96\pi^2}10^{\frac{7(n+1)}{n}}
   \frac{m_{\chi}^{\frac{3n-1}{n}}
   H_{\rm END}^{\frac{1-n}{n}}}{M_{\rm pl}^2},
\end{eqnarray}
and recalling  that the maximum reheating temperature is given by
$T_{\rm reh}^{\rm max}(n)\cong 5\times 10^{-1}\Theta_1^{\frac{n}{2(n-1)}}\sqrt{ H_{\rm END}M_{\rm pl}}$,
we get:
\begin{eqnarray}
 T_{\rm reh}^{\rm max}(n)
&&
 \leq
    5\times 10^{-1}\left[\frac{1}{(96\pi^2)^n}\left(\frac{10^7m_{\chi}}{M_{\rm pl}}\right)^{n+1}
    \right]^{\frac{1}{2(n-1)}}m_{\chi} \nonumber\\ 
    && \leq 2\times 10\left(\frac{10^4m_{\chi}}{M_{\rm pl}}\right)^{\frac{n+1}{2(n-1)}}m_{\chi} 
    ,
\end{eqnarray}
and thus,
\begin{eqnarray}
 T_{\rm reh}^{\rm max}(n)&&\leq T_{\rm reh}^{\rm max}(\infty)\leq
    \frac{5}{4\pi}\times 10^{2}
    \sqrt{\frac{5m_{\chi}}{3M_{\rm pl}}}m_{\chi} \nonumber\\
&& \leq 2\times 10^{-9}M_{\rm pl}\sim 5\times 10^9 \mbox{ GeV},
        \end{eqnarray}
where we have taken  $m_{\chi}\leq 10^{-7}M_{\rm pl}$. Of course, this is a very high upper bound, based on the constraint made in   (\ref{betabound}). 
However,  in all models the relevant modes always satisfy  $|\beta_k|\ll 1$ what results in a lower upper bound, and in fact,  the maximum reheating temperature  always overpass the gravitino problem, that is, it is lower than $10^9$ GeV (see ~\cite{Ellis:1982yb} for details).

\subsection{Numerical results}
Having in mind these qualitative considerations, 
we will  consider the potential (\ref{potential})
with $n=3$.  As we have already seen in the formulas (\ref{END}) and (\ref{lambda}) of the previous Section,  when
$n=3$, we have $H_{\rm END}\cong 9\times 10^{-7} M_{\rm pl}$, $\varphi_{\rm END}\cong 1.8 M_{\rm pl}$ and taking into account that at the end of inflation $\dot{\varphi}^2_{\rm END}=V_{\rm END}$ we find $\dot{\varphi}_{\rm END}\cong 3\times 10^{-6} M_{\rm pl}^2$.
Then, 
taking  these initial conditions, we  numerically solve the conservation equation
$\ddot{\varphi}+3H(t)\dot{\varphi}+\partial_{\varphi}V(\varphi)=0$, where
$H(t)=\frac{1}{\sqrt{3}M_{\rm pl}}
\sqrt{\dot{\varphi}^2/2+V(\varphi)}
$. Having the evolution of the inflaton field we also  have the evolution of the Hubble rate, and thus integrating
$H(t)=\dot{a}(t)/a(t)$, we obtain the evolution of the scale factor. 

Once we have obtained the evolution of the scale factor, 
choosing  $m_{\chi}\cong 10^{-2}H_{\rm END}$
we have numerically computed the Bogoliubov coefficients for different modes by solving the following differential equations:
\begin{subequations}
\begin{align}
   && \text{Re}\{\ddot{\alpha}_k\}(t)=\text{Re}\{\dot{\alpha}_k\}(t)
   \left(
\frac{\dot{H}(t)\omega_k^2(t)+2{H}^2(t)k^2}{H(t)\omega_k^2(t)}
   \right)-
   \nonumber\\&&2\,\text{Im}\{\dot{\alpha}_k\}(t)\frac{\omega_k(t)}{a(t)}+\text{Re}\{\alpha_k\}(t)\left(
   \frac{H(t)(\omega_k^2(t)-k^2)}
   {2\omega_k^2(t)}\right)^2,   
\end{align}
\begin{align}
   && \text{Im}\{\ddot{\alpha}_k\}(t)=\text{Im}\{\dot{\alpha}_k\}(t)
    \left(
\frac{\dot{H}(t)\omega_k^2(t)+2{H}^2(t)k^2}{H(t)\omega_k^2(t)}
   \right)+   
   \nonumber\\&&2\,\text{Re}\{\dot{\alpha}_k\}(t)\frac{\omega_k(t)}{a(t)}+\text{Im}\{\alpha_k\}(t)\left(
   \frac{H(t)(\omega_k^2(t)-k^2)}
   {2\omega_k^2(t)}\right)^2,   
\end{align}
\end{subequations}
obtained from (\ref{Bogoliubov}) and  using that $|\beta_k(t)|^2=|\alpha_k(t)|^2-1$.
We have been able to verify that the value of $|\beta_k|^2$ attains a non-trivial value for modes $k<e^{5}a_{\rm END}H_{\rm END}$ and the contribution of the modes satisfying $k\leq e^{-3}a_{\rm END}H_{\rm END}$ to the value of $ \langle \rho_{\rm END}\rangle$ is negligible.
That is, we have  numerically checked that the relevant modes that contribute to the reheating temperature are in the range:
\begin{eqnarray}
    e^{-3}a_{\rm END}H_{\rm END}\leq k\leq 
    e^{5}a_{\rm END}H_{\rm END}.    \end{eqnarray}

\begin{figure}[ht]
    \centering
\includegraphics[width=0.9\linewidth]{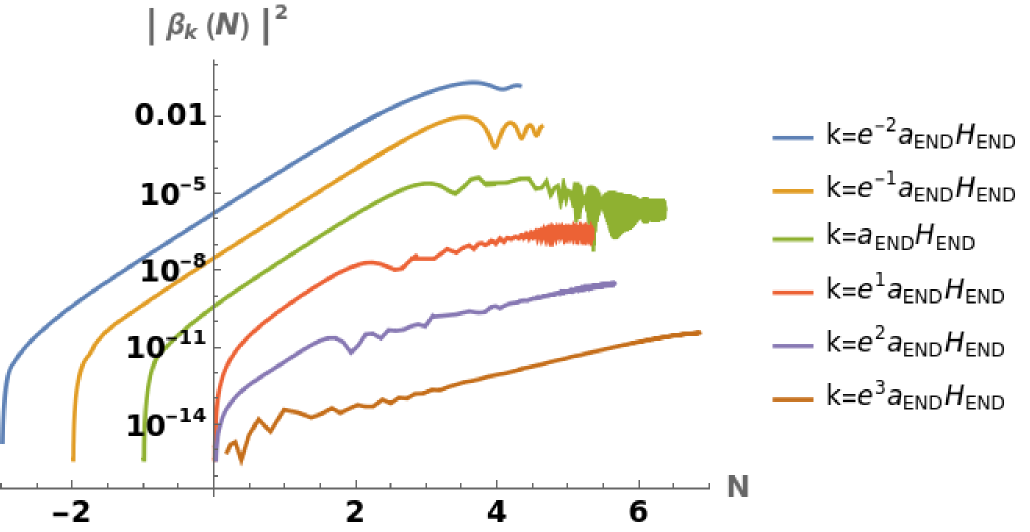}
\caption{Stabilisation of the value of $|\beta_k|^2$ 
for $m_{\chi}=10^{-2}H_{\rm END}$ in function of the number of $e$-folds after the end of inflation for different modes.}
    \label{fig:betes}
\end{figure}

In Fig. 
\ref{fig:betes} we show the stabilisation of the value of $|\beta_k|^2$ for some of the contribution modes in function of the $e$-folds after the end of inflation.

Finally, these values let us approximate the value of  $\langle \rho_{\rm END}\rangle$ via a Riemann sum as:
\begin{eqnarray}
    \langle \rho_{\rm END}\rangle\cong 3\times 10^{-3}H_{\rm END}^2m_{\chi}^2,
\end{eqnarray}
which is of the same order as the result obtained in \cite{Kolb:2023ydq} (see the formula (\ref{particle_kolb})). 
Therefore, 
\begin{eqnarray}\Theta_1\cong 10^{-3}\left(\frac{m_{\chi}}{M_{\rm pl}}\right)^2=10^{-7}\left(\frac{H_{\rm END}}{M_{\rm pl}}\right)^2,
\end{eqnarray}
and the maximum reheating temperature is given by:
\begin{align}
    T_{\rm reh}^{\rm max}(3)\cong 3\times 10^{-6} \frac{H_{\rm END}^2}{M_{\rm pl}}\cong 2\times 10^{-18}M_{\rm pl}\sim 5 \mbox{ GeV}. 
\end{align}

\section{ Analytic results: Born approximation}
\label{sec-5}

In this section we will follow the theory developed in ~\cite{Shtanov:1994ce}.  We consider an auxiliary scalar field $\Phi$,  defined by the relation
$V_n(\Phi(t))=\rho_{\rm B}(t)$ where $V_n$ is defined in (\ref{potential}). Now during the oscillation the potential satisfies the equation 
\begin{eqnarray}
    \dot{\Phi}\cong -\frac{3}{n+1}H\Phi,
\end{eqnarray}
whose solution is given by  
\begin{eqnarray}
    \Phi(t)=\Phi_{\rm END}\left(
  \frac{a_{\rm END}}{a(t)}  \right)^{\frac{3}{n+1}}.
\end{eqnarray}
On the other hand, the scale factor evolves as \cite{Ema:2015dka}:
\begin{eqnarray}
a(t)\cong \bar{a}(t)\left(1-
    \frac{1}{4(n+1)}\frac{\varphi^2}{M_{\rm pl}^2}
    \right),
\end{eqnarray}
where $\bar{a}(t)=a_{\rm END}\left(t/t_{\rm END}\right)^{
    \frac{n+1}{3n}}$. Next we make the approximation
$\bar{a}(t)=a_{\rm END}$ and
$a^2(t)\cong a_{\rm END}^2\left(1-
 \frac{1}{2(n+1)}\frac{\varphi^2(t)}{M_{\rm pl}^2}\right)$, which  means that the mode-frequencies can be approximated by 
\begin{eqnarray}
    \omega_k^2(t)\cong
    k^2+a_{\rm END}^2m_{\chi}^2
-\frac{a_{\rm END}^2m_{\chi}^2}{2(n+1)}\frac{\varphi^2(t)}{M_{\rm pl}^2},    
\end{eqnarray}
where we can see that this approximation is equivalent,
at Lagrangian level, 
to a quadratic interaction 
$-\frac{a_{\rm END}^2m_{\chi}^2}{2(n+1)
M_{\rm pl}^2}{\varphi^2(t)}\chi^2(t)$,
between the inflaton and the massive scalar quantum field $\chi$.  Then, for this frequency, the creation rate of $\chi$-particles, satisfying $m_{\chi}<\omega(t)$,  is \cite{Shtanov:1994ce}
\begin{eqnarray}\label{decayrate}
    \Gamma_{\chi}(t)\sim \frac{m_{\chi}^4}{16\pi(n+1)^2\omega(t)
    M_{\rm pl}^4}\Phi^2(t),
\end{eqnarray}
and the leading frequency $\omega$ is approximately 
\begin{eqnarray}
    \omega(t)\sim \frac{\sqrt{\rho_B(t)}}{\Phi(t)}
    \Longrightarrow
    \omega_{\rm END}\sim 
\frac{\sqrt{3}H_{\rm END}M_{\rm pl}}{\varphi_{\rm END}},
\end{eqnarray}
where we have used that
$\Phi_{\rm END}\sim \varphi_{\rm END}$, because at the end of reheating, 
$\rho_{\rm END}=V(\Phi_{\rm END})=\frac{3}{2}V(\varphi_{\rm END})$.
Note also that, in the quadratic case
$V(\varphi)=\frac{1}{2}m_{\varphi}^2\varphi^2$, one gets $\omega_{\rm END}\sim m_{\varphi}$,  which is the expected result in the quadratic case. 

\begin{remark}
    It is important to realize that we have started with the Bogoliubov approach, i.e., considering the coupling of the massive scalar quantum field with the Ricci scalar,
    and after the Born approximation we have shown that it is equivalent to Boltzmann approach with the decay rate appearing in (\ref{decayrate}). Our approach is not equivalent to the one performed in \cite{Clery:2021bwz,Garcia:2020wiy,Garcia:2020eof}
    where the authors assume that the inflaton field is directly coupled to a light quantum field (scalars and fermions of the Standard Model) and use the usual form of the decay rate of the 
inflaton field in pairs of fermions and in pairs of scalars
\cite{Garcia:2020wiy}.
\end{remark}

\begin{table*}
\centering
 \title{table}\begin{tabular}{|c|c|c|c|}
 \hline
 & Reheating Temperature & Upper bound& Viable masses\\
 \hline
 & & & \\
 $n=3$& $T_{\rm reh}^{\rm max}\cong  10^4\left(\frac{m_{\chi}}{M_{\rm pl}}\right)^3M_{\rm pl}$& $<  10^{-17}M_{\rm pl}\sim 20 \mbox{GeV} $ & $10^{10}\mbox{ GeV}< m_{\chi}<
2\times 10^{11}\mbox{ GeV}$    \\
& & & \\
\hline 
& & & \\
$n=4$ &  $T_{\rm reh}^{\rm max}\cong 2\times 10^3\left(\frac{m_{\chi}}{M_{\rm pl}}\right)^{8/3}M_{\rm pl} $ & $<  3\times10^{-16}M_{\rm pl}\sim 7\times 10^2\mbox{GeV}$
 &  $10^{9}\mbox{ GeV}< m_{\chi}<
2\times 10^{11}\mbox{ GeV}$  \\
& & & \\
\hline 
& & & \\
$n=5$& $T_{\rm reh}^{\rm max}\cong 6\times 10^2\left(\frac{m_{\chi}}{M_{\rm pl}}\right)^{5/2}M_{\rm pl}$ & $<  2\times10^{-15}M_{\rm pl}\sim 5\times 10^3\mbox{GeV}$
 & $5\times 10^{8}\mbox{ GeV}< m_{\chi}<
2\times 10^{11}\mbox{ GeV}$   \\
& & & \\
\hline
& & & \\
$n=6$ & $T_{\rm reh}^{\rm max}\cong 4\times 10^2\left(\frac{m_{\chi}}{M_{\rm pl}}\right)^{12/5}M_{\rm pl}$ & $<6\times10^{-15}M_{\rm pl}\sim 10^4\mbox{GeV}$ & $2\times 10^{8}\mbox{ GeV}< m_{\chi}<
2\times 10^{11}\mbox{ GeV}$ \\
& & & \\
\hline
& & & \\
$n=7$ & $T_{\rm reh}^{\rm max}\cong 2\times 10^2\left(\frac{m_{\chi}}{M_{\rm pl}}\right)^{7/3}M_{\rm pl}$  & 
 $<  10^{-14}M_{\rm pl}\sim 2\times 10^4\mbox{GeV}$
 & $ 2\times 10^{8}\mbox{ GeV}< m_{\chi}<
2\times 10^{11}\mbox{ GeV}$\\
& & & \\
\hline
\end{tabular}
\caption{For masses satisfying $m_{\chi}<10^{-7}M_{\rm pl}$, we show the maximum reheating temperature 
and viable values of the mass
for different values of the parameter $n$. }
\label{tab:temperature}
\end{table*}
Next, we use the 
Boltzmann equation for massive particles 
\begin{eqnarray}
\frac{d\langle\rho(t)\rangle}{dt}+3H(t)\langle\rho(t)\rangle=\Gamma_{\chi}(t)\rho_{\rm B}(t),
\end{eqnarray}
which after one Hubble time  after the end of inflation can be approximated by 
\begin{eqnarray}
    \langle \rho_{\rm END}\rangle\sim
    \frac{\Gamma_{\chi,\rm END}}{H_{\rm END}}\rho_{\rm B,\rm END},
\end{eqnarray}
and thus,
\begin{eqnarray}\label{Theta1}
\Theta_1&&=\frac{\langle\rho_{\rm END}\rangle}{\rho_{\rm B,\rm END}}\cong 
\frac{\Gamma_{\chi,\rm END}}{H_{\rm END}} \nonumber\\
&& \cong
\frac{1}{16\sqrt{3}\pi (n+1)^2}
\left(\frac{m_{\chi}}{M_{\rm pl}}\right)^4\frac{\varphi_{\rm END}^3}{H_{\rm END}^2M_{\rm pl}}.
\end{eqnarray}
On the other hand, in order to obtain $\varphi_{\rm END}$ and $H_{\rm END}$,  we solve $\epsilon\equiv\frac{M_{\rm pl}^2}{2}\left(\frac{dV_n}{d \varphi}/V_n\right)^2=1$ obtaining:
\begin{eqnarray}\label{END}
&&\varphi_{\rm END}=-\sqrt{\frac{3}{2}}\ln\left(\frac{2\sqrt{3}n-3}{4n^2-3} \right)M_{\rm pl}
\Longrightarrow\nonumber\\
    && 
H_{\rm END}=\sqrt{\frac{\lambda}{2}}
    \left(1- \frac{2\sqrt{3}n-3}{4n^2-3}    \right)^nM_{\rm pl}.
\end{eqnarray}
And to find the value of $\lambda$ we need to relate the value of main slow roll parameter at the horizon crossing, with the spectral index of scalar perturbation. To do it, note that
$H_*^2\cong \frac{\lambda}{3}M_{\rm pl}^2$. Next, we calculate the slow roll parameters at the horizon crossing
\begin{eqnarray}
    \epsilon_*\cong \frac{4n^2}{3}
    e^{-2\sqrt{\frac{2}{3}}\frac{\varphi_*}{M_{\rm pl}}},\qquad
    \eta_*\cong -\frac{4n}{3}e^{-\sqrt{\frac{2}{3}}\frac{\varphi_*}{M_{\rm pl}}},\end{eqnarray}
and using the relation $1-n_s\cong 6\epsilon_*-2\eta_*$, we find
\begin{eqnarray}
    1-n_s\cong \frac{8n}{3}e^{-\sqrt{\frac{2}{3}}\frac{\varphi_*}{M_{\rm pl}}},\end{eqnarray}
and thus,
\begin{eqnarray}
    \epsilon_*=\frac{3}{16}(1-n_s)^2.
\end{eqnarray}
Finally, using the formula of the power spectrum of scalar perturbations, 
we get
\begin{eqnarray}\label{lambda}
    \lambda\sim 3\pi^2(1-n_s)^2 10^{-9}\cong 5\times 10^{-11},
\end{eqnarray}
where as usual, we have taken $n_s=0.96$ as a typical value which is inferred by the Planck 2018  collaboration~\cite{Planck:2018jri}.
Therefore, from (\ref{Theta1}), (\ref{END}) and (\ref{lambda}) we arrive at
\begin{align}
   && \Theta_1\cong  10^9\frac{1}{(n+1)^2}
    \left(\frac{m_{\chi}}{M_{\rm pl}}\right)^4  
    \ln^3\left(\frac{4n^2-3}{2\sqrt{3}n-3} \right)\nonumber\\ &&\times
    \left(1- \frac{2\sqrt{3}n-3}{4n^2-3}  \right)^{-2n},    
\end{align}
and the maximum reheating temperature is given by:
\begin{align}
T_{\rm reh}^{\rm max}(n)\cong
4\times 10^{\frac{2n+7}{2(n-1)}}
\frac{1}{(1+n)^{\frac{n}{n-1}}}
    \left(\frac{m_{\chi}}{M_{\rm pl}}\right)^{\frac{2n}{n-1}}\nonumber\\ 
    \times \left(\ln\left(
    \frac
    {4n^2-3}{2\sqrt{3}n-3} \right)\right)^{
    \frac{3n}{2(n-1)}}    
    \left(1- \frac{2\sqrt{3}n-3}{4n^2-3}  \right)^{-\frac{n(n+1)}{n-1}} M_{\rm pl}.     
    \end{align}
In Table~\ref{tab:temperature}
we show, for different values of $n$
and masses satisfying $m_{\chi}\ll\omega_{\rm END}\sim 10^{-6}M_{\rm pl}$, the expression of the maximum reheating temperature, its bound and the range of viable values of the mass. From the values of the Table~\ref{tab:temperature}, we can deduce that, for moderate values of $n$,  as 
$n$ increases, the maximum reheating temperature increases, and the range of viable values for the mass also becomes larger.

A final remark is in order: By performing the Born approximation, we have obtained an interaction term between the inflaton and a massive scalar quantum field, which induces the production of heavy particles. The evolution of their energy density is governed by the Boltzmann equation, and these particles must decay into lighter ones to reheat the universe. This approach differs from that of \cite{Garcia:2020wiy,Kaneta:2022gug,Garcia:2020eof}, where the authors use an interaction between the inflaton and light particles, thus avoiding the need for the decay of the produced particles.

\section{Conclusions}
\label{sec-summary}

In the present work, we have studied the reheating temperature for inflationary potentials where gravitational reheating, via production of massive particles conformally coupled with gravity,  is a viable mechanism. We have demonstrated that during the oscillations of the inflaton, achieving viable gravitational reheating requires the Universe to have an effective EoS parameter greater than
$1/3$, because this ensures that the energy density of the inflaton decreases rapidly enough to avoid re-domination after the initial reheating phase. This condition is critical for maintaining the consistency of the post-inflationary Universe with the well-established $\Lambda$CDM model.

We have also derived the formulae for the reheating temperature in two scenarios: when the decay of the heavy massive particles occurs before and after the domination of the inflaton's energy density. Additionally, we have established bounds for the maximum reheating temperature, which corresponds to the reheating temperature when the decay occurs at the onset of the domination of the energy density of the produced particles.
In particular, we have seen that for masses smaller than $10^{-7} M_{\rm pl}$, the reheating temperature is below $10^9$ GeV, thereby avoiding the gravitino problem. Additionally, we have constrained the masses that lead to a viable reheating temperature.

Our study contributes to a deeper understanding of gravitational reheating, offering insights into the necessary conditions and constraints for successful reheating in the context of different inflationary models. By providing explicit formulas and bounds, we aim to guide future research in refining these models and ensuring their compatibility with the observed Universe.

\section*{Acknowledgments} 
We thank the referee for some useful comments which improved the manuscript.  JdH is supported by the Spanish grants PID2021-123903NB-I00 and RED2022-134784-T
funded by MCIN/AEI/10.13039/501100011033 and by ERDF ``A way of making Europe''.
LAS is supported by an LMS Early Career Fellowship. SP acknowledges the financial support from the Department of Science and Technology (DST), Govt. of India under the Scheme   ``Fund for Improvement of S\&T Infrastructure (FIST)'' (File No. SR/FST/MS-I/2019/41).

\section*{Appendix-A: The decay process}

To study in detail the decay process, we use the dynamics described by the Boltzmann equations  (\ref{Boltzmann}), where the constant decay rate
 $\Gamma$,
 represents the decay of heavy particles into light fermions. Specifically, 
   $\Gamma \sim \frac{h^2 m_{\chi}}{8\pi}$, where
$h$ is a dimensionless constant
\cite{Kofman:1994rk}.

We choose the following as the solution for the energy density of the massive particles:
\begin{eqnarray}
    \langle\rho(t)\rangle=
    \langle\rho_{\rm st}\rangle\left(
  \frac{a_{\rm st}}{a(t)}  \right)^3
  e^{-\Gamma(t-t_{\rm st})},  \qquad t\geq t_{\rm st},
\end{eqnarray}
where the subscript ``{\rm st}'', as previously explained, denotes the time at which the $\beta$-Bogoliubov coefficient stabilizes, i.e., the time when particle production has ceased. We can see that this solution models a decay that begins when the particles are produced, which occurs a few 
$e$-folds after the end of inflation.

Inserting this solution into the second equation of (\ref{Boltzmann}) and imposing once again that the decay begins when all the heavy particles have been created, we obtain:
\begin{eqnarray}
    \langle\rho_{\rm r}(t)\rangle=\langle\rho_{\rm st}\rangle\left(
  \frac{a_{\rm st}}{a(t)}  \right)^4
  \int_{t_{\rm st}}^t \frac{a(s)}{a_{\rm st}}\Gamma
  e^{-\Gamma(s-t_{\rm st})}ds.
\end{eqnarray}

We define the time at which decay ends, denoted as $t_{\rm dec}$,  as the time when $\Gamma(t_{\rm dec}-t_{\rm st})\sim 1$, which gives
$t_{\rm dec}\sim t_{\rm st}+\frac{1}{\Gamma}$, and we want
to examine the evolution of the decay products when $t\gg t_{\rm dec}$. 
Assuming that the background dominates, we will have
$a(s)\cong a_{\rm st}\left(\frac{s}{t_{\rm st}} \right)^{\frac{n+1}{3n}}$, and taking into account that 
\begin{eqnarray}
    \int_{t_{\rm st}}^{\infty}
\left(s/t_{\rm st} \right)^{\frac{n+1}{3n}}\Gamma
e^{-\Gamma(s-t_{\rm st})}ds\nonumber\\\cong 
\left( \frac{H_{\rm st}}{\Gamma}\right)^{\frac{n+1}{3n}}\Gamma_{\rm Euler}
\left( \frac{4n+1}{3n} \right)\cong \left( \frac{H_{\rm st}}{\Gamma}\right)^{\frac{n+1}{3n}}\nonumber\\=\left( \frac{H_{\rm END}}{\Gamma}\right)^{\frac{n+1}{3n}}
\frac{a_{\rm END}}{a_{\rm st}},
\end{eqnarray}
where we have denoted by $\Gamma_{\rm Euler}$ the Euler's Gamma function, we  conclude that
 after the decay, i.e., for $t>t_{\rm dec}$,  we can safely make the approximation:
\begin{eqnarray}
\langle\rho_{\rm r}(t)\rangle\cong\langle\rho_{\rm st}\rangle\left( \frac{H_{\rm END}}{\Gamma}\right)^{\frac{n+1}{3n}}
\frac{a_{\rm END}}{a_{\rm st}}
\left(
  \frac{a_{\rm st}}{a(t)}  \right)^4\nonumber\\
  \cong \langle\rho_{\rm END}\rangle\left( \frac{H_{\rm END}}{\Gamma}\right)^{\frac{n+1}{3n}}\left(
  \frac{a_{\rm END}}{a(t)}  \right)^4  .
\end{eqnarray}

\
On the other hand, 
since the evolution of the background is given by:
\begin{align}
    \rho_{\rm B}(t)=\rho_{\rm B, st}\left( \frac{a_{\rm st}}{a(t)}\right)^{\frac{6n}{n+1}}=
    \rho_{\rm B, END}\left( \frac{a_{\rm END}}{a(t)}\right)^{\frac{6n}{n+1}},
    \end{align}
and  the 
 universe becomes reheated when
$\rho_{\rm B}\sim \langle\rho_{\rm r} \rangle$, we get:
\begin{eqnarray}
    \Theta_1\equiv 
\frac{\langle\rho_{\rm END}\rangle}{\rho_{\rm B, \rm END}}= \left( \frac{\Gamma}{H_{\rm END}}\right)^{\frac{n+1}{3n}}    \left(\frac{a_{\rm END}}{a_{\rm reh}} \right)^
    {\frac{2(n-2)}{n+1}},
\end{eqnarray}
and thus:
\begin{eqnarray}
   \langle\rho_{\rm r, \rm reh}\rangle\cong
   \rho_{\rm B, \rm END} 
   \left(\frac{H_{\rm END}}{\Gamma} \right)^{\frac{n+1}{n-2}}
   \Theta_1^{\frac{3n}{n-2}}.
\end{eqnarray}

Therefore, from the
 Stefan-Boltzmann law, we obtain the following reheating temperature:
\begin{eqnarray}
    T_{\rm reh}(n)=\left( \frac{30}{\pi^2 g_{\rm reh}}\right)^{1/4}\langle\rho_{\rm r, \rm reh}\rangle^{1/4}=\nonumber\\
\left( \frac{90}{\pi^2 g_{\rm reh}}\right)^{1/4}
\left(\frac{H_{\rm END}}{\Gamma} \right)^{\frac{n+1}{4(n-2)}}
\Theta_1^{\frac{3n}{4(n-2)}}\sqrt{H_{\rm END}M_{\rm pl}}\nonumber\\
\cong 5\times 10^{-1}
\left(\frac{H_{\rm END}}{\Gamma} \right)^{\frac{n+1}{4(n-2)}}\Theta_1^{\frac{3n}{4(n-2)}}\sqrt{H_{\rm END}M_{\rm pl}},\end{eqnarray}
which coincides with (\ref{temperature}) when $\rho_{\rm B, \rm dec}\gg \langle\rho_{\rm dec}\rangle$.

\

Finally, we need to determine the range of values for the decay rate 
$\Gamma$. Since the decay occurs well after the creation of the massive particles,  we  have $\Gamma\ll H_{\rm st}$. On the other hand, we have assumed that by the end of the decay, the energy density of the background dominates, i.e.,
$\langle \rho_{\rm r,\rm  dec}\rangle\ll \rho_{\rm B, \rm dec}\cong 3M_{\rm pl}^2\Gamma^2$. To get the bound, we use that the relation 
$\rho_{\rm B, \rm dec}\cong 3 M_{\rm pl}^2\Gamma^2$, leads to:
\begin{eqnarray}
    \left(\frac{a_{\rm END}}{a_{\rm dec}} \right)^4\cong 
    \left( \frac{\Gamma}{H_{\rm END}}\right)^{\frac{4(n+1)}{3n}},
\end{eqnarray}
and thus, inserting it into the relation: 
$\langle \rho_{\rm r,\rm  dec}\rangle\ll  3M_{\rm pl}^2\Gamma^2$, we conclude that:
\begin{eqnarray}
    \Theta_1^{\frac{n}{n-1}}    \ll \frac{\Gamma}{H_{\rm END}}\ll \frac{H_{\rm st}}{H_{\rm END}}.
\end{eqnarray}

Therefore, the maximum temperature is reached when 
$\Theta_1^{\frac{n}{n-1}}    \cong \frac{\Gamma}{H_{\rm END}}$, i.e., when the end of the decay occurs close to the onset of the radiation epoch, 
yielding:
\begin{eqnarray}
    T_{\rm reh}^{\rm max}(n)
\cong 5\times 10^{-1}
\Theta_1^{\frac{n}{2(n-1)}}\sqrt{H_{\rm END}M_{\rm pl}},\end{eqnarray}
which, of course, coincides with (\ref{Tem_max}).

\bibliography{references}

\begin{thebibliography}{120}%
\makeatletter
\providecommand \@ifxundefined [1]{%
 \@ifx{#1\undefined}
}%
\providecommand \@ifnum [1]{%
 \ifnum #1\expandafter \@firstoftwo
 \else \expandafter \@secondoftwo
 \fi
}%
\providecommand \@ifx [1]{%
 \ifx #1\expandafter \@firstoftwo
 \else \expandafter \@secondoftwo
 \fi
}%
\providecommand \natexlab [1]{#1}%
\providecommand \enquote  [1]{``#1''}%
\providecommand \bibnamefont  [1]{#1}%
\providecommand \bibfnamefont [1]{#1}%
\providecommand \citenamefont [1]{#1}%
\providecommand \href@noop [0]{\@secondoftwo}%
\providecommand \href [0]{\begingroup \@sanitize@url \@href}%
\providecommand \@href[1]{\@@startlink{#1}\@@href}%
\providecommand \@@href[1]{\endgroup#1\@@endlink}%
\providecommand \@sanitize@url [0]{\catcode `\\12\catcode `\$12\catcode
  `\&12\catcode `\#12\catcode `\^12\catcode `\_12\catcode `\%12\relax}%
\providecommand \@@startlink[1]{}%
\providecommand \@@endlink[0]{}%
\providecommand \url  [0]{\begingroup\@sanitize@url \@url }%
\providecommand \@url [1]{\endgroup\@href {#1}{\urlprefix }}%
\providecommand \urlprefix  [0]{URL }%
\providecommand \Eprint [0]{\href }%
\providecommand \doibase [0]{http://dx.doi.org/}%
\providecommand \selectlanguage [0]{\@gobble}%
\providecommand \bibinfo  [0]{\@secondoftwo}%
\providecommand \bibfield  [0]{\@secondoftwo}%
\providecommand \translation [1]{[#1]}%
\providecommand \BibitemOpen [0]{}%
\providecommand \bibitemStop [0]{}%
\providecommand \bibitemNoStop [0]{.\EOS\space}%
\providecommand \EOS [0]{\spacefactor3000\relax}%
\providecommand \BibitemShut  [1]{\csname bibitem#1\endcsname}%
\let\auto@bib@innerbib\@empty
\bibitem [{\citenamefont {Guth}(1981)}]{Guth:1980zm}%
  \BibitemOpen
  \bibfield  {author} {\bibinfo {author} {\bibfnamefont {A.~H.}\ \bibnamefont
  {Guth}},\ }\href {\doibase 10.1103/PhysRevD.23.347} {\bibfield  {journal}
  {\bibinfo  {journal} {Phys. Rev. D}\ }\textbf {\bibinfo {volume} {23}},\
  \bibinfo {pages} {347} (\bibinfo {year} {1981})}\BibitemShut {NoStop}%
\bibitem [{\citenamefont {Sato}(1981)}]{Sato:1980yn}%
  \BibitemOpen
  \bibfield  {author} {\bibinfo {author} {\bibfnamefont {K.}~\bibnamefont
  {Sato}},\ }\href@noop {} {\bibfield  {journal} {\bibinfo  {journal} {Mon.
  Not. Roy. Astron. Soc.}\ }\textbf {\bibinfo {volume} {195}},\ \bibinfo
  {pages} {467} (\bibinfo {year} {1981})}\BibitemShut {NoStop}%
\bibitem [{\citenamefont {Linde}(1982)}]{Linde:1981mu}%
  \BibitemOpen
  \bibfield  {author} {\bibinfo {author} {\bibfnamefont {A.~D.}\ \bibnamefont
  {Linde}},\ }\href {\doibase 10.1016/0370-2693(82)91219-9} {\bibfield
  {journal} {\bibinfo  {journal} {Phys. Lett. B}\ }\textbf {\bibinfo {volume}
  {108}},\ \bibinfo {pages} {389} (\bibinfo {year} {1982})}\BibitemShut
  {NoStop}%
\bibitem [{\citenamefont {Starobinsky}(1980)}]{Starobinsky:1980te}%
  \BibitemOpen
  \bibfield  {author} {\bibinfo {author} {\bibfnamefont {A.~A.}\ \bibnamefont
  {Starobinsky}},\ }\href {\doibase 10.1016/0370-2693(80)90670-X} {\bibfield
  {journal} {\bibinfo  {journal} {Phys. Lett. B}\ }\textbf {\bibinfo {volume}
  {91}},\ \bibinfo {pages} {99} (\bibinfo {year} {1980})}\BibitemShut {NoStop}%
\bibitem [{\citenamefont {Barrow}\ and\ \citenamefont
  {Turner}(1981)}]{Barrow:1981pa}%
  \BibitemOpen
  \bibfield  {author} {\bibinfo {author} {\bibfnamefont {J.~D.}\ \bibnamefont
  {Barrow}}\ and\ \bibinfo {author} {\bibfnamefont {M.~S.}\ \bibnamefont
  {Turner}},\ }\href {\doibase 10.1038/292035a0} {\bibfield  {journal}
  {\bibinfo  {journal} {Nature}\ }\textbf {\bibinfo {volume} {292}},\ \bibinfo
  {pages} {35} (\bibinfo {year} {1981})}\BibitemShut {NoStop}%
\bibitem [{\citenamefont {Lucchin}\ and\ \citenamefont
  {Matarrese}(1985)}]{Lucchin:1984yf}%
  \BibitemOpen
  \bibfield  {author} {\bibinfo {author} {\bibfnamefont {F.}~\bibnamefont
  {Lucchin}}\ and\ \bibinfo {author} {\bibfnamefont {S.}~\bibnamefont
  {Matarrese}},\ }\href {\doibase 10.1103/PhysRevD.32.1316} {\bibfield
  {journal} {\bibinfo  {journal} {Phys. Rev. D}\ }\textbf {\bibinfo {volume}
  {32}},\ \bibinfo {pages} {1316} (\bibinfo {year} {1985})}\BibitemShut
  {NoStop}%
\bibitem [{\citenamefont {Hawking}(1985)}]{Hawking:1984kj}%
  \BibitemOpen
  \bibfield  {author} {\bibinfo {author} {\bibfnamefont {S.~W.}\ \bibnamefont
  {Hawking}},\ }\href {\doibase 10.1016/0370-2693(85)90989-X} {\bibfield
  {journal} {\bibinfo  {journal} {Phys. Lett. B}\ }\textbf {\bibinfo {volume}
  {150}},\ \bibinfo {pages} {339} (\bibinfo {year} {1985})}\BibitemShut
  {NoStop}%
\bibitem [{\citenamefont {Lyth}(1985)}]{Lyth:1984gv}%
  \BibitemOpen
  \bibfield  {author} {\bibinfo {author} {\bibfnamefont {D.~H.}\ \bibnamefont
  {Lyth}},\ }\href {\doibase 10.1103/PhysRevD.31.1792} {\bibfield  {journal}
  {\bibinfo  {journal} {Phys. Rev. D}\ }\textbf {\bibinfo {volume} {31}},\
  \bibinfo {pages} {1792} (\bibinfo {year} {1985})}\BibitemShut {NoStop}%
\bibitem [{\citenamefont {Belinsky}\ \emph {et~al.}(1985)\citenamefont
  {Belinsky}, \citenamefont {Khalatnikov}, \citenamefont {Grishchuk},\ and\
  \citenamefont {Zeldovich}}]{Belinsky:1985zd}%
  \BibitemOpen
  \bibfield  {author} {\bibinfo {author} {\bibfnamefont {V.~A.}\ \bibnamefont
  {Belinsky}}, \bibinfo {author} {\bibfnamefont {I.~M.}\ \bibnamefont
  {Khalatnikov}}, \bibinfo {author} {\bibfnamefont {L.~P.}\ \bibnamefont
  {Grishchuk}}, \ and\ \bibinfo {author} {\bibfnamefont {Y.~B.}\ \bibnamefont
  {Zeldovich}},\ }\href {\doibase 10.1016/0370-2693(85)90644-6} {\bibfield
  {journal} {\bibinfo  {journal} {Phys. Lett. B}\ }\textbf {\bibinfo {volume}
  {155}},\ \bibinfo {pages} {232} (\bibinfo {year} {1985})}\BibitemShut
  {NoStop}%
\bibitem [{\citenamefont {Khalfin}(1986)}]{Khalfin:1986ui}%
  \BibitemOpen
  \bibfield  {author} {\bibinfo {author} {\bibfnamefont {L.~A.}\ \bibnamefont
  {Khalfin}},\ }\href@noop {} {\bibfield  {journal} {\bibinfo  {journal} {Sov.
  Phys. JETP}\ }\textbf {\bibinfo {volume} {64}},\ \bibinfo {pages} {673}
  (\bibinfo {year} {1986})}\BibitemShut {NoStop}%
\bibitem [{\citenamefont {Silk}\ and\ \citenamefont
  {Turner}(1987)}]{Silk:1986vc}%
  \BibitemOpen
  \bibfield  {author} {\bibinfo {author} {\bibfnamefont {J.}~\bibnamefont
  {Silk}}\ and\ \bibinfo {author} {\bibfnamefont {M.~S.}\ \bibnamefont
  {Turner}},\ }\href {\doibase 10.1103/PhysRevD.35.419} {\bibfield  {journal}
  {\bibinfo  {journal} {Phys. Rev. D}\ }\textbf {\bibinfo {volume} {35}},\
  \bibinfo {pages} {419} (\bibinfo {year} {1987})}\BibitemShut {NoStop}%
\bibitem [{\citenamefont {Mijic}\ \emph {et~al.}(1986)\citenamefont {Mijic},
  \citenamefont {Morris},\ and\ \citenamefont {Suen}}]{Mijic:1986iv}%
  \BibitemOpen
  \bibfield  {author} {\bibinfo {author} {\bibfnamefont {M.~B.}\ \bibnamefont
  {Mijic}}, \bibinfo {author} {\bibfnamefont {M.~S.}\ \bibnamefont {Morris}}, \
  and\ \bibinfo {author} {\bibfnamefont {W.-M.}\ \bibnamefont {Suen}},\ }\href
  {\doibase 10.1103/PhysRevD.34.2934} {\bibfield  {journal} {\bibinfo
  {journal} {Phys. Rev. D}\ }\textbf {\bibinfo {volume} {34}},\ \bibinfo
  {pages} {2934} (\bibinfo {year} {1986})}\BibitemShut {NoStop}%
\bibitem [{\citenamefont {Burd}\ and\ \citenamefont
  {Barrow}(1988)}]{Burd:1988ss}%
  \BibitemOpen
  \bibfield  {author} {\bibinfo {author} {\bibfnamefont {A.~B.}\ \bibnamefont
  {Burd}}\ and\ \bibinfo {author} {\bibfnamefont {J.~D.}\ \bibnamefont
  {Barrow}},\ }\href {\doibase 10.1016/0550-3213(88)90135-6} {\bibfield
  {journal} {\bibinfo  {journal} {Nucl. Phys. B}\ }\textbf {\bibinfo {volume}
  {308}},\ \bibinfo {pages} {929} (\bibinfo {year} {1988})},\ \bibinfo {note}
  {[Erratum: Nucl.Phys.B 324, 276--276 (1989)]}\BibitemShut {NoStop}%
\bibitem [{\citenamefont {Olive}(1990)}]{Olive:1989nu}%
  \BibitemOpen
  \bibfield  {author} {\bibinfo {author} {\bibfnamefont {K.~A.}\ \bibnamefont
  {Olive}},\ }\href {\doibase 10.1016/0370-1573(90)90144-Q} {\bibfield
  {journal} {\bibinfo  {journal} {Phys. Rept.}\ }\textbf {\bibinfo {volume}
  {190}},\ \bibinfo {pages} {307} (\bibinfo {year} {1990})}\BibitemShut
  {NoStop}%
\bibitem [{\citenamefont {Ford}(1989)}]{Ford:1989me}%
  \BibitemOpen
  \bibfield  {author} {\bibinfo {author} {\bibfnamefont {L.~H.}\ \bibnamefont
  {Ford}},\ }\href {\doibase 10.1103/PhysRevD.40.967} {\bibfield  {journal}
  {\bibinfo  {journal} {Phys. Rev. D}\ }\textbf {\bibinfo {volume} {40}},\
  \bibinfo {pages} {967} (\bibinfo {year} {1989})}\BibitemShut {NoStop}%
\bibitem [{\citenamefont {Adams}\ \emph {et~al.}(1991)\citenamefont {Adams},
  \citenamefont {Freese},\ and\ \citenamefont {Guth}}]{Adams:1990pn}%
  \BibitemOpen
  \bibfield  {author} {\bibinfo {author} {\bibfnamefont {F.~C.}\ \bibnamefont
  {Adams}}, \bibinfo {author} {\bibfnamefont {K.}~\bibnamefont {Freese}}, \
  and\ \bibinfo {author} {\bibfnamefont {A.~H.}\ \bibnamefont {Guth}},\ }\href
  {\doibase 10.1103/PhysRevD.43.965} {\bibfield  {journal} {\bibinfo  {journal}
  {Phys. Rev. D}\ }\textbf {\bibinfo {volume} {43}},\ \bibinfo {pages} {965}
  (\bibinfo {year} {1991})}\BibitemShut {NoStop}%
\bibitem [{\citenamefont {Freese}\ \emph {et~al.}(1990)\citenamefont {Freese},
  \citenamefont {Frieman},\ and\ \citenamefont {Olinto}}]{Freese:1990rb}%
  \BibitemOpen
  \bibfield  {author} {\bibinfo {author} {\bibfnamefont {K.}~\bibnamefont
  {Freese}}, \bibinfo {author} {\bibfnamefont {J.~A.}\ \bibnamefont {Frieman}},
  \ and\ \bibinfo {author} {\bibfnamefont {A.~V.}\ \bibnamefont {Olinto}},\
  }\href {\doibase 10.1103/PhysRevLett.65.3233} {\bibfield  {journal} {\bibinfo
   {journal} {Phys. Rev. Lett.}\ }\textbf {\bibinfo {volume} {65}},\ \bibinfo
  {pages} {3233} (\bibinfo {year} {1990})}\BibitemShut {NoStop}%
\bibitem [{\citenamefont {Wang}(1991)}]{Wang:1991ww}%
  \BibitemOpen
  \bibfield  {author} {\bibinfo {author} {\bibfnamefont {Y.}~\bibnamefont
  {Wang}},\ }\href {\doibase 10.1103/PhysRevD.44.991} {\bibfield  {journal}
  {\bibinfo  {journal} {Phys. Rev. D}\ }\textbf {\bibinfo {volume} {44}},\
  \bibinfo {pages} {991} (\bibinfo {year} {1991})}\BibitemShut {NoStop}%
\bibitem [{\citenamefont {Linde}(1994)}]{Linde:1993cn}%
  \BibitemOpen
  \bibfield  {author} {\bibinfo {author} {\bibfnamefont {A.~D.}\ \bibnamefont
  {Linde}},\ }\href {\doibase 10.1103/PhysRevD.49.748} {\bibfield  {journal}
  {\bibinfo  {journal} {Phys. Rev. D}\ }\textbf {\bibinfo {volume} {49}},\
  \bibinfo {pages} {748} (\bibinfo {year} {1994})},\ \Eprint
  {http://arxiv.org/abs/astro-ph/9307002} {arXiv:astro-ph/9307002} \BibitemShut
  {NoStop}%
\bibitem [{\citenamefont {Barrow}(1993)}]{Barrow:1993hn}%
  \BibitemOpen
  \bibfield  {author} {\bibinfo {author} {\bibfnamefont {J.~D.}\ \bibnamefont
  {Barrow}},\ }\href {\doibase 10.1103/PhysRevD.48.1585} {\bibfield  {journal}
  {\bibinfo  {journal} {Phys. Rev. D}\ }\textbf {\bibinfo {volume} {48}},\
  \bibinfo {pages} {1585} (\bibinfo {year} {1993})}\BibitemShut {NoStop}%
\bibitem [{\citenamefont {Barrow}(1994)}]{Barrow:1994nt}%
  \BibitemOpen
  \bibfield  {author} {\bibinfo {author} {\bibfnamefont {J.~D.}\ \bibnamefont
  {Barrow}},\ }\href {\doibase 10.1103/PhysRevD.49.3055} {\bibfield  {journal}
  {\bibinfo  {journal} {Phys. Rev. D}\ }\textbf {\bibinfo {volume} {49}},\
  \bibinfo {pages} {3055} (\bibinfo {year} {1994})}\BibitemShut {NoStop}%
\bibitem [{\citenamefont {Vilenkin}(1994)}]{Vilenkin:1994pv}%
  \BibitemOpen
  \bibfield  {author} {\bibinfo {author} {\bibfnamefont {A.}~\bibnamefont
  {Vilenkin}},\ }\href {\doibase 10.1103/PhysRevLett.72.3137} {\bibfield
  {journal} {\bibinfo  {journal} {Phys. Rev. Lett.}\ }\textbf {\bibinfo
  {volume} {72}},\ \bibinfo {pages} {3137} (\bibinfo {year} {1994})},\ \Eprint
  {http://arxiv.org/abs/hep-th/9402085} {arXiv:hep-th/9402085} \BibitemShut
  {NoStop}%
\bibitem [{\citenamefont {Peter}\ \emph {et~al.}(1994)\citenamefont {Peter},
  \citenamefont {Polarski},\ and\ \citenamefont {Starobinsky}}]{Peter:1994dx}%
  \BibitemOpen
  \bibfield  {author} {\bibinfo {author} {\bibfnamefont {P.}~\bibnamefont
  {Peter}}, \bibinfo {author} {\bibfnamefont {D.}~\bibnamefont {Polarski}}, \
  and\ \bibinfo {author} {\bibfnamefont {A.~A.}\ \bibnamefont {Starobinsky}},\
  }\href {\doibase 10.1103/PhysRevD.50.4827} {\bibfield  {journal} {\bibinfo
  {journal} {Phys. Rev. D}\ }\textbf {\bibinfo {volume} {50}},\ \bibinfo
  {pages} {4827} (\bibinfo {year} {1994})},\ \Eprint
  {http://arxiv.org/abs/astro-ph/9403037} {arXiv:astro-ph/9403037} \BibitemShut
  {NoStop}%
\bibitem [{\citenamefont {Sasaki}\ and\ \citenamefont
  {Stewart}(1996)}]{Sasaki:1995aw}%
  \BibitemOpen
  \bibfield  {author} {\bibinfo {author} {\bibfnamefont {M.}~\bibnamefont
  {Sasaki}}\ and\ \bibinfo {author} {\bibfnamefont {E.~D.}\ \bibnamefont
  {Stewart}},\ }\href {\doibase 10.1143/PTP.95.71} {\bibfield  {journal}
  {\bibinfo  {journal} {Prog. Theor. Phys.}\ }\textbf {\bibinfo {volume}
  {95}},\ \bibinfo {pages} {71} (\bibinfo {year} {1996})},\ \Eprint
  {http://arxiv.org/abs/astro-ph/9507001} {arXiv:astro-ph/9507001} \BibitemShut
  {NoStop}%
\bibitem [{\citenamefont {Barrow}\ and\ \citenamefont
  {Parsons}(1995)}]{Barrow:1995xb}%
  \BibitemOpen
  \bibfield  {author} {\bibinfo {author} {\bibfnamefont {J.~D.}\ \bibnamefont
  {Barrow}}\ and\ \bibinfo {author} {\bibfnamefont {P.}~\bibnamefont
  {Parsons}},\ }\href {\doibase 10.1103/PhysRevD.52.5576} {\bibfield  {journal}
  {\bibinfo  {journal} {Phys. Rev. D}\ }\textbf {\bibinfo {volume} {52}},\
  \bibinfo {pages} {5576} (\bibinfo {year} {1995})},\ \Eprint
  {http://arxiv.org/abs/astro-ph/9506049} {arXiv:astro-ph/9506049} \BibitemShut
  {NoStop}%
\bibitem [{\citenamefont {Lidsey}\ \emph {et~al.}(1997)\citenamefont {Lidsey},
  \citenamefont {Liddle}, \citenamefont {Kolb}, \citenamefont {Copeland},
  \citenamefont {Barreiro},\ and\ \citenamefont {Abney}}]{Lidsey:1995np}%
  \BibitemOpen
  \bibfield  {author} {\bibinfo {author} {\bibfnamefont {J.~E.}\ \bibnamefont
  {Lidsey}}, \bibinfo {author} {\bibfnamefont {A.~R.}\ \bibnamefont {Liddle}},
  \bibinfo {author} {\bibfnamefont {E.~W.}\ \bibnamefont {Kolb}}, \bibinfo
  {author} {\bibfnamefont {E.~J.}\ \bibnamefont {Copeland}}, \bibinfo {author}
  {\bibfnamefont {T.}~\bibnamefont {Barreiro}}, \ and\ \bibinfo {author}
  {\bibfnamefont {M.}~\bibnamefont {Abney}},\ }\href {\doibase
  10.1103/RevModPhys.69.373} {\bibfield  {journal} {\bibinfo  {journal} {Rev.
  Mod. Phys.}\ }\textbf {\bibinfo {volume} {69}},\ \bibinfo {pages} {373}
  (\bibinfo {year} {1997})},\ \Eprint {http://arxiv.org/abs/astro-ph/9508078}
  {arXiv:astro-ph/9508078} \BibitemShut {NoStop}%
\bibitem [{\citenamefont {Parsons}\ and\ \citenamefont
  {Barrow}(1995)}]{Parsons:1995ew}%
  \BibitemOpen
  \bibfield  {author} {\bibinfo {author} {\bibfnamefont {P.}~\bibnamefont
  {Parsons}}\ and\ \bibinfo {author} {\bibfnamefont {J.~D.}\ \bibnamefont
  {Barrow}},\ }\href {\doibase 10.1103/PhysRevD.51.6757} {\bibfield  {journal}
  {\bibinfo  {journal} {Phys. Rev. D}\ }\textbf {\bibinfo {volume} {51}},\
  \bibinfo {pages} {6757} (\bibinfo {year} {1995})},\ \Eprint
  {http://arxiv.org/abs/astro-ph/9501086} {arXiv:astro-ph/9501086} \BibitemShut
  {NoStop}%
\bibitem [{\citenamefont {Liddle}(1998)}]{Liddle:1998wc}%
  \BibitemOpen
  \bibfield  {author} {\bibinfo {author} {\bibfnamefont {A.~R.}\ \bibnamefont
  {Liddle}},\ }\href {\doibase 10.1016/S0370-1573(98)00055-6} {\bibfield
  {journal} {\bibinfo  {journal} {Phys. Rept.}\ }\textbf {\bibinfo {volume}
  {307}},\ \bibinfo {pages} {53} (\bibinfo {year} {1998})},\ \Eprint
  {http://arxiv.org/abs/astro-ph/9801148} {arXiv:astro-ph/9801148} \BibitemShut
  {NoStop}%
\bibitem [{\citenamefont {Guth}(2000)}]{Guth:2000ka}%
  \BibitemOpen
  \bibfield  {author} {\bibinfo {author} {\bibfnamefont {A.~H.}\ \bibnamefont
  {Guth}},\ }\href {\doibase 10.1016/S0370-1573(00)00037-5} {\bibfield
  {journal} {\bibinfo  {journal} {Phys. Rept.}\ }\textbf {\bibinfo {volume}
  {333}},\ \bibinfo {pages} {555} (\bibinfo {year} {2000})},\ \Eprint
  {http://arxiv.org/abs/astro-ph/0002156} {arXiv:astro-ph/0002156} \BibitemShut
  {NoStop}%
\bibitem [{\citenamefont {Riotto}(2003)}]{Riotto:2002yw}%
  \BibitemOpen
  \bibfield  {author} {\bibinfo {author} {\bibfnamefont {A.}~\bibnamefont
  {Riotto}},\ }\href@noop {} {\bibfield  {journal} {\bibinfo  {journal} {ICTP
  Lect. Notes Ser.}\ }\textbf {\bibinfo {volume} {14}},\ \bibinfo {pages} {317}
  (\bibinfo {year} {2003})},\ \Eprint {http://arxiv.org/abs/hep-ph/0210162}
  {arXiv:hep-ph/0210162} \BibitemShut {NoStop}%
\bibitem [{\citenamefont {Feinstein}(2002)}]{Feinstein:2002aj}%
  \BibitemOpen
  \bibfield  {author} {\bibinfo {author} {\bibfnamefont {A.}~\bibnamefont
  {Feinstein}},\ }\href {\doibase 10.1103/PhysRevD.66.063511} {\bibfield
  {journal} {\bibinfo  {journal} {Phys. Rev. D}\ }\textbf {\bibinfo {volume}
  {66}},\ \bibinfo {pages} {063511} (\bibinfo {year} {2002})},\ \Eprint
  {http://arxiv.org/abs/hep-th/0204140} {arXiv:hep-th/0204140} \BibitemShut
  {NoStop}%
\bibitem [{\citenamefont {Boubekeur}\ and\ \citenamefont
  {Lyth}(2005)}]{Boubekeur:2005zm}%
  \BibitemOpen
  \bibfield  {author} {\bibinfo {author} {\bibfnamefont {L.}~\bibnamefont
  {Boubekeur}}\ and\ \bibinfo {author} {\bibfnamefont {D.~H.}\ \bibnamefont
  {Lyth}},\ }\href {\doibase 10.1088/1475-7516/2005/07/010} {\bibfield
  {journal} {\bibinfo  {journal} {JCAP}\ }\textbf {\bibinfo {volume} {07}},\
  \bibinfo {pages} {010} (\bibinfo {year} {2005})},\ \Eprint
  {http://arxiv.org/abs/hep-ph/0502047} {arXiv:hep-ph/0502047} \BibitemShut
  {NoStop}%
\bibitem [{\citenamefont {Conlon}\ and\ \citenamefont
  {Quevedo}(2006)}]{Conlon:2005jm}%
  \BibitemOpen
  \bibfield  {author} {\bibinfo {author} {\bibfnamefont {J.~P.}\ \bibnamefont
  {Conlon}}\ and\ \bibinfo {author} {\bibfnamefont {F.}~\bibnamefont
  {Quevedo}},\ }\href {\doibase 10.1088/1126-6708/2006/01/146} {\bibfield
  {journal} {\bibinfo  {journal} {JHEP}\ }\textbf {\bibinfo {volume} {01}},\
  \bibinfo {pages} {146} (\bibinfo {year} {2006})},\ \Eprint
  {http://arxiv.org/abs/hep-th/0509012} {arXiv:hep-th/0509012} \BibitemShut
  {NoStop}%
\bibitem [{\citenamefont {Ferraro}\ and\ \citenamefont
  {Fiorini}(2007)}]{Ferraro:2006jd}%
  \BibitemOpen
  \bibfield  {author} {\bibinfo {author} {\bibfnamefont {R.}~\bibnamefont
  {Ferraro}}\ and\ \bibinfo {author} {\bibfnamefont {F.}~\bibnamefont
  {Fiorini}},\ }\href {\doibase 10.1103/PhysRevD.75.084031} {\bibfield
  {journal} {\bibinfo  {journal} {Phys. Rev. D}\ }\textbf {\bibinfo {volume}
  {75}},\ \bibinfo {pages} {084031} (\bibinfo {year} {2007})},\ \Eprint
  {http://arxiv.org/abs/gr-qc/0610067} {arXiv:gr-qc/0610067} \BibitemShut
  {NoStop}%
\bibitem [{\citenamefont {Cheung}\ \emph {et~al.}(2008)\citenamefont {Cheung},
  \citenamefont {Creminelli}, \citenamefont {Fitzpatrick}, \citenamefont
  {Kaplan},\ and\ \citenamefont {Senatore}}]{Cheung:2007st}%
  \BibitemOpen
  \bibfield  {author} {\bibinfo {author} {\bibfnamefont {C.}~\bibnamefont
  {Cheung}}, \bibinfo {author} {\bibfnamefont {P.}~\bibnamefont {Creminelli}},
  \bibinfo {author} {\bibfnamefont {A.~L.}\ \bibnamefont {Fitzpatrick}},
  \bibinfo {author} {\bibfnamefont {J.}~\bibnamefont {Kaplan}}, \ and\ \bibinfo
  {author} {\bibfnamefont {L.}~\bibnamefont {Senatore}},\ }\href {\doibase
  10.1088/1126-6708/2008/03/014} {\bibfield  {journal} {\bibinfo  {journal}
  {JHEP}\ }\textbf {\bibinfo {volume} {03}},\ \bibinfo {pages} {014} (\bibinfo
  {year} {2008})},\ \Eprint {http://arxiv.org/abs/0709.0293} {arXiv:0709.0293
  [hep-th]} \BibitemShut {NoStop}%
\bibitem [{\citenamefont {Baumann}\ and\ \citenamefont
  {Peiris}(2009)}]{Baumann:2008bn}%
  \BibitemOpen
  \bibfield  {author} {\bibinfo {author} {\bibfnamefont {D.}~\bibnamefont
  {Baumann}}\ and\ \bibinfo {author} {\bibfnamefont {H.~V.}\ \bibnamefont
  {Peiris}},\ }\href {\doibase 10.1166/asl.2009.1019} {\bibfield  {journal}
  {\bibinfo  {journal} {Adv. Sci. Lett.}\ }\textbf {\bibinfo {volume} {2}},\
  \bibinfo {pages} {105} (\bibinfo {year} {2009})},\ \Eprint
  {http://arxiv.org/abs/0810.3022} {arXiv:0810.3022 [astro-ph]} \BibitemShut
  {NoStop}%
\bibitem [{\citenamefont {Pal}\ \emph {et~al.}(2010)\citenamefont {Pal},
  \citenamefont {Pal},\ and\ \citenamefont {Basu}}]{Pal:2009sd}%
  \BibitemOpen
  \bibfield  {author} {\bibinfo {author} {\bibfnamefont {B.~K.}\ \bibnamefont
  {Pal}}, \bibinfo {author} {\bibfnamefont {S.}~\bibnamefont {Pal}}, \ and\
  \bibinfo {author} {\bibfnamefont {B.}~\bibnamefont {Basu}},\ }\href {\doibase
  10.1088/1475-7516/2010/01/029} {\bibfield  {journal} {\bibinfo  {journal}
  {JCAP}\ }\textbf {\bibinfo {volume} {01}},\ \bibinfo {pages} {029} (\bibinfo
  {year} {2010})},\ \Eprint {http://arxiv.org/abs/0908.2302} {arXiv:0908.2302
  [hep-th]} \BibitemShut {NoStop}%
\bibitem [{\citenamefont {Martin}\ \emph
  {et~al.}(2014{\natexlab{a}})\citenamefont {Martin}, \citenamefont {Ringeval},
  \citenamefont {Trotta},\ and\ \citenamefont {Vennin}}]{Martin:2013nzq}%
  \BibitemOpen
  \bibfield  {author} {\bibinfo {author} {\bibfnamefont {J.}~\bibnamefont
  {Martin}}, \bibinfo {author} {\bibfnamefont {C.}~\bibnamefont {Ringeval}},
  \bibinfo {author} {\bibfnamefont {R.}~\bibnamefont {Trotta}}, \ and\ \bibinfo
  {author} {\bibfnamefont {V.}~\bibnamefont {Vennin}},\ }\href {\doibase
  10.1088/1475-7516/2014/03/039} {\bibfield  {journal} {\bibinfo  {journal}
  {JCAP}\ }\textbf {\bibinfo {volume} {03}},\ \bibinfo {pages} {039} (\bibinfo
  {year} {2014}{\natexlab{a}})},\ \Eprint {http://arxiv.org/abs/1312.3529}
  {arXiv:1312.3529 [astro-ph.CO]} \BibitemShut {NoStop}%
\bibitem [{\citenamefont {Martin}\ \emph
  {et~al.}(2014{\natexlab{b}})\citenamefont {Martin}, \citenamefont
  {Ringeval},\ and\ \citenamefont {Vennin}}]{Martin:2013tda}%
  \BibitemOpen
  \bibfield  {author} {\bibinfo {author} {\bibfnamefont {J.}~\bibnamefont
  {Martin}}, \bibinfo {author} {\bibfnamefont {C.}~\bibnamefont {Ringeval}}, \
  and\ \bibinfo {author} {\bibfnamefont {V.}~\bibnamefont {Vennin}},\ }\href
  {\doibase 10.1016/j.dark.2014.01.003} {\bibfield  {journal} {\bibinfo
  {journal} {Phys. Dark Univ.}\ }\textbf {\bibinfo {volume} {5-6}},\ \bibinfo
  {pages} {75} (\bibinfo {year} {2014}{\natexlab{b}})},\ \Eprint
  {http://arxiv.org/abs/1303.3787} {arXiv:1303.3787 [astro-ph.CO]} \BibitemShut
  {NoStop}%
\bibitem [{\citenamefont {Sebastiani}\ \emph {et~al.}(2014)\citenamefont
  {Sebastiani}, \citenamefont {Cognola}, \citenamefont {Myrzakulov},
  \citenamefont {Odintsov},\ and\ \citenamefont
  {Zerbini}}]{Sebastiani:2013eqa}%
  \BibitemOpen
  \bibfield  {author} {\bibinfo {author} {\bibfnamefont {L.}~\bibnamefont
  {Sebastiani}}, \bibinfo {author} {\bibfnamefont {G.}~\bibnamefont {Cognola}},
  \bibinfo {author} {\bibfnamefont {R.}~\bibnamefont {Myrzakulov}}, \bibinfo
  {author} {\bibfnamefont {S.~D.}\ \bibnamefont {Odintsov}}, \ and\ \bibinfo
  {author} {\bibfnamefont {S.}~\bibnamefont {Zerbini}},\ }\href {\doibase
  10.1103/PhysRevD.89.023518} {\bibfield  {journal} {\bibinfo  {journal} {Phys.
  Rev. D}\ }\textbf {\bibinfo {volume} {89}},\ \bibinfo {pages} {023518}
  (\bibinfo {year} {2014})},\ \Eprint {http://arxiv.org/abs/1311.0744}
  {arXiv:1311.0744 [gr-qc]} \BibitemShut {NoStop}%
\bibitem [{\citenamefont {Hamada}\ \emph {et~al.}(2014)\citenamefont {Hamada},
  \citenamefont {Kawai}, \citenamefont {Oda},\ and\ \citenamefont
  {Park}}]{Hamada:2014iga}%
  \BibitemOpen
  \bibfield  {author} {\bibinfo {author} {\bibfnamefont {Y.}~\bibnamefont
  {Hamada}}, \bibinfo {author} {\bibfnamefont {H.}~\bibnamefont {Kawai}},
  \bibinfo {author} {\bibfnamefont {K.-y.}\ \bibnamefont {Oda}}, \ and\
  \bibinfo {author} {\bibfnamefont {S.~C.}\ \bibnamefont {Park}},\ }\href
  {\doibase 10.1103/PhysRevLett.112.241301} {\bibfield  {journal} {\bibinfo
  {journal} {Phys. Rev. Lett.}\ }\textbf {\bibinfo {volume} {112}},\ \bibinfo
  {pages} {241301} (\bibinfo {year} {2014})},\ \Eprint
  {http://arxiv.org/abs/1403.5043} {arXiv:1403.5043 [hep-ph]} \BibitemShut
  {NoStop}%
\bibitem [{\citenamefont {Freese}\ and\ \citenamefont
  {Kinney}(2015)}]{Freese:2014nla}%
  \BibitemOpen
  \bibfield  {author} {\bibinfo {author} {\bibfnamefont {K.}~\bibnamefont
  {Freese}}\ and\ \bibinfo {author} {\bibfnamefont {W.~H.}\ \bibnamefont
  {Kinney}},\ }\href {\doibase 10.1088/1475-7516/2015/03/044} {\bibfield
  {journal} {\bibinfo  {journal} {JCAP}\ }\textbf {\bibinfo {volume} {03}},\
  \bibinfo {pages} {044} (\bibinfo {year} {2015})},\ \Eprint
  {http://arxiv.org/abs/1403.5277} {arXiv:1403.5277 [astro-ph.CO]} \BibitemShut
  {NoStop}%
\bibitem [{\citenamefont {Rubio}(2019)}]{Rubio:2018ogq}%
  \BibitemOpen
  \bibfield  {author} {\bibinfo {author} {\bibfnamefont {J.}~\bibnamefont
  {Rubio}},\ }\href {\doibase 10.3389/fspas.2018.00050} {\bibfield  {journal}
  {\bibinfo  {journal} {Front. Astron. Space Sci.}\ }\textbf {\bibinfo {volume}
  {5}},\ \bibinfo {pages} {50} (\bibinfo {year} {2019})},\ \Eprint
  {http://arxiv.org/abs/1807.02376} {arXiv:1807.02376 [hep-ph]} \BibitemShut
  {NoStop}%
\bibitem [{\citenamefont {Odintsov}\ \emph {et~al.}(2023)\citenamefont
  {Odintsov}, \citenamefont {Oikonomou}, \citenamefont {Giannakoudi},
  \citenamefont {Fronimos},\ and\ \citenamefont
  {Lymperiadou}}]{Odintsov:2023weg}%
  \BibitemOpen
  \bibfield  {author} {\bibinfo {author} {\bibfnamefont {S.~D.}\ \bibnamefont
  {Odintsov}}, \bibinfo {author} {\bibfnamefont {V.~K.}\ \bibnamefont
  {Oikonomou}}, \bibinfo {author} {\bibfnamefont {I.}~\bibnamefont
  {Giannakoudi}}, \bibinfo {author} {\bibfnamefont {F.~P.}\ \bibnamefont
  {Fronimos}}, \ and\ \bibinfo {author} {\bibfnamefont {E.~C.}\ \bibnamefont
  {Lymperiadou}},\ }\href {\doibase 10.3390/sym15091701} {\bibfield  {journal}
  {\bibinfo  {journal} {Symmetry}\ }\textbf {\bibinfo {volume} {15}},\ \bibinfo
  {pages} {1701} (\bibinfo {year} {2023})},\ \Eprint
  {http://arxiv.org/abs/2307.16308} {arXiv:2307.16308 [gr-qc]} \BibitemShut
  {NoStop}%
\bibitem [{\citenamefont {Giar\`e}\ \emph {et~al.}(2023)\citenamefont
  {Giar\`e}, \citenamefont {Pan}, \citenamefont {Di~Valentino}, \citenamefont
  {Yang}, \citenamefont {de~Haro},\ and\ \citenamefont
  {Melchiorri}}]{Giare:2023wzl}%
  \BibitemOpen
  \bibfield  {author} {\bibinfo {author} {\bibfnamefont {W.}~\bibnamefont
  {Giar\`e}}, \bibinfo {author} {\bibfnamefont {S.}~\bibnamefont {Pan}},
  \bibinfo {author} {\bibfnamefont {E.}~\bibnamefont {Di~Valentino}}, \bibinfo
  {author} {\bibfnamefont {W.}~\bibnamefont {Yang}}, \bibinfo {author}
  {\bibfnamefont {J.}~\bibnamefont {de~Haro}}, \ and\ \bibinfo {author}
  {\bibfnamefont {A.}~\bibnamefont {Melchiorri}},\ }\href {\doibase
  10.1088/1475-7516/2023/09/019} {\bibfield  {journal} {\bibinfo  {journal}
  {JCAP}\ }\textbf {\bibinfo {volume} {09}},\ \bibinfo {pages} {019} (\bibinfo
  {year} {2023})},\ \Eprint {http://arxiv.org/abs/2305.15378} {arXiv:2305.15378
  [astro-ph.CO]} \BibitemShut {NoStop}%
\bibitem [{\citenamefont {Jinno}\ \emph {et~al.}(2024)\citenamefont {Jinno},
  \citenamefont {Kohri}, \citenamefont {Moroi}, \citenamefont {Takahashi},\
  and\ \citenamefont {Hazumi}}]{Jinno:2023bpc}%
  \BibitemOpen
  \bibfield  {author} {\bibinfo {author} {\bibfnamefont {R.}~\bibnamefont
  {Jinno}}, \bibinfo {author} {\bibfnamefont {K.}~\bibnamefont {Kohri}},
  \bibinfo {author} {\bibfnamefont {T.}~\bibnamefont {Moroi}}, \bibinfo
  {author} {\bibfnamefont {T.}~\bibnamefont {Takahashi}}, \ and\ \bibinfo
  {author} {\bibfnamefont {M.}~\bibnamefont {Hazumi}},\ }\href {\doibase
  10.1088/1475-7516/2024/03/011} {\bibfield  {journal} {\bibinfo  {journal}
  {JCAP}\ }\textbf {\bibinfo {volume} {03}},\ \bibinfo {pages} {011} (\bibinfo
  {year} {2024})},\ \Eprint {http://arxiv.org/abs/2310.08158} {arXiv:2310.08158
  [astro-ph.CO]} \BibitemShut {NoStop}%
\bibitem [{\citenamefont {Kofman}\ \emph {et~al.}(1994)\citenamefont {Kofman},
  \citenamefont {Linde},\ and\ \citenamefont {Starobinsky}}]{Kofman:1994rk}%
  \BibitemOpen
  \bibfield  {author} {\bibinfo {author} {\bibfnamefont {L.}~\bibnamefont
  {Kofman}}, \bibinfo {author} {\bibfnamefont {A.~D.}\ \bibnamefont {Linde}}, \
  and\ \bibinfo {author} {\bibfnamefont {A.~A.}\ \bibnamefont {Starobinsky}},\
  }\href {\doibase 10.1103/PhysRevLett.73.3195} {\bibfield  {journal} {\bibinfo
   {journal} {Phys. Rev. Lett.}\ }\textbf {\bibinfo {volume} {73}},\ \bibinfo
  {pages} {3195} (\bibinfo {year} {1994})},\ \Eprint
  {http://arxiv.org/abs/hep-th/9405187} {arXiv:hep-th/9405187} \BibitemShut
  {NoStop}%
\bibitem [{\citenamefont {Kofman}\ \emph {et~al.}(1997)\citenamefont {Kofman},
  \citenamefont {Linde},\ and\ \citenamefont {Starobinsky}}]{Kofman:1997yn}%
  \BibitemOpen
  \bibfield  {author} {\bibinfo {author} {\bibfnamefont {L.}~\bibnamefont
  {Kofman}}, \bibinfo {author} {\bibfnamefont {A.~D.}\ \bibnamefont {Linde}}, \
  and\ \bibinfo {author} {\bibfnamefont {A.~A.}\ \bibnamefont {Starobinsky}},\
  }\href {\doibase 10.1103/PhysRevD.56.3258} {\bibfield  {journal} {\bibinfo
  {journal} {Phys. Rev. D}\ }\textbf {\bibinfo {volume} {56}},\ \bibinfo
  {pages} {3258} (\bibinfo {year} {1997})},\ \Eprint
  {http://arxiv.org/abs/hep-ph/9704452} {arXiv:hep-ph/9704452} \BibitemShut
  {NoStop}%
\bibitem [{\citenamefont {Felder}\ \emph {et~al.}(1999)\citenamefont {Felder},
  \citenamefont {Kofman},\ and\ \citenamefont {Linde}}]{Felder:1998vq}%
  \BibitemOpen
  \bibfield  {author} {\bibinfo {author} {\bibfnamefont {G.~N.}\ \bibnamefont
  {Felder}}, \bibinfo {author} {\bibfnamefont {L.}~\bibnamefont {Kofman}}, \
  and\ \bibinfo {author} {\bibfnamefont {A.~D.}\ \bibnamefont {Linde}},\ }\href
  {\doibase 10.1103/PhysRevD.59.123523} {\bibfield  {journal} {\bibinfo
  {journal} {Phys. Rev. D}\ }\textbf {\bibinfo {volume} {59}},\ \bibinfo
  {pages} {123523} (\bibinfo {year} {1999})},\ \Eprint
  {http://arxiv.org/abs/hep-ph/9812289} {arXiv:hep-ph/9812289} \BibitemShut
  {NoStop}%
\bibitem [{\citenamefont {Bassett}\ \emph {et~al.}(2006)\citenamefont
  {Bassett}, \citenamefont {Tsujikawa},\ and\ \citenamefont
  {Wands}}]{Bassett:2005xm}%
  \BibitemOpen
  \bibfield  {author} {\bibinfo {author} {\bibfnamefont {B.~A.}\ \bibnamefont
  {Bassett}}, \bibinfo {author} {\bibfnamefont {S.}~\bibnamefont {Tsujikawa}},
  \ and\ \bibinfo {author} {\bibfnamefont {D.}~\bibnamefont {Wands}},\ }\href
  {\doibase 10.1103/RevModPhys.78.537} {\bibfield  {journal} {\bibinfo
  {journal} {Rev. Mod. Phys.}\ }\textbf {\bibinfo {volume} {78}},\ \bibinfo
  {pages} {537} (\bibinfo {year} {2006})},\ \Eprint
  {http://arxiv.org/abs/astro-ph/0507632} {arXiv:astro-ph/0507632} \BibitemShut
  {NoStop}%
\bibitem [{\citenamefont {Allahverdi}\ \emph {et~al.}(2010)\citenamefont
  {Allahverdi}, \citenamefont {Brandenberger}, \citenamefont {Cyr-Racine},\
  and\ \citenamefont {Mazumdar}}]{Allahverdi:2010xz}%
  \BibitemOpen
  \bibfield  {author} {\bibinfo {author} {\bibfnamefont {R.}~\bibnamefont
  {Allahverdi}}, \bibinfo {author} {\bibfnamefont {R.}~\bibnamefont
  {Brandenberger}}, \bibinfo {author} {\bibfnamefont {F.-Y.}\ \bibnamefont
  {Cyr-Racine}}, \ and\ \bibinfo {author} {\bibfnamefont {A.}~\bibnamefont
  {Mazumdar}},\ }\href {\doibase 10.1146/annurev.nucl.012809.104511} {\bibfield
   {journal} {\bibinfo  {journal} {Ann. Rev. Nucl. Part. Sci.}\ }\textbf
  {\bibinfo {volume} {60}},\ \bibinfo {pages} {27} (\bibinfo {year} {2010})},\
  \Eprint {http://arxiv.org/abs/1001.2600} {arXiv:1001.2600 [hep-th]}
  \BibitemShut {NoStop}%
\bibitem [{\citenamefont {Martin}\ \emph {et~al.}(2015)\citenamefont {Martin},
  \citenamefont {Ringeval},\ and\ \citenamefont {Vennin}}]{Martin:2014nya}%
  \BibitemOpen
  \bibfield  {author} {\bibinfo {author} {\bibfnamefont {J.}~\bibnamefont
  {Martin}}, \bibinfo {author} {\bibfnamefont {C.}~\bibnamefont {Ringeval}}, \
  and\ \bibinfo {author} {\bibfnamefont {V.}~\bibnamefont {Vennin}},\ }\href
  {\doibase 10.1103/PhysRevLett.114.081303} {\bibfield  {journal} {\bibinfo
  {journal} {Phys. Rev. Lett.}\ }\textbf {\bibinfo {volume} {114}},\ \bibinfo
  {pages} {081303} (\bibinfo {year} {2015})},\ \Eprint
  {http://arxiv.org/abs/1410.7958} {arXiv:1410.7958 [astro-ph.CO]} \BibitemShut
  {NoStop}%
\bibitem [{\citenamefont {Ford}(1987)}]{Ford:1986sy}%
  \BibitemOpen
  \bibfield  {author} {\bibinfo {author} {\bibfnamefont {L.~H.}\ \bibnamefont
  {Ford}},\ }\href {\doibase 10.1103/PhysRevD.35.2955} {\bibfield  {journal}
  {\bibinfo  {journal} {Phys. Rev. D}\ }\textbf {\bibinfo {volume} {35}},\
  \bibinfo {pages} {2955} (\bibinfo {year} {1987})}\BibitemShut {NoStop}%
\bibitem [{\citenamefont {Spokoiny}(1993)}]{Spokoiny:1993kt}%
  \BibitemOpen
  \bibfield  {author} {\bibinfo {author} {\bibfnamefont {B.}~\bibnamefont
  {Spokoiny}},\ }\href {\doibase 10.1016/0370-2693(93)90155-B} {\bibfield
  {journal} {\bibinfo  {journal} {Phys. Lett. B}\ }\textbf {\bibinfo {volume}
  {315}},\ \bibinfo {pages} {40} (\bibinfo {year} {1993})},\ \Eprint
  {http://arxiv.org/abs/gr-qc/9306008} {arXiv:gr-qc/9306008} \BibitemShut
  {NoStop}%
\bibitem [{\citenamefont {Lankinen}\ \emph {et~al.}(2020)\citenamefont
  {Lankinen}, \citenamefont {Kerppo},\ and\ \citenamefont
  {Vilja}}]{Lankinen:2019ifa}%
  \BibitemOpen
  \bibfield  {author} {\bibinfo {author} {\bibfnamefont {J.}~\bibnamefont
  {Lankinen}}, \bibinfo {author} {\bibfnamefont {O.}~\bibnamefont {Kerppo}}, \
  and\ \bibinfo {author} {\bibfnamefont {I.}~\bibnamefont {Vilja}},\ }\href
  {\doibase 10.1103/PhysRevD.101.063529} {\bibfield  {journal} {\bibinfo
  {journal} {Phys. Rev. D}\ }\textbf {\bibinfo {volume} {101}},\ \bibinfo
  {pages} {063529} (\bibinfo {year} {2020})},\ \Eprint
  {http://arxiv.org/abs/1910.07520} {arXiv:1910.07520 [gr-qc]} \BibitemShut
  {NoStop}%
\bibitem [{\citenamefont {Dorsch}\ \emph {et~al.}(2024)\citenamefont {Dorsch},
  \citenamefont {Miranda},\ and\ \citenamefont {Yokomizo}}]{Dorsch:2024nan}%
  \BibitemOpen
  \bibfield  {author} {\bibinfo {author} {\bibfnamefont {G.~C.}\ \bibnamefont
  {Dorsch}}, \bibinfo {author} {\bibfnamefont {L.}~\bibnamefont {Miranda}}, \
  and\ \bibinfo {author} {\bibfnamefont {N.}~\bibnamefont {Yokomizo}},\
  }\href@noop {} {\  (\bibinfo {year} {2024})},\ \Eprint
  {http://arxiv.org/abs/2406.04161} {arXiv:2406.04161 [gr-qc]} \BibitemShut
  {NoStop}%
\bibitem [{\citenamefont {Peebles}\ and\ \citenamefont
  {Vilenkin}(1999)}]{Peebles:1998qn}%
  \BibitemOpen
  \bibfield  {author} {\bibinfo {author} {\bibfnamefont {P.~J.~E.}\
  \bibnamefont {Peebles}}\ and\ \bibinfo {author} {\bibfnamefont
  {A.}~\bibnamefont {Vilenkin}},\ }\href {\doibase 10.1103/PhysRevD.59.063505}
  {\bibfield  {journal} {\bibinfo  {journal} {Phys. Rev. D}\ }\textbf {\bibinfo
  {volume} {59}},\ \bibinfo {pages} {063505} (\bibinfo {year} {1999})},\
  \Eprint {http://arxiv.org/abs/astro-ph/9810509} {arXiv:astro-ph/9810509}
  \BibitemShut {NoStop}%
\bibitem [{\citenamefont {Chun}\ \emph {et~al.}(2009)\citenamefont {Chun},
  \citenamefont {Scopel},\ and\ \citenamefont {Zaballa}}]{Chun:2009yu}%
  \BibitemOpen
  \bibfield  {author} {\bibinfo {author} {\bibfnamefont {E.~J.}\ \bibnamefont
  {Chun}}, \bibinfo {author} {\bibfnamefont {S.}~\bibnamefont {Scopel}}, \ and\
  \bibinfo {author} {\bibfnamefont {I.}~\bibnamefont {Zaballa}},\ }\href
  {\doibase 10.1088/1475-7516/2009/07/022} {\bibfield  {journal} {\bibinfo
  {journal} {JCAP}\ }\textbf {\bibinfo {volume} {07}},\ \bibinfo {pages} {022}
  (\bibinfo {year} {2009})},\ \Eprint {http://arxiv.org/abs/0904.0675}
  {arXiv:0904.0675 [hep-ph]} \BibitemShut {NoStop}%
\bibitem [{\citenamefont {Lankinen}\ and\ \citenamefont
  {Vilja}(2017)}]{Lankinen:2016ile}%
  \BibitemOpen
  \bibfield  {author} {\bibinfo {author} {\bibfnamefont {J.}~\bibnamefont
  {Lankinen}}\ and\ \bibinfo {author} {\bibfnamefont {I.}~\bibnamefont
  {Vilja}},\ }\href {\doibase 10.1088/1475-7516/2017/08/025} {\bibfield
  {journal} {\bibinfo  {journal} {JCAP}\ }\textbf {\bibinfo {volume} {08}},\
  \bibinfo {pages} {025} (\bibinfo {year} {2017})},\ \Eprint
  {http://arxiv.org/abs/1612.02586} {arXiv:1612.02586 [gr-qc]} \BibitemShut
  {NoStop}%
\bibitem [{\citenamefont {De~Haro}\ and\ \citenamefont
  {Arest\'e~Sal\'o}(2017)}]{DeHaro:2017abf}%
  \BibitemOpen
  \bibfield  {author} {\bibinfo {author} {\bibfnamefont {J.}~\bibnamefont
  {De~Haro}}\ and\ \bibinfo {author} {\bibfnamefont {L.}~\bibnamefont
  {Arest\'e~Sal\'o}},\ }\href {\doibase 10.1103/PhysRevD.95.123501} {\bibfield
  {journal} {\bibinfo  {journal} {Phys. Rev. D}\ }\textbf {\bibinfo {volume}
  {95}},\ \bibinfo {pages} {123501} (\bibinfo {year} {2017})},\ \Eprint
  {http://arxiv.org/abs/1702.04212} {arXiv:1702.04212 [gr-qc]} \BibitemShut
  {NoStop}%
\bibitem [{\citenamefont {Arest\'e~Sal\'o}\ and\ \citenamefont
  {de~Haro}(2017)}]{AresteSalo:2017lkv}%
  \BibitemOpen
  \bibfield  {author} {\bibinfo {author} {\bibfnamefont {L.}~\bibnamefont
  {Arest\'e~Sal\'o}}\ and\ \bibinfo {author} {\bibfnamefont {J.}~\bibnamefont
  {de~Haro}},\ }\href {\doibase 10.1140/epjc/s10052-017-5337-0} {\bibfield
  {journal} {\bibinfo  {journal} {Eur. Phys. J. C}\ }\textbf {\bibinfo {volume}
  {77}},\ \bibinfo {pages} {798} (\bibinfo {year} {2017})},\ \Eprint
  {http://arxiv.org/abs/1707.02810} {arXiv:1707.02810 [gr-qc]} \BibitemShut
  {NoStop}%
\bibitem [{\citenamefont {Hashiba}\ and\ \citenamefont
  {Yokoyama}(2019)}]{Hashiba:2018iff}%
  \BibitemOpen
  \bibfield  {author} {\bibinfo {author} {\bibfnamefont {S.}~\bibnamefont
  {Hashiba}}\ and\ \bibinfo {author} {\bibfnamefont {J.}~\bibnamefont
  {Yokoyama}},\ }\href {\doibase 10.1088/1475-7516/2019/01/028} {\bibfield
  {journal} {\bibinfo  {journal} {JCAP}\ }\textbf {\bibinfo {volume} {01}},\
  \bibinfo {pages} {028} (\bibinfo {year} {2019})},\ \Eprint
  {http://arxiv.org/abs/1809.05410} {arXiv:1809.05410 [gr-qc]} \BibitemShut
  {NoStop}%
\bibitem [{\citenamefont {Haro}\ \emph {et~al.}(2019)\citenamefont {Haro},
  \citenamefont {Yang},\ and\ \citenamefont {Pan}}]{Haro:2018zdb}%
  \BibitemOpen
  \bibfield  {author} {\bibinfo {author} {\bibfnamefont {J.}~\bibnamefont
  {Haro}}, \bibinfo {author} {\bibfnamefont {W.}~\bibnamefont {Yang}}, \ and\
  \bibinfo {author} {\bibfnamefont {S.}~\bibnamefont {Pan}},\ }\href {\doibase
  10.1088/1475-7516/2019/01/023} {\bibfield  {journal} {\bibinfo  {journal}
  {JCAP}\ }\textbf {\bibinfo {volume} {01}},\ \bibinfo {pages} {023} (\bibinfo
  {year} {2019})},\ \Eprint {http://arxiv.org/abs/1811.07371} {arXiv:1811.07371
  [gr-qc]} \BibitemShut {NoStop}%
\bibitem [{\citenamefont {de~Haro}\ \emph {et~al.}(2019)\citenamefont
  {de~Haro}, \citenamefont {Pan},\ and\ \citenamefont
  {Arest\'e~Sal\'o}}]{deHaro:2019oki}%
  \BibitemOpen
  \bibfield  {author} {\bibinfo {author} {\bibfnamefont {J.}~\bibnamefont
  {de~Haro}}, \bibinfo {author} {\bibfnamefont {S.}~\bibnamefont {Pan}}, \ and\
  \bibinfo {author} {\bibfnamefont {L.}~\bibnamefont {Arest\'e~Sal\'o}},\
  }\href {\doibase 10.1088/1475-7516/2019/06/056} {\bibfield  {journal}
  {\bibinfo  {journal} {JCAP}\ }\textbf {\bibinfo {volume} {06}},\ \bibinfo
  {pages} {056} (\bibinfo {year} {2019})},\ \Eprint
  {http://arxiv.org/abs/1903.01181} {arXiv:1903.01181 [gr-qc]} \BibitemShut
  {NoStop}%
\bibitem [{\citenamefont {Sal\'o}\ and\ \citenamefont
  {de~Haro}(2021)}]{Salo:2021vdv}%
  \BibitemOpen
  \bibfield  {author} {\bibinfo {author} {\bibfnamefont {L.~A.}\ \bibnamefont
  {Sal\'o}}\ and\ \bibinfo {author} {\bibfnamefont {J.}~\bibnamefont
  {de~Haro}},\ }\href {\doibase 10.1103/PhysRevD.104.083544} {\bibfield
  {journal} {\bibinfo  {journal} {Phys. Rev. D}\ }\textbf {\bibinfo {volume}
  {104}},\ \bibinfo {pages} {083544} (\bibinfo {year} {2021})},\ \Eprint
  {http://arxiv.org/abs/2108.10795} {arXiv:2108.10795 [gr-qc]} \BibitemShut
  {NoStop}%
\bibitem [{\citenamefont {de~Haro}\ and\ \citenamefont
  {Arest\'e~Sal\'o}(2023)}]{deHaro:2022ukj}%
  \BibitemOpen
  \bibfield  {author} {\bibinfo {author} {\bibfnamefont {J.}~\bibnamefont
  {de~Haro}}\ and\ \bibinfo {author} {\bibfnamefont {L.}~\bibnamefont
  {Arest\'e~Sal\'o}},\ }\href {\doibase 10.1103/PhysRevD.107.063542} {\bibfield
   {journal} {\bibinfo  {journal} {Phys. Rev. D}\ }\textbf {\bibinfo {volume}
  {107}},\ \bibinfo {pages} {063542} (\bibinfo {year} {2023})},\ \Eprint
  {http://arxiv.org/abs/2212.01276} {arXiv:2212.01276 [gr-qc]} \BibitemShut
  {NoStop}%
\bibitem [{\citenamefont {de~Haro}(2024)}]{deHaro:2023gho}%
  \BibitemOpen
  \bibfield  {author} {\bibinfo {author} {\bibfnamefont {J.}~\bibnamefont
  {de~Haro}},\ }\href {\doibase 10.1103/PhysRevD.109.023517} {\bibfield
  {journal} {\bibinfo  {journal} {Phys. Rev. D}\ }\textbf {\bibinfo {volume}
  {109}},\ \bibinfo {pages} {023517} (\bibinfo {year} {2024})},\ \Eprint
  {http://arxiv.org/abs/2310.02245} {arXiv:2310.02245 [gr-qc]} \BibitemShut
  {NoStop}%
\bibitem [{\citenamefont {Giovannini}(1999)}]{Giovannini:1999bh}%
  \BibitemOpen
  \bibfield  {author} {\bibinfo {author} {\bibfnamefont {M.}~\bibnamefont
  {Giovannini}},\ }\href {\doibase 10.1103/PhysRevD.60.123511} {\bibfield
  {journal} {\bibinfo  {journal} {Phys. Rev. D}\ }\textbf {\bibinfo {volume}
  {60}},\ \bibinfo {pages} {123511} (\bibinfo {year} {1999})},\ \Eprint
  {http://arxiv.org/abs/astro-ph/9903004} {arXiv:astro-ph/9903004} \BibitemShut
  {NoStop}%
\bibitem [{\citenamefont {Peloso}\ and\ \citenamefont
  {Rosati}(1999)}]{Peloso:1999dm}%
  \BibitemOpen
  \bibfield  {author} {\bibinfo {author} {\bibfnamefont {M.}~\bibnamefont
  {Peloso}}\ and\ \bibinfo {author} {\bibfnamefont {F.}~\bibnamefont
  {Rosati}},\ }\href {\doibase 10.1088/1126-6708/1999/12/026} {\bibfield
  {journal} {\bibinfo  {journal} {JHEP}\ }\textbf {\bibinfo {volume} {12}},\
  \bibinfo {pages} {026} (\bibinfo {year} {1999})},\ \Eprint
  {http://arxiv.org/abs/hep-ph/9908271} {arXiv:hep-ph/9908271} \BibitemShut
  {NoStop}%
\bibitem [{\citenamefont {Kaganovich}(2001)}]{Kaganovich:2000fc}%
  \BibitemOpen
  \bibfield  {author} {\bibinfo {author} {\bibfnamefont {A.~B.}\ \bibnamefont
  {Kaganovich}},\ }\href {\doibase 10.1103/PhysRevD.63.025022} {\bibfield
  {journal} {\bibinfo  {journal} {Phys. Rev. D}\ }\textbf {\bibinfo {volume}
  {63}},\ \bibinfo {pages} {025022} (\bibinfo {year} {2001})},\ \Eprint
  {http://arxiv.org/abs/hep-th/0007144} {arXiv:hep-th/0007144} \BibitemShut
  {NoStop}%
\bibitem [{\citenamefont {Yahiro}\ \emph {et~al.}(2002)\citenamefont {Yahiro},
  \citenamefont {Mathews}, \citenamefont {Ichiki}, \citenamefont {Kajino},\
  and\ \citenamefont {Orito}}]{Yahiro:2001uh}%
  \BibitemOpen
  \bibfield  {author} {\bibinfo {author} {\bibfnamefont {M.}~\bibnamefont
  {Yahiro}}, \bibinfo {author} {\bibfnamefont {G.~J.}\ \bibnamefont {Mathews}},
  \bibinfo {author} {\bibfnamefont {K.}~\bibnamefont {Ichiki}}, \bibinfo
  {author} {\bibfnamefont {T.}~\bibnamefont {Kajino}}, \ and\ \bibinfo {author}
  {\bibfnamefont {M.}~\bibnamefont {Orito}},\ }\href {\doibase
  10.1103/PhysRevD.65.063502} {\bibfield  {journal} {\bibinfo  {journal} {Phys.
  Rev. D}\ }\textbf {\bibinfo {volume} {65}},\ \bibinfo {pages} {063502}
  (\bibinfo {year} {2002})},\ \Eprint {http://arxiv.org/abs/astro-ph/0106349}
  {arXiv:astro-ph/0106349} \BibitemShut {NoStop}%
\bibitem [{\citenamefont {Dimopoulos}\ and\ \citenamefont
  {Valle}(2002)}]{Dimopoulos:2001ix}%
  \BibitemOpen
  \bibfield  {author} {\bibinfo {author} {\bibfnamefont {K.}~\bibnamefont
  {Dimopoulos}}\ and\ \bibinfo {author} {\bibfnamefont {J.~W.~F.}\ \bibnamefont
  {Valle}},\ }\href {\doibase 10.1016/S0927-6505(02)00115-9} {\bibfield
  {journal} {\bibinfo  {journal} {Astropart. Phys.}\ }\textbf {\bibinfo
  {volume} {18}},\ \bibinfo {pages} {287} (\bibinfo {year} {2002})},\ \Eprint
  {http://arxiv.org/abs/astro-ph/0111417} {arXiv:astro-ph/0111417} \BibitemShut
  {NoStop}%
\bibitem [{\citenamefont {Giovannini}(2003)}]{Giovannini:2003jw}%
  \BibitemOpen
  \bibfield  {author} {\bibinfo {author} {\bibfnamefont {M.}~\bibnamefont
  {Giovannini}},\ }\href {\doibase 10.1103/PhysRevD.67.123512} {\bibfield
  {journal} {\bibinfo  {journal} {Phys. Rev. D}\ }\textbf {\bibinfo {volume}
  {67}},\ \bibinfo {pages} {123512} (\bibinfo {year} {2003})},\ \Eprint
  {http://arxiv.org/abs/hep-ph/0301264} {arXiv:hep-ph/0301264} \BibitemShut
  {NoStop}%
\bibitem [{\citenamefont {Sami}\ and\ \citenamefont
  {Sahni}(2004)}]{Sami:2004xk}%
  \BibitemOpen
  \bibfield  {author} {\bibinfo {author} {\bibfnamefont {M.}~\bibnamefont
  {Sami}}\ and\ \bibinfo {author} {\bibfnamefont {V.}~\bibnamefont {Sahni}},\
  }\href {\doibase 10.1103/PhysRevD.70.083513} {\bibfield  {journal} {\bibinfo
  {journal} {Phys. Rev. D}\ }\textbf {\bibinfo {volume} {70}},\ \bibinfo
  {pages} {083513} (\bibinfo {year} {2004})},\ \Eprint
  {http://arxiv.org/abs/hep-th/0402086} {arXiv:hep-th/0402086} \BibitemShut
  {NoStop}%
\bibitem [{\citenamefont {Rosenfeld}\ and\ \citenamefont
  {Frieman}(2005)}]{Rosenfeld:2005mt}%
  \BibitemOpen
  \bibfield  {author} {\bibinfo {author} {\bibfnamefont {R.}~\bibnamefont
  {Rosenfeld}}\ and\ \bibinfo {author} {\bibfnamefont {J.~A.}\ \bibnamefont
  {Frieman}},\ }\href {\doibase 10.1088/1475-7516/2005/09/003} {\bibfield
  {journal} {\bibinfo  {journal} {JCAP}\ }\textbf {\bibinfo {volume} {09}},\
  \bibinfo {pages} {003} (\bibinfo {year} {2005})},\ \Eprint
  {http://arxiv.org/abs/astro-ph/0504191} {arXiv:astro-ph/0504191} \BibitemShut
  {NoStop}%
\bibitem [{\citenamefont {Rosenfeld}\ and\ \citenamefont
  {Frieman}(2007)}]{Rosenfeld:2006hs}%
  \BibitemOpen
  \bibfield  {author} {\bibinfo {author} {\bibfnamefont {R.}~\bibnamefont
  {Rosenfeld}}\ and\ \bibinfo {author} {\bibfnamefont {J.~A.}\ \bibnamefont
  {Frieman}},\ }\href {\doibase 10.1103/PhysRevD.75.043513} {\bibfield
  {journal} {\bibinfo  {journal} {Phys. Rev. D}\ }\textbf {\bibinfo {volume}
  {75}},\ \bibinfo {pages} {043513} (\bibinfo {year} {2007})},\ \Eprint
  {http://arxiv.org/abs/astro-ph/0611241} {arXiv:astro-ph/0611241} \BibitemShut
  {NoStop}%
\bibitem [{\citenamefont {Neupane}(2008)}]{Neupane:2007mu}%
  \BibitemOpen
  \bibfield  {author} {\bibinfo {author} {\bibfnamefont {I.~P.}\ \bibnamefont
  {Neupane}},\ }\href {\doibase 10.1088/0264-9381/25/12/125013} {\bibfield
  {journal} {\bibinfo  {journal} {Class. Quant. Grav.}\ }\textbf {\bibinfo
  {volume} {25}},\ \bibinfo {pages} {125013} (\bibinfo {year} {2008})},\
  \Eprint {http://arxiv.org/abs/0706.2654} {arXiv:0706.2654 [hep-th]}
  \BibitemShut {NoStop}%
\bibitem [{\citenamefont {Bento}\ \emph {et~al.}(2008)\citenamefont {Bento},
  \citenamefont {Felipe},\ and\ \citenamefont {Santos}}]{Bento:2008yx}%
  \BibitemOpen
  \bibfield  {author} {\bibinfo {author} {\bibfnamefont {M.~C.}\ \bibnamefont
  {Bento}}, \bibinfo {author} {\bibfnamefont {R.~G.}\ \bibnamefont {Felipe}}, \
  and\ \bibinfo {author} {\bibfnamefont {N.~M.~C.}\ \bibnamefont {Santos}},\
  }\href {\doibase 10.1103/PhysRevD.77.123512} {\bibfield  {journal} {\bibinfo
  {journal} {Phys. Rev. D}\ }\textbf {\bibinfo {volume} {77}},\ \bibinfo
  {pages} {123512} (\bibinfo {year} {2008})},\ \Eprint
  {http://arxiv.org/abs/0801.3450} {arXiv:0801.3450 [astro-ph]} \BibitemShut
  {NoStop}%
\bibitem [{\citenamefont {Bento}\ \emph {et~al.}(2009)\citenamefont {Bento},
  \citenamefont {Gonzalez~Felipe},\ and\ \citenamefont
  {Santos}}]{Bento:2009zz}%
  \BibitemOpen
  \bibfield  {author} {\bibinfo {author} {\bibfnamefont {M.~C.}\ \bibnamefont
  {Bento}}, \bibinfo {author} {\bibfnamefont {R.}~\bibnamefont
  {Gonzalez~Felipe}}, \ and\ \bibinfo {author} {\bibfnamefont {N.~M.~C.}\
  \bibnamefont {Santos}},\ }\href {\doibase 10.1142/S0217751X09045145}
  {\bibfield  {journal} {\bibinfo  {journal} {Int. J. Mod. Phys. A}\ }\textbf
  {\bibinfo {volume} {24}},\ \bibinfo {pages} {1639} (\bibinfo {year}
  {2009})}\BibitemShut {NoStop}%
\bibitem [{\citenamefont {Hossain}\ \emph {et~al.}(2014)\citenamefont
  {Hossain}, \citenamefont {Myrzakulov}, \citenamefont {Sami},\ and\
  \citenamefont {Saridakis}}]{Hossain:2014coa}%
  \BibitemOpen
  \bibfield  {author} {\bibinfo {author} {\bibfnamefont {M.~W.}\ \bibnamefont
  {Hossain}}, \bibinfo {author} {\bibfnamefont {R.}~\bibnamefont {Myrzakulov}},
  \bibinfo {author} {\bibfnamefont {M.}~\bibnamefont {Sami}}, \ and\ \bibinfo
  {author} {\bibfnamefont {E.~N.}\ \bibnamefont {Saridakis}},\ }\href {\doibase
  10.1103/PhysRevD.89.123513} {\bibfield  {journal} {\bibinfo  {journal} {Phys.
  Rev. D}\ }\textbf {\bibinfo {volume} {89}},\ \bibinfo {pages} {123513}
  (\bibinfo {year} {2014})},\ \Eprint {http://arxiv.org/abs/1404.1445}
  {arXiv:1404.1445 [gr-qc]} \BibitemShut {NoStop}%
\bibitem [{\citenamefont {de~Haro}\ \emph
  {et~al.}(2016{\natexlab{a}})\citenamefont {de~Haro}, \citenamefont
  {Amor\'os},\ and\ \citenamefont {Pan}}]{deHaro:2016hpl}%
  \BibitemOpen
  \bibfield  {author} {\bibinfo {author} {\bibfnamefont {J.}~\bibnamefont
  {de~Haro}}, \bibinfo {author} {\bibfnamefont {J.}~\bibnamefont {Amor\'os}}, \
  and\ \bibinfo {author} {\bibfnamefont {S.}~\bibnamefont {Pan}},\ }\href
  {\doibase 10.1103/PhysRevD.93.084018} {\bibfield  {journal} {\bibinfo
  {journal} {Phys. Rev. D}\ }\textbf {\bibinfo {volume} {93}},\ \bibinfo
  {pages} {084018} (\bibinfo {year} {2016}{\natexlab{a}})},\ \Eprint
  {http://arxiv.org/abs/1601.08175} {arXiv:1601.08175 [gr-qc]} \BibitemShut
  {NoStop}%
\bibitem [{\citenamefont {de~Haro}\ and\ \citenamefont
  {Elizalde}(2016)}]{deHaro:2016hsh}%
  \BibitemOpen
  \bibfield  {author} {\bibinfo {author} {\bibfnamefont {J.}~\bibnamefont
  {de~Haro}}\ and\ \bibinfo {author} {\bibfnamefont {E.}~\bibnamefont
  {Elizalde}},\ }\href {\doibase 10.1007/s10714-016-2072-z} {\bibfield
  {journal} {\bibinfo  {journal} {Gen. Rel. Grav.}\ }\textbf {\bibinfo {volume}
  {48}},\ \bibinfo {pages} {77} (\bibinfo {year} {2016})},\ \Eprint
  {http://arxiv.org/abs/1602.03433} {arXiv:1602.03433 [gr-qc]} \BibitemShut
  {NoStop}%
\bibitem [{\citenamefont {de~Haro}(2017)}]{deHaro:2016ftq}%
  \BibitemOpen
  \bibfield  {author} {\bibinfo {author} {\bibfnamefont {J.}~\bibnamefont
  {de~Haro}},\ }\href {\doibase 10.1007/s10714-016-2173-8} {\bibfield
  {journal} {\bibinfo  {journal} {Gen. Rel. Grav.}\ }\textbf {\bibinfo {volume}
  {49}},\ \bibinfo {pages} {6} (\bibinfo {year} {2017})},\ \Eprint
  {http://arxiv.org/abs/1602.07138} {arXiv:1602.07138 [gr-qc]} \BibitemShut
  {NoStop}%
\bibitem [{\citenamefont {de~Haro}\ \emph
  {et~al.}(2016{\natexlab{b}})\citenamefont {de~Haro}, \citenamefont
  {Amor\'os},\ and\ \citenamefont {Pan}}]{deHaro:2016cdm}%
  \BibitemOpen
  \bibfield  {author} {\bibinfo {author} {\bibfnamefont {J.}~\bibnamefont
  {de~Haro}}, \bibinfo {author} {\bibfnamefont {J.}~\bibnamefont {Amor\'os}}, \
  and\ \bibinfo {author} {\bibfnamefont {S.}~\bibnamefont {Pan}},\ }\href
  {\doibase 10.1103/PhysRevD.94.064060} {\bibfield  {journal} {\bibinfo
  {journal} {Phys. Rev. D}\ }\textbf {\bibinfo {volume} {94}},\ \bibinfo
  {pages} {064060} (\bibinfo {year} {2016}{\natexlab{b}})},\ \Eprint
  {http://arxiv.org/abs/1607.06726} {arXiv:1607.06726 [gr-qc]} \BibitemShut
  {NoStop}%
\bibitem [{\citenamefont {Geng}\ \emph {et~al.}(2017)\citenamefont {Geng},
  \citenamefont {Lee}, \citenamefont {Sami}, \citenamefont {Saridakis},\ and\
  \citenamefont {Starobinsky}}]{Geng:2017mic}%
  \BibitemOpen
  \bibfield  {author} {\bibinfo {author} {\bibfnamefont {C.-Q.}\ \bibnamefont
  {Geng}}, \bibinfo {author} {\bibfnamefont {C.-C.}\ \bibnamefont {Lee}},
  \bibinfo {author} {\bibfnamefont {M.}~\bibnamefont {Sami}}, \bibinfo {author}
  {\bibfnamefont {E.~N.}\ \bibnamefont {Saridakis}}, \ and\ \bibinfo {author}
  {\bibfnamefont {A.~A.}\ \bibnamefont {Starobinsky}},\ }\href {\doibase
  10.1088/1475-7516/2017/06/011} {\bibfield  {journal} {\bibinfo  {journal}
  {JCAP}\ }\textbf {\bibinfo {volume} {06}},\ \bibinfo {pages} {011} (\bibinfo
  {year} {2017})},\ \Eprint {http://arxiv.org/abs/1705.01329} {arXiv:1705.01329
  [gr-qc]} \BibitemShut {NoStop}%
\bibitem [{\citenamefont {Dimopoulos}\ and\ \citenamefont
  {Owen}(2017)}]{Dimopoulos:2017zvq}%
  \BibitemOpen
  \bibfield  {author} {\bibinfo {author} {\bibfnamefont {K.}~\bibnamefont
  {Dimopoulos}}\ and\ \bibinfo {author} {\bibfnamefont {C.}~\bibnamefont
  {Owen}},\ }\href {\doibase 10.1088/1475-7516/2017/06/027} {\bibfield
  {journal} {\bibinfo  {journal} {JCAP}\ }\textbf {\bibinfo {volume} {06}},\
  \bibinfo {pages} {027} (\bibinfo {year} {2017})},\ \Eprint
  {http://arxiv.org/abs/1703.00305} {arXiv:1703.00305 [gr-qc]} \BibitemShut
  {NoStop}%
\bibitem [{\citenamefont {Agarwal}\ \emph {et~al.}(2017)\citenamefont
  {Agarwal}, \citenamefont {Myrzakulov}, \citenamefont {Sami},\ and\
  \citenamefont {Singh}}]{Agarwal:2017wxo}%
  \BibitemOpen
  \bibfield  {author} {\bibinfo {author} {\bibfnamefont {A.}~\bibnamefont
  {Agarwal}}, \bibinfo {author} {\bibfnamefont {R.}~\bibnamefont {Myrzakulov}},
  \bibinfo {author} {\bibfnamefont {M.}~\bibnamefont {Sami}}, \ and\ \bibinfo
  {author} {\bibfnamefont {N.~K.}\ \bibnamefont {Singh}},\ }\href {\doibase
  10.1016/j.physletb.2017.04.066} {\bibfield  {journal} {\bibinfo  {journal}
  {Phys. Lett. B}\ }\textbf {\bibinfo {volume} {770}},\ \bibinfo {pages} {200}
  (\bibinfo {year} {2017})},\ \Eprint {http://arxiv.org/abs/1708.00156}
  {arXiv:1708.00156 [gr-qc]} \BibitemShut {NoStop}%
\bibitem [{\citenamefont {Haro}\ and\ \citenamefont
  {Pan}(2018)}]{Haro:2015ljc}%
  \BibitemOpen
  \bibfield  {author} {\bibinfo {author} {\bibfnamefont {J.}~\bibnamefont
  {Haro}}\ and\ \bibinfo {author} {\bibfnamefont {S.}~\bibnamefont {Pan}},\
  }\href {\doibase 10.1142/S0218271818500529} {\bibfield  {journal} {\bibinfo
  {journal} {Int. J. Mod. Phys. D}\ }\textbf {\bibinfo {volume} {27}},\
  \bibinfo {pages} {1850052} (\bibinfo {year} {2018})},\ \Eprint
  {http://arxiv.org/abs/1512.03033} {arXiv:1512.03033 [gr-qc]} \BibitemShut
  {NoStop}%
\bibitem [{\citenamefont {Haro}\ \emph {et~al.}(2020)\citenamefont {Haro},
  \citenamefont {Amor\'os},\ and\ \citenamefont {Pan}}]{Haro:2019peq}%
  \BibitemOpen
  \bibfield  {author} {\bibinfo {author} {\bibfnamefont {J.}~\bibnamefont
  {Haro}}, \bibinfo {author} {\bibfnamefont {J.}~\bibnamefont {Amor\'os}}, \
  and\ \bibinfo {author} {\bibfnamefont {S.}~\bibnamefont {Pan}},\ }\href
  {\doibase 10.1140/epjc/s10052-020-7950-6} {\bibfield  {journal} {\bibinfo
  {journal} {Eur. Phys. J. C}\ }\textbf {\bibinfo {volume} {80}},\ \bibinfo
  {pages} {404} (\bibinfo {year} {2020})},\ \Eprint
  {http://arxiv.org/abs/1908.01516} {arXiv:1908.01516 [gr-qc]} \BibitemShut
  {NoStop}%
\bibitem [{\citenamefont {Ahmad}\ \emph {et~al.}(2019)\citenamefont {Ahmad},
  \citenamefont {De~Felice}, \citenamefont {Jaman}, \citenamefont
  {Kuroyanagi},\ and\ \citenamefont {Sami}}]{Ahmad:2019jbm}%
  \BibitemOpen
  \bibfield  {author} {\bibinfo {author} {\bibfnamefont {S.}~\bibnamefont
  {Ahmad}}, \bibinfo {author} {\bibfnamefont {A.}~\bibnamefont {De~Felice}},
  \bibinfo {author} {\bibfnamefont {N.}~\bibnamefont {Jaman}}, \bibinfo
  {author} {\bibfnamefont {S.}~\bibnamefont {Kuroyanagi}}, \ and\ \bibinfo
  {author} {\bibfnamefont {M.}~\bibnamefont {Sami}},\ }\href {\doibase
  10.1103/PhysRevD.100.103525} {\bibfield  {journal} {\bibinfo  {journal}
  {Phys. Rev. D}\ }\textbf {\bibinfo {volume} {100}},\ \bibinfo {pages}
  {103525} (\bibinfo {year} {2019})},\ \Eprint
  {http://arxiv.org/abs/1908.03742} {arXiv:1908.03742 [gr-qc]} \BibitemShut
  {NoStop}%
\bibitem [{\citenamefont {Akrami}\ \emph {et~al.}(2021)\citenamefont {Akrami},
  \citenamefont {Casas}, \citenamefont {Deng},\ and\ \citenamefont
  {Vardanyan}}]{Akrami:2020zxw}%
  \BibitemOpen
  \bibfield  {author} {\bibinfo {author} {\bibfnamefont {Y.}~\bibnamefont
  {Akrami}}, \bibinfo {author} {\bibfnamefont {S.}~\bibnamefont {Casas}},
  \bibinfo {author} {\bibfnamefont {S.}~\bibnamefont {Deng}}, \ and\ \bibinfo
  {author} {\bibfnamefont {V.}~\bibnamefont {Vardanyan}},\ }\href {\doibase
  10.1088/1475-7516/2021/04/006} {\bibfield  {journal} {\bibinfo  {journal}
  {JCAP}\ }\textbf {\bibinfo {volume} {04}},\ \bibinfo {pages} {006} (\bibinfo
  {year} {2021})},\ \Eprint {http://arxiv.org/abs/2010.15822} {arXiv:2010.15822
  [astro-ph.CO]} \BibitemShut {NoStop}%
\bibitem [{\citenamefont {Arest\'e~Sal\'o}\ \emph
  {et~al.}(2021{\natexlab{a}})\citenamefont {Arest\'e~Sal\'o}, \citenamefont
  {Benisty}, \citenamefont {Guendelman},\ and\ \citenamefont
  {Haro}}]{AresteSalo:2021lmp}%
  \BibitemOpen
  \bibfield  {author} {\bibinfo {author} {\bibfnamefont {L.}~\bibnamefont
  {Arest\'e~Sal\'o}}, \bibinfo {author} {\bibfnamefont {D.}~\bibnamefont
  {Benisty}}, \bibinfo {author} {\bibfnamefont {E.~I.}\ \bibnamefont
  {Guendelman}}, \ and\ \bibinfo {author} {\bibfnamefont {J.~d.}\ \bibnamefont
  {Haro}},\ }\href {\doibase 10.1088/1475-7516/2021/07/007} {\bibfield
  {journal} {\bibinfo  {journal} {JCAP}\ }\textbf {\bibinfo {volume} {07}},\
  \bibinfo {pages} {007} (\bibinfo {year} {2021}{\natexlab{a}})},\ \Eprint
  {http://arxiv.org/abs/2102.09514} {arXiv:2102.09514 [astro-ph.CO]}
  \BibitemShut {NoStop}%
\bibitem [{\citenamefont {Arest\'e~Sal\'o}\ \emph
  {et~al.}(2021{\natexlab{b}})\citenamefont {Arest\'e~Sal\'o}, \citenamefont
  {Benisty}, \citenamefont {Guendelman},\ and\ \citenamefont
  {de~Haro}}]{AresteSalo:2021wgb}%
  \BibitemOpen
  \bibfield  {author} {\bibinfo {author} {\bibfnamefont {L.}~\bibnamefont
  {Arest\'e~Sal\'o}}, \bibinfo {author} {\bibfnamefont {D.}~\bibnamefont
  {Benisty}}, \bibinfo {author} {\bibfnamefont {E.~I.}\ \bibnamefont
  {Guendelman}}, \ and\ \bibinfo {author} {\bibfnamefont {J.}~\bibnamefont
  {de~Haro}},\ }\href {\doibase 10.1103/PhysRevD.103.123535} {\bibfield
  {journal} {\bibinfo  {journal} {Phys. Rev. D}\ }\textbf {\bibinfo {volume}
  {103}},\ \bibinfo {pages} {123535} (\bibinfo {year} {2021}{\natexlab{b}})},\
  \Eprint {http://arxiv.org/abs/2103.07892} {arXiv:2103.07892 [astro-ph.CO]}
  \BibitemShut {NoStop}%
\bibitem [{\citenamefont {de~Haro}\ and\ \citenamefont
  {Sal\'o}(2021)}]{deHaro:2021swo}%
  \BibitemOpen
  \bibfield  {author} {\bibinfo {author} {\bibfnamefont {J.}~\bibnamefont
  {de~Haro}}\ and\ \bibinfo {author} {\bibfnamefont {L.~A.}\ \bibnamefont
  {Sal\'o}},\ }\href {\doibase 10.3390/galaxies9040073} {\bibfield  {journal}
  {\bibinfo  {journal} {Galaxies}\ }\textbf {\bibinfo {volume} {9}},\ \bibinfo
  {pages} {73} (\bibinfo {year} {2021})},\ \Eprint
  {http://arxiv.org/abs/2108.11144} {arXiv:2108.11144 [gr-qc]} \BibitemShut
  {NoStop}%
\bibitem [{\citenamefont {Dimopoulos}\ \emph {et~al.}(2022)\citenamefont
  {Dimopoulos}, \citenamefont {Karam}, \citenamefont {S\'anchez~L\'opez},\ and\
  \citenamefont {Tomberg}}]{Dimopoulos:2022rdp}%
  \BibitemOpen
  \bibfield  {author} {\bibinfo {author} {\bibfnamefont {K.}~\bibnamefont
  {Dimopoulos}}, \bibinfo {author} {\bibfnamefont {A.}~\bibnamefont {Karam}},
  \bibinfo {author} {\bibfnamefont {S.}~\bibnamefont {S\'anchez~L\'opez}}, \
  and\ \bibinfo {author} {\bibfnamefont {E.}~\bibnamefont {Tomberg}},\ }\href
  {\doibase 10.1088/1475-7516/2022/10/076} {\bibfield  {journal} {\bibinfo
  {journal} {JCAP}\ }\textbf {\bibinfo {volume} {10}},\ \bibinfo {pages} {076}
  (\bibinfo {year} {2022})},\ \Eprint {http://arxiv.org/abs/2206.14117}
  {arXiv:2206.14117 [gr-qc]} \BibitemShut {NoStop}%
\bibitem [{\citenamefont {Alho}\ and\ \citenamefont
  {Uggla}(2023)}]{Alho:2023pkl}%
  \BibitemOpen
  \bibfield  {author} {\bibinfo {author} {\bibfnamefont {A.}~\bibnamefont
  {Alho}}\ and\ \bibinfo {author} {\bibfnamefont {C.}~\bibnamefont {Uggla}},\
  }\href {\doibase 10.1088/1475-7516/2023/11/083} {\bibfield  {journal}
  {\bibinfo  {journal} {JCAP}\ }\textbf {\bibinfo {volume} {11}},\ \bibinfo
  {pages} {083} (\bibinfo {year} {2023})},\ \Eprint
  {http://arxiv.org/abs/2306.15326} {arXiv:2306.15326 [gr-qc]} \BibitemShut
  {NoStop}%
\bibitem [{\citenamefont {Das}\ \emph {et~al.}(2023)\citenamefont {Das},
  \citenamefont {Jaman},\ and\ \citenamefont {Sami}}]{Das:2023nmm}%
  \BibitemOpen
  \bibfield  {author} {\bibinfo {author} {\bibfnamefont {B.}~\bibnamefont
  {Das}}, \bibinfo {author} {\bibfnamefont {N.}~\bibnamefont {Jaman}}, \ and\
  \bibinfo {author} {\bibfnamefont {M.}~\bibnamefont {Sami}},\ }\href {\doibase
  10.1103/PhysRevD.108.103510} {\bibfield  {journal} {\bibinfo  {journal}
  {Phys. Rev. D}\ }\textbf {\bibinfo {volume} {108}},\ \bibinfo {pages}
  {103510} (\bibinfo {year} {2023})},\ \Eprint
  {http://arxiv.org/abs/2307.12913} {arXiv:2307.12913 [gr-qc]} \BibitemShut
  {NoStop}%
\bibitem [{\citenamefont {Inagaki}\ and\ \citenamefont
  {Taniguchi}(2023)}]{Inagaki:2023mxv}%
  \BibitemOpen
  \bibfield  {author} {\bibinfo {author} {\bibfnamefont {T.}~\bibnamefont
  {Inagaki}}\ and\ \bibinfo {author} {\bibfnamefont {M.}~\bibnamefont
  {Taniguchi}},\ }\href@noop {} {\bibfield  {journal} {\bibinfo  {journal}
  {2312.11776}\ } (\bibinfo {year} {2023})}\BibitemShut {NoStop}%
\bibitem [{\citenamefont {Giar\`e}\ \emph {et~al.}(2024)\citenamefont
  {Giar\`e}, \citenamefont {Di~Valentino}, \citenamefont {Linder},\ and\
  \citenamefont {Specogna}}]{Giare:2024sdl}%
  \BibitemOpen
  \bibfield  {author} {\bibinfo {author} {\bibfnamefont {W.}~\bibnamefont
  {Giar\`e}}, \bibinfo {author} {\bibfnamefont {E.}~\bibnamefont
  {Di~Valentino}}, \bibinfo {author} {\bibfnamefont {E.~V.}\ \bibnamefont
  {Linder}}, \ and\ \bibinfo {author} {\bibfnamefont {E.}~\bibnamefont
  {Specogna}},\ }\href@noop {} {\bibfield  {journal} {\bibinfo  {journal}
  {2402.01560}\ } (\bibinfo {year} {2024})}\BibitemShut {NoStop}%
\bibitem [{\citenamefont {Kaneta}\ \emph
  {et~al.}(2022{\natexlab{a}})\citenamefont {Kaneta}, \citenamefont {Lee},\
  and\ \citenamefont {Oda}}]{Kaneta2022}%
  \BibitemOpen
  \bibfield  {author} {\bibinfo {author} {\bibfnamefont {K.}~\bibnamefont
  {Kaneta}}, \bibinfo {author} {\bibfnamefont {S.~M.}\ \bibnamefont {Lee}}, \
  and\ \bibinfo {author} {\bibfnamefont {K.}~\bibnamefont {Oda}},\ }\href
  {\doibase https://doi.org/10.1088/1475-7516/2022/09/018} {\bibfield
  {journal} {\bibinfo  {journal} {JCAP}\ }\textbf {\bibinfo {volume} {09}},\
  \bibinfo {pages} {018} (\bibinfo {year} {2022}{\natexlab{a}})},\ \Eprint
  {http://arxiv.org/abs/2206.10929} {arXiv:2206.10929 [astro-ph]} \BibitemShut
  {NoStop}%
\bibitem [{\citenamefont {Ema}\ \emph {et~al.}(2018)\citenamefont {Ema},
  \citenamefont {Nakayama},\ and\ \citenamefont {Tang}}]{Ema:2018ucl}%
  \BibitemOpen
  \bibfield  {author} {\bibinfo {author} {\bibfnamefont {Y.}~\bibnamefont
  {Ema}}, \bibinfo {author} {\bibfnamefont {K.}~\bibnamefont {Nakayama}}, \
  and\ \bibinfo {author} {\bibfnamefont {Y.}~\bibnamefont {Tang}},\ }\href
  {\doibase 10.1007/JHEP09(2018)135} {\bibfield  {journal} {\bibinfo  {journal}
  {JHEP}\ }\textbf {\bibinfo {volume} {09}},\ \bibinfo {pages} {135} (\bibinfo
  {year} {2018})},\ \Eprint {http://arxiv.org/abs/1804.07471} {arXiv:1804.07471
  [hep-ph]} \BibitemShut {NoStop}%
\bibitem [{\citenamefont {Kolb}\ and\ \citenamefont
  {Long}(2023)}]{Kolb:2023ydq}%
  \BibitemOpen
  \bibfield  {author} {\bibinfo {author} {\bibfnamefont {E.~W.}\ \bibnamefont
  {Kolb}}\ and\ \bibinfo {author} {\bibfnamefont {A.~J.}\ \bibnamefont
  {Long}},\ }\href@noop {} {\bibfield  {journal} {\bibinfo  {journal}
  {2312.09042}\ } (\bibinfo {year} {2023})}\BibitemShut {NoStop}%
\bibitem [{\citenamefont {Cléry}\ \emph {et~al.}(2022)\citenamefont {Cléry},
  \citenamefont {Mambrini}, \citenamefont {Olive},\ and\ \citenamefont
  {Verner}}]{Clery2022}%
  \BibitemOpen
  \bibfield  {author} {\bibinfo {author} {\bibfnamefont {S.}~\bibnamefont
  {Cléry}}, \bibinfo {author} {\bibfnamefont {Y.}~\bibnamefont {Mambrini}},
  \bibinfo {author} {\bibfnamefont {K.~A.}\ \bibnamefont {Olive}}, \ and\
  \bibinfo {author} {\bibfnamefont {S.}~\bibnamefont {Verner}},\ }\href
  {\doibase 10.1103/PhysRevD.105.075005} {\bibfield  {journal} {\bibinfo
  {journal} {Phys. Rev. D}\ }\textbf {\bibinfo {volume} {105}},\ \bibinfo
  {pages} {07005} (\bibinfo {year} {2022})},\ \Eprint
  {http://arxiv.org/abs/2112.15214} {arXiv:2112.15214 [hep-ph]} \BibitemShut
  {NoStop}%
\bibitem [{\citenamefont {M.~A. G.~Garcia}\ and\ \citenamefont
  {Olive}(2021)}]{Kaneta2021}%
  \BibitemOpen
  \bibfield  {author} {\bibinfo {author} {\bibfnamefont {Y.~M.}\ \bibnamefont
  {M.~A. G.~Garcia}, \bibfnamefont {K.~Kaneta}}\ and\ \bibinfo {author}
  {\bibfnamefont {K.~A.}\ \bibnamefont {Olive}},\ }\href {\doibase
  https://doi.org/10.1088/1475-7516/2021/04/012} {\bibfield  {journal}
  {\bibinfo  {journal} {JCAP}\ }\textbf {\bibinfo {volume} {04}},\ \bibinfo
  {pages} {012} (\bibinfo {year} {2021})},\ \Eprint
  {http://arxiv.org/abs/2012.10756} {arXiv:2012.10756 [hep-ph]} \BibitemShut
  {NoStop}%
\bibitem [{\citenamefont {M.~A. G.~Garcia}\ and\ \citenamefont
  {Olive}(2020)}]{Kaneta2020}%
  \BibitemOpen
  \bibfield  {author} {\bibinfo {author} {\bibfnamefont {Y.~M.}\ \bibnamefont
  {M.~A. G.~Garcia}, \bibfnamefont {K.~Kaneta}}\ and\ \bibinfo {author}
  {\bibfnamefont {K.~A.}\ \bibnamefont {Olive}},\ }\href {\doibase
  https://doi.org/10.1103/PhysRevD.101.123507} {\bibfield  {journal} {\bibinfo
  {journal} {Phys. Rev. D}\ }\textbf {\bibinfo {volume} {101}},\ \bibinfo
  {pages} {123507} (\bibinfo {year} {2020})},\ \Eprint
  {http://arxiv.org/abs/2004.08404} {arXiv:2004.08404 [hep-ph]} \BibitemShut
  {NoStop}%
\bibitem [{\citenamefont {Drewes}\ \emph {et~al.}(2017)\citenamefont {Drewes},
  \citenamefont {Kang},\ and\ \citenamefont {Mun}}]{Drewes:2017fmn}%
  \BibitemOpen
  \bibfield  {author} {\bibinfo {author} {\bibfnamefont {M.}~\bibnamefont
  {Drewes}}, \bibinfo {author} {\bibfnamefont {J.~U.}\ \bibnamefont {Kang}}, \
  and\ \bibinfo {author} {\bibfnamefont {U.~R.}\ \bibnamefont {Mun}},\ }\href
  {\doibase 10.1007/JHEP11(2017)072} {\bibfield  {journal} {\bibinfo  {journal}
  {JHEP}\ }\textbf {\bibinfo {volume} {11}},\ \bibinfo {pages} {072} (\bibinfo
  {year} {2017})},\ \Eprint {http://arxiv.org/abs/1708.01197} {arXiv:1708.01197
  [astro-ph.CO]} \BibitemShut {NoStop}%
\bibitem [{\citenamefont {Turner}(1983)}]{Turner:1983he}%
  \BibitemOpen
  \bibfield  {author} {\bibinfo {author} {\bibfnamefont {M.~S.}\ \bibnamefont
  {Turner}},\ }\href {\doibase 10.1103/PhysRevD.28.1243} {\bibfield  {journal}
  {\bibinfo  {journal} {Phys. Rev. D}\ }\textbf {\bibinfo {volume} {28}},\
  \bibinfo {pages} {1243} (\bibinfo {year} {1983})}\BibitemShut {NoStop}%
\bibitem [{\citenamefont {Birrell}\ and\ \citenamefont
  {Davies}(1982)}]{birrell_davies_1982}%
  \BibitemOpen
  \bibfield  {author} {\bibinfo {author} {\bibfnamefont {N.~D.}\ \bibnamefont
  {Birrell}}\ and\ \bibinfo {author} {\bibfnamefont {P.~C.~W.}\ \bibnamefont
  {Davies}},\ }\href {\doibase 10.1017/CBO9780511622632} {\emph {\bibinfo
  {title} {Quantum Fields in Curved Space}}},\ Cambridge Monographs on
  Mathematical Physics\ (\bibinfo  {publisher} {Cambridge University Press},\
  \bibinfo {year} {1982})\BibitemShut {NoStop}%
\bibitem [{\citenamefont {Zeldovich}\ and\ \citenamefont
  {Starobinsky}(1971)}]{Zeldovich:1971mw}%
  \BibitemOpen
  \bibfield  {author} {\bibinfo {author} {\bibfnamefont {Y.~B.}\ \bibnamefont
  {Zeldovich}}\ and\ \bibinfo {author} {\bibfnamefont {A.~A.}\ \bibnamefont
  {Starobinsky}},\ }\href@noop {} {\bibfield  {journal} {\bibinfo  {journal}
  {Zh. Eksp. Teor. Fiz.}\ }\textbf {\bibinfo {volume} {61}},\ \bibinfo {pages}
  {2161} (\bibinfo {year} {1971})}\BibitemShut {NoStop}%
\bibitem [{\citenamefont {Ema}\ \emph {et~al.}(2015)\citenamefont {Ema},
  \citenamefont {Jinno}, \citenamefont {Mukaida},\ and\ \citenamefont
  {Nakayama}}]{Ema:2015dka}%
  \BibitemOpen
  \bibfield  {author} {\bibinfo {author} {\bibfnamefont {Y.}~\bibnamefont
  {Ema}}, \bibinfo {author} {\bibfnamefont {R.}~\bibnamefont {Jinno}}, \bibinfo
  {author} {\bibfnamefont {K.}~\bibnamefont {Mukaida}}, \ and\ \bibinfo
  {author} {\bibfnamefont {K.}~\bibnamefont {Nakayama}},\ }\href {\doibase
  10.1088/1475-7516/2015/05/038} {\bibfield  {journal} {\bibinfo  {journal}
  {JCAP}\ }\textbf {\bibinfo {volume} {05}},\ \bibinfo {pages} {038} (\bibinfo
  {year} {2015})},\ \Eprint {http://arxiv.org/abs/1502.02475} {arXiv:1502.02475
  [hep-ph]} \BibitemShut {NoStop}%
\bibitem [{\citenamefont {Chung}\ \emph {et~al.}(2019)\citenamefont {Chung},
  \citenamefont {Kolb},\ and\ \citenamefont {Long}}]{Chung:2018ayg}%
  \BibitemOpen
  \bibfield  {author} {\bibinfo {author} {\bibfnamefont {D.~J.~H.}\
  \bibnamefont {Chung}}, \bibinfo {author} {\bibfnamefont {E.~W.}\ \bibnamefont
  {Kolb}}, \ and\ \bibinfo {author} {\bibfnamefont {A.~J.}\ \bibnamefont
  {Long}},\ }\href {\doibase 10.1007/JHEP01(2019)189} {\bibfield  {journal}
  {\bibinfo  {journal} {JHEP}\ }\textbf {\bibinfo {volume} {01}},\ \bibinfo
  {pages} {189} (\bibinfo {year} {2019})},\ \Eprint
  {http://arxiv.org/abs/1812.00211} {arXiv:1812.00211 [hep-ph]} \BibitemShut
  {NoStop}%
\bibitem [{\citenamefont {Ellis}\ \emph {et~al.}(1982)\citenamefont {Ellis},
  \citenamefont {Linde},\ and\ \citenamefont {Nanopoulos}}]{Ellis:1982yb}%
  \BibitemOpen
  \bibfield  {author} {\bibinfo {author} {\bibfnamefont {J.~R.}\ \bibnamefont
  {Ellis}}, \bibinfo {author} {\bibfnamefont {A.~D.}\ \bibnamefont {Linde}}, \
  and\ \bibinfo {author} {\bibfnamefont {D.~V.}\ \bibnamefont {Nanopoulos}},\
  }\href {\doibase 10.1016/0370-2693(82)90601-3} {\bibfield  {journal}
  {\bibinfo  {journal} {Phys. Lett. B}\ }\textbf {\bibinfo {volume} {118}},\
  \bibinfo {pages} {59} (\bibinfo {year} {1982})}\BibitemShut {NoStop}%
\bibitem [{\citenamefont {Flores}\ and\ \citenamefont
  {Perez-Gonzalez}(2024)}]{Flores:2024lzv}%
  \BibitemOpen
  \bibfield  {author} {\bibinfo {author} {\bibfnamefont {M.~M.}\ \bibnamefont
  {Flores}}\ and\ \bibinfo {author} {\bibfnamefont {Y.~F.}\ \bibnamefont
  {Perez-Gonzalez}},\ }\href {\doibase 10.1103/PhysRevD.109.115017} {\bibfield
  {journal} {\bibinfo  {journal} {Phys. Rev. D}\ }\textbf {\bibinfo {volume}
  {109}},\ \bibinfo {pages} {115017} (\bibinfo {year} {2024})},\ \Eprint
  {http://arxiv.org/abs/2404.06530} {arXiv:2404.06530 [hep-ph]} \BibitemShut
  {NoStop}%
\bibitem [{\citenamefont {Ling}\ and\ \citenamefont
  {Long}(2021)}]{Ling:2021zlj}%
  \BibitemOpen
  \bibfield  {author} {\bibinfo {author} {\bibfnamefont {S.}~\bibnamefont
  {Ling}}\ and\ \bibinfo {author} {\bibfnamefont {A.~J.}\ \bibnamefont
  {Long}},\ }\href {\doibase 10.1103/PhysRevD.103.103532} {\bibfield  {journal}
  {\bibinfo  {journal} {Phys. Rev. D}\ }\textbf {\bibinfo {volume} {103}},\
  \bibinfo {pages} {103532} (\bibinfo {year} {2021})},\ \Eprint
  {http://arxiv.org/abs/2101.11621} {arXiv:2101.11621 [astro-ph.CO]}
  \BibitemShut {NoStop}%
\bibitem [{\citenamefont {Shtanov}\ \emph {et~al.}(1995)\citenamefont
  {Shtanov}, \citenamefont {Traschen},\ and\ \citenamefont
  {Brandenberger}}]{Shtanov:1994ce}%
  \BibitemOpen
  \bibfield  {author} {\bibinfo {author} {\bibfnamefont {Y.}~\bibnamefont
  {Shtanov}}, \bibinfo {author} {\bibfnamefont {J.~H.}\ \bibnamefont
  {Traschen}}, \ and\ \bibinfo {author} {\bibfnamefont {R.~H.}\ \bibnamefont
  {Brandenberger}},\ }\href {\doibase 10.1103/PhysRevD.51.5438} {\bibfield
  {journal} {\bibinfo  {journal} {Phys. Rev. D}\ }\textbf {\bibinfo {volume}
  {51}},\ \bibinfo {pages} {5438} (\bibinfo {year} {1995})},\ \Eprint
  {http://arxiv.org/abs/hep-ph/9407247} {arXiv:hep-ph/9407247} \BibitemShut
  {NoStop}%
\bibitem [{\citenamefont {Clery}\ \emph {et~al.}(2022)\citenamefont {Clery},
  \citenamefont {Mambrini}, \citenamefont {Olive},\ and\ \citenamefont
  {Verner}}]{Clery:2021bwz}%
  \BibitemOpen
  \bibfield  {author} {\bibinfo {author} {\bibfnamefont {S.}~\bibnamefont
  {Clery}}, \bibinfo {author} {\bibfnamefont {Y.}~\bibnamefont {Mambrini}},
  \bibinfo {author} {\bibfnamefont {K.~A.}\ \bibnamefont {Olive}}, \ and\
  \bibinfo {author} {\bibfnamefont {S.}~\bibnamefont {Verner}},\ }\href
  {\doibase 10.1103/PhysRevD.105.075005} {\bibfield  {journal} {\bibinfo
  {journal} {Phys. Rev. D}\ }\textbf {\bibinfo {volume} {105}},\ \bibinfo
  {pages} {075005} (\bibinfo {year} {2022})},\ \Eprint
  {http://arxiv.org/abs/2112.15214} {arXiv:2112.15214 [hep-ph]} \BibitemShut
  {NoStop}%
\bibitem [{\citenamefont {Garcia}\ \emph {et~al.}(2021)\citenamefont {Garcia},
  \citenamefont {Kaneta}, \citenamefont {Mambrini},\ and\ \citenamefont
  {Olive}}]{Garcia:2020wiy}%
  \BibitemOpen
  \bibfield  {author} {\bibinfo {author} {\bibfnamefont {M.~A.~G.}\
  \bibnamefont {Garcia}}, \bibinfo {author} {\bibfnamefont {K.}~\bibnamefont
  {Kaneta}}, \bibinfo {author} {\bibfnamefont {Y.}~\bibnamefont {Mambrini}}, \
  and\ \bibinfo {author} {\bibfnamefont {K.~A.}\ \bibnamefont {Olive}},\ }\href
  {\doibase 10.1088/1475-7516/2021/04/012} {\bibfield  {journal} {\bibinfo
  {journal} {JCAP}\ }\textbf {\bibinfo {volume} {04}},\ \bibinfo {pages} {012}
  (\bibinfo {year} {2021})},\ \Eprint {http://arxiv.org/abs/2012.10756}
  {arXiv:2012.10756 [hep-ph]} \BibitemShut {NoStop}%
\bibitem [{\citenamefont {Garcia}\ \emph {et~al.}(2020)\citenamefont {Garcia},
  \citenamefont {Kaneta}, \citenamefont {Mambrini},\ and\ \citenamefont
  {Olive}}]{Garcia:2020eof}%
  \BibitemOpen
  \bibfield  {author} {\bibinfo {author} {\bibfnamefont {M.~A.~G.}\
  \bibnamefont {Garcia}}, \bibinfo {author} {\bibfnamefont {K.}~\bibnamefont
  {Kaneta}}, \bibinfo {author} {\bibfnamefont {Y.}~\bibnamefont {Mambrini}}, \
  and\ \bibinfo {author} {\bibfnamefont {K.~A.}\ \bibnamefont {Olive}},\ }\href
  {\doibase 10.1103/PhysRevD.101.123507} {\bibfield  {journal} {\bibinfo
  {journal} {Phys. Rev. D}\ }\textbf {\bibinfo {volume} {101}},\ \bibinfo
  {pages} {123507} (\bibinfo {year} {2020})},\ \Eprint
  {http://arxiv.org/abs/2004.08404} {arXiv:2004.08404 [hep-ph]} \BibitemShut
  {NoStop}%
\bibitem [{\citenamefont {Akrami}\ \emph {et~al.}(2020)\citenamefont {Akrami}
  \emph {et~al.}}]{Planck:2018jri}%
  \BibitemOpen
  \bibfield  {author} {\bibinfo {author} {\bibfnamefont {Y.}~\bibnamefont
  {Akrami}} \emph {et~al.} (\bibinfo {collaboration} {Planck}),\ }\href
  {\doibase 10.1051/0004-6361/201833887} {\bibfield  {journal} {\bibinfo
  {journal} {Astron. Astrophys.}\ }\textbf {\bibinfo {volume} {641}},\ \bibinfo
  {pages} {A10} (\bibinfo {year} {2020})},\ \Eprint
  {http://arxiv.org/abs/1807.06211} {arXiv:1807.06211 [astro-ph.CO]}
  \BibitemShut {NoStop}%
\bibitem [{\citenamefont {Kaneta}\ \emph
  {et~al.}(2022{\natexlab{b}})\citenamefont {Kaneta}, \citenamefont {Lee},\
  and\ \citenamefont {Oda}}]{Kaneta:2022gug}%
  \BibitemOpen
  \bibfield  {author} {\bibinfo {author} {\bibfnamefont {K.}~\bibnamefont
  {Kaneta}}, \bibinfo {author} {\bibfnamefont {S.~M.}\ \bibnamefont {Lee}}, \
  and\ \bibinfo {author} {\bibfnamefont {K.-y.}\ \bibnamefont {Oda}},\ }\href
  {\doibase 10.1088/1475-7516/2022/09/018} {\bibfield  {journal} {\bibinfo
  {journal} {JCAP}\ }\textbf {\bibinfo {volume} {09}},\ \bibinfo {pages} {018}
  (\bibinfo {year} {2022}{\natexlab{b}})},\ \Eprint
  {http://arxiv.org/abs/2206.10929} {arXiv:2206.10929 [astro-ph.CO]}
  \BibitemShut {NoStop}%
\end{thebibliography}%

\end{document}